\documentclass[fleqn,usenatbib]{mnras}

\usepackage{newtxtext,newtxmath}
\usepackage[T1]{fontenc}

\DeclareRobustCommand{\VAN}[3]{#2}
\let\VANthebibliography\thebibliography
\def\thebibliography{\DeclareRobustCommand{\VAN}[3]{##3}\VANthebibliography}

\usepackage{graphicx}	% Including figure files
\usepackage{amsmath}	% Advanced maths commands
\usepackage{newtxtext,newtxmath}
\usepackage{pdfpages}
\usepackage{float}
\usepackage{tabularx}
\usepackage{comment}
\usepackage[normalem]{ulem}
\usepackage[mathlines, pagewise]{lineno}
\usepackage{colortbl}

\usepackage{subcaption}
\usepackage{graphicx}	% Including figure files
\usepackage{amsmath}	% Advanced maths commands
\usepackage{hyperref}
\usepackage{physics}
\usepackage{arydshln}

\title[Large-scale AGN feedback in XFABLE]{The case for large-scale AGN feedback in galaxy formation simulations: insights from XFABLE}

\author[L. Bigwood et al.]{
Leah Bigwood,$^{1,2}$\thanks{E-mail: lmb224@cam.ac.uk}
Martin A. Bourne,$^{1,2,3}$
Vid Iršič,$^{2,4,5,6,7}$
Alexandra Amon,$^{8,2}$
Debora Sijacki$^{1,2}$
\\
$^{1}$Institute of Astronomy, University of Cambridge, Madingley Road, Cambridge, CB3 0HA, UK\\
$^{2}$Kavli Institute for Cosmology (KICC), University of Cambridge, Madingley Road, Cambridge CB3 0HA, UK\\
$^{3}$Centre for Astrophysics Research, Department of Physics, Astronomy and Mathematics, University of Hertfordshire, College Lane, Hatfield,
AL10 9AB, UK\\
$^{4}$SISSA - International School for Advanced Studies, Via Bonomea 265I-34136 Trieste, Italy\\
$^{5}$IFPU, Institute for Fundamental Physics of the Universe, Via Beirut 2, I-34151 Trieste, Italy\\
$^{6}$INFN, Sezione di Trieste, Via Valerio 2, I-34127 Trieste, Italy\\
$^{7}$INAF - Osservatorio Astronomico di Trieste, Via G. B. Tiepolo 11, I-34143 Trieste, Italy\\
$^{8}$Department of Astrophysical Sciences, Princeton University, Peyton Hall, Princeton, NJ 08544, USA
}

\date{submitted to MNRAS}

\pubyear{2024}

\begin{document}
\label{firstpage}
\pagerange{\pageref{firstpage}--\pageref{lastpage}}
\maketitle
 
\begin{abstract}
While cosmological simulations of galaxy formation have reached maturity, able to reproduce many fundamental galaxy and halo properties, no consensus has yet been reached on the impact of `baryonic feedback' on the non-linear matter power spectrum. This severely limits the precision of (and potentially biases) small-scale cosmological constraints obtained from weak lensing and galaxy surveys. Recent observational evidence indicates that `baryonic feedback' may be more extreme than commonly assumed in current cosmological hydrodynamical simulations. In this paper, we therefore explore a range of empirical AGN feedback models, within the FABLE simulation suite, with different parameterizations as a function of cosmic time, host halo properties, and/or spatial location where feedback energy is thermalized. We demonstrate that an AGN radio-mode feedback acting in a larger population of black holes, with jets thermalizing at relatively large cluster-centric distances, as exemplified by our XFABLE model, is in good agreement with the latest weak lensing + kSZ constraints across all k-scales. Furthermore, XFABLE maintains good agreement with the galaxy stellar mass function, gas fraction measurements, and all key galaxy group and cluster properties, including scaling relations and ICM radial profiles. Our work highlights the pressing need to model black hole accretion and feedback physics with a greater level of realism, including relativistic, magnetized jets in full cosmological simulations. Finally, we discuss how a range of complementary observational probes in the near future will enable us to constrain AGN feedback models, and therefore reduce `baryonic feedback' modelling uncertainty for the upcoming era of large cosmological surveys. 
\end{abstract}


%%%%%%%%%%%%%%%%%%%%%%%%%%%%%%%%%%%%%%%%%%%%%%%%%%

%%%%%%%%%%%%%%%%% BODY OF PAPER %%%%%%%%%%%%%%%%%%

\section{Introduction}\label{sec:intro}

The $\Lambda$CDM model of cosmology has proven extremely successful when stress-tested against observations over a remarkable span of cosmic history, from low redshift measurements of the expansion history probed by baryonic acoustic oscillations \citep[e.g.][]{DESIBAO2024} and growth of structure \citep[e.g.][]{DESIRSD2024} to the accurate measurements of anisotropies and lensing of the cosmic microwave background \citep[CMB;][]{Planck2018, Pan2023, Madhavacheri2024}. The $\Lambda$CDM model assumes the Universe comprises three main components: dark energy in the form of a cosmological constant ($\Lambda$), which drives an accelerated expansion of the Universe, cold dark matter (CDM), which interacts only gravitationally, and the ordinary baryonic matter, which is, in principle, observable. However, it is challenging to map baryons onto the underlying dark matter distribution due to the complex physical processes that regulate baryons' properties, such as gas radiative cooling and heating, star formation and associated stellar feedback, as well as black hole accretion and feedback physics -- which are often referred to by the umbrella term `baryonic feedback' \citep{Semboloni2011, vanDaalen2011, Vogelsberger2020}. These processes influence the total matter distribution through the gas heating and cooling, the ejection and redistribution of gas (within and) beyond the virial radii of groups and clusters, and the back-reaction on the CDM distribution. Therefore, tests of the $\Lambda$CDM model on relatively small, non-linear scales, such as through measurements of weak galaxy lensing, require accurate models of how `baryonic feedback' impacts the overall matter distribution \citep{Chisari2019, Schneider2019, Amon2022, Preston2023}.

Hydrodynamical simulations have implemented active galactic nucleus (AGN) feedback models to demonstrate that it is necessary to regulate the star formation rate in massive galaxies and prevent overcooling of gas in groups and clusters (see \citet{Sijacki2007}, \citet{McCarthy2010}, \citet{Fabjan2010} for early studies, and a recent review by \citet{BourneYang2023}).  The impact of AGN feedback on the matter distribution is to suppress the matter power spectrum by up to $\sim 30\%$ on the non-linear scales with respect to a dark matter-only scenario; however, the simulations significantly differ in their predictions for the amplitude and scale dependence of the suppression at scales $0.1~ h\,\mathrm{Mpc}^{-1}<k<10~h\,\mathrm{Mpc}^{-1}$ \citep{McCarthy2017, Springel2018, Chisari2019, vandaalen:2020, Schaye2023, Pakmor2023, Schaller2024, Martin-Alvarez2024, Gebhardt2024}.
These inconsistent predictions can be attributed to a number of factors.  
The modelling choices of the astrophysical feedback processes can have a large outcome on the predicted matter power spectrum suppression \citep{vanDaalen2011, vandaalen:2020, Pandey2023}, in addition to the adopted box size, resolution and hydrodynamical scheme by different studies. Despite the differences in feedback modelling (and the resulting matter power spectrum suppression),  the state-of-the-art hydrodynamical simulations give reasonably similar matches to other observations, such as the galaxy stellar mass function (GSMF) and X-ray observations of cluster gas mass fractions, albeit noting a significant range of observed gas fractions at a given halo mass. 
  
The current level of uncertainty in feedback modelling stands as the limiting factor for the precision of cosmological constraints from weak galaxy lensing \citep{AmonDES2022, DESKIDS2023}.  Beyond that, `baryonic feedback' may have a role in the so-called `$S_8$ tension' associated with the $\Lambda$CDM model.  Over the last decade, discrepancies in the measurements of the clustering amplitude parameter\footnote{Here, $\Omega_{\rm m}$ is the ratio of the present day matter density to the
critical density of the Universe and $\sigma_8$ is the root mean square linear amplitude of
the matter fluctuation spectrum in spheres of radius $8~h^{-1}$Mpc extrapolated to the present day.}, $S_8=\sigma_8(\Omega_{\rm m} / 0.3)^{0.5}$  by weak galaxy lensing surveys with respect to 
\textit{Planck} $\Lambda$CDM best-fit cosmology have persisted. \citet{Amon2022} and \citet{Preston2023} hypothesized that `baryonic feedback' could be responsible if it had a stronger impact on the non-linear matter distribution than that predicted by many state-of-the-art hydrodynamical simulations. Indeed, the proposal for more extreme feedback has been supported by recent evidence from cosmic shear and stacked kinetic Sunyaev-Zeldovich measurements \citep[kSZ,][]{Bigwood2024}, measurements of the kSZ effect \citep{McCarthy2024, Hadzhiyska2024}, cross-correlations of weak lensing with diffuse X-ray and thermal SZ \citep[tSZ,][]{Ferreira2024, LaPosta2024} and measurements of the tSZ effect, including the power spectrum (\citealt{Ruan2015,Crichton2016,Chowdhury2017}, Efstathiou \& McCarthy in prep.). However, it remains a challenge to identify a physical mechanism to produce stronger feedback that remains in accord with galaxy group and cluster X-ray data, not only in terms of gas fractions but also of spatially resolved intracluster medium (ICM) properties.

Feedback effects are one of a number of ‘sub-grid’ processes that occur below the resolution scale of cosmological simulations and are therefore modelled through empirical prescriptions that aim to capture the complex small-scale physics. In a widely adopted but simplistic picture, AGN feedback is often modelled using two primary modes, dependent on the accretion rate of the supermassive black hole (SMBH), or more specifically the Eddington ratio \citep{Sijacki2007}. The quasar-mode (or `thermal-mode') acts at high SMBH accretion rates and is often attributed to high-velocity quasar-driven winds directly impacting the host galaxy and circumgalactic medium (CGM) \citep{Harrison2014, Mullaney2013}. The radio-mode (or `kinetic-mode') is instead associated with inefficient SMBH accretion and launches AGN jets impacting the CGM and the ICM. These jets inflate expanding bubbles, displacing the hot gas and leaving cavities and shock-fronts detectable in X-ray images of galaxy groups and clusters \citep{Fabian2012}. Some hydrodynamical simulations distinguish between the two modes by imposing both thermal and kinetic outflows, with others opting for a purely thermal feedback model, regardless of the SMBH accretion state. It should be noted however that observationally this picture is less clear, both in terms of the roles that different forms of feedback (i.e., radiation, winds and jets) play in galaxy evolution and under what conditions they are produced, with radio jets being observed in systems undergoing accretion at high, as well as low Eddington ratios \citep[see e.g.][for reviews]{Hardcastle2020, Hlavacek-Larrond2022}.

The choice of sub-grid parameters utilised to model feedback processes in simulations also plays a role in the varying predictions of the matter power spectrum suppression. 
Although physical arguments can be used to narrow the range of plausible parameter values of the feedback models, they are typically poorly constrained and often resolution-dependent. As such, calibrating to external observational datasets is required.  Observations of the galaxy stellar mass function, star formation history, and stellar sizes are all frequently used to guide hydrodynamical simulations. However, the hot gas mass fractions of groups and clusters, measured using X-ray observations, are generally the key benchmark of the AGN feedback model efficacy \citet[e.g.][]{McCarthy2017, Henden2018, Schaye2023}.  
                           
In this paper, we explore the potential for more extreme AGN feedback in hydrodynamical simulations. Using the FABLE simulation framework \citep{Henden2018, Henden2019, Henden2020} as a test-bed, we explore an extensive number of modifications to the AGN feedback model in FABLE, to demonstrate that it is possible to produce the more extreme matter power spectrum suppression required to resolve the $S_8$ tension and remain consistent with new observational weak lensing, tSZ and kSZ constraints, whilst still maintaining consistency with key galaxy and cluster observables typically used to calibrate simulations, as exemplified by our new empirical AGN feedback model, XFABLE. We stress that it is crucial that the potential degeneracies within hydrodynamical simulations are understood if weak lensing analyses are to continue utilising them to calibrate their `baryonic feedback' mitigation.

The paper is structured as follows. In Section~\ref{sec:comparehydrosims} we motivate the study by discussing the spread in the suppression of the matter power spectrum predicted by a range of hydrodynamical simulations, despite each providing a good fit to GSMF and cluster gas mass fractions observations. Section~\ref{sec:methods} describes the basic properties of the simulation suite we utilise. It also details the computation of a number of galaxy, group and cluster properties from the simulation outputs to allow for comparison to observations. Section~\ref{sec:modelsoverview} describes the key modifications to the FABLE AGN feedback model we test in this work, including XFABLE, and the motivations behind the models. In Section~\ref{sec: obs} we compare the predictions made by each of our key simulation boxes to a range of observational measurements.  Finally, in Section~\ref{sec:outlook} we summarise our findings from the simulation suite and discuss the outlook for XFABLE.

\section{The uncertainty in simulating baryonic feedback and implications for $S_{\rm 8}$ tension}\label{sec:comparehydrosims}

\begin{figure*}
\centering
\includegraphics[height=5.5cm,keepaspectratio]{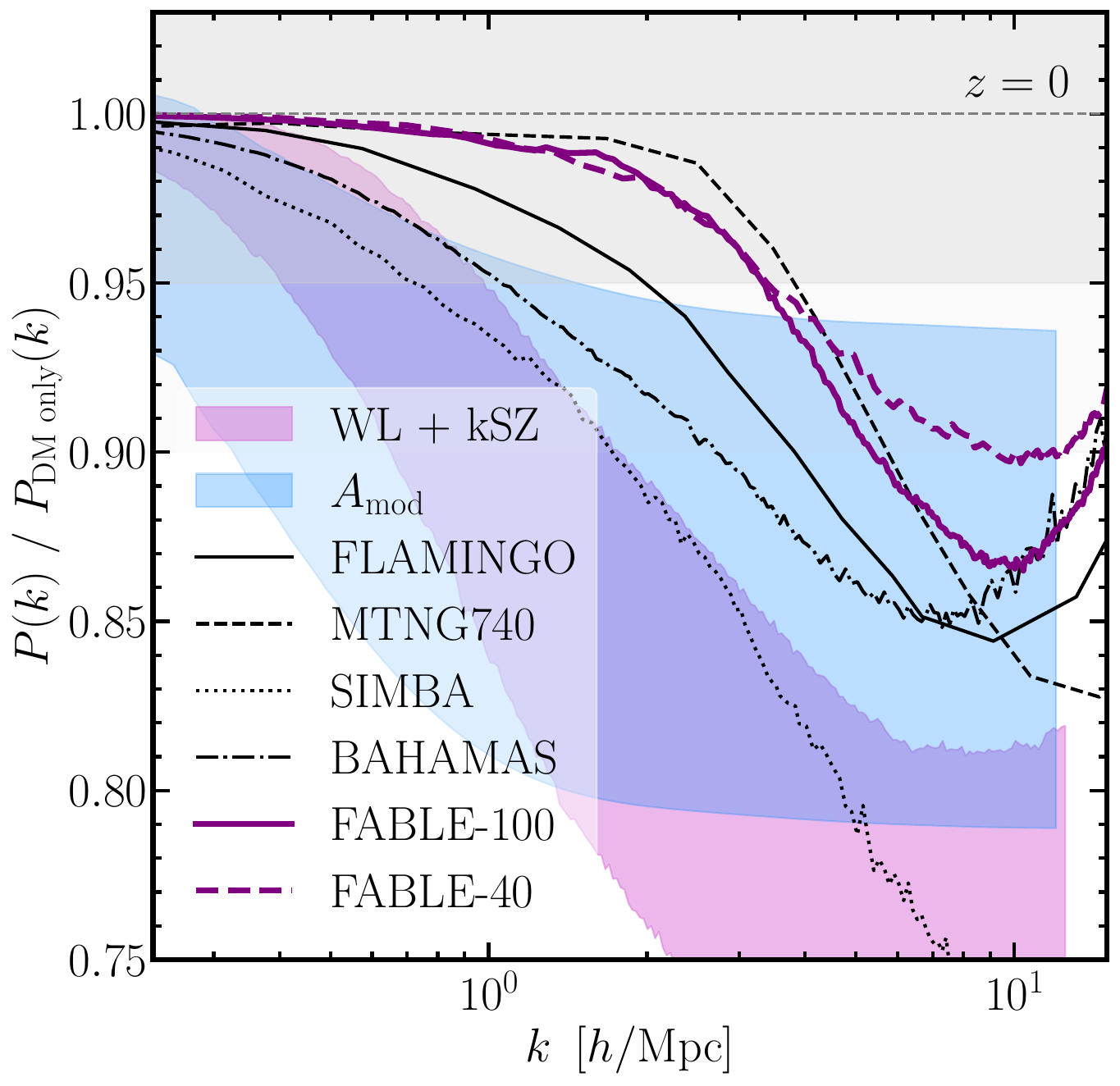}\hfill
\includegraphics[height=5.5cm,keepaspectratio]{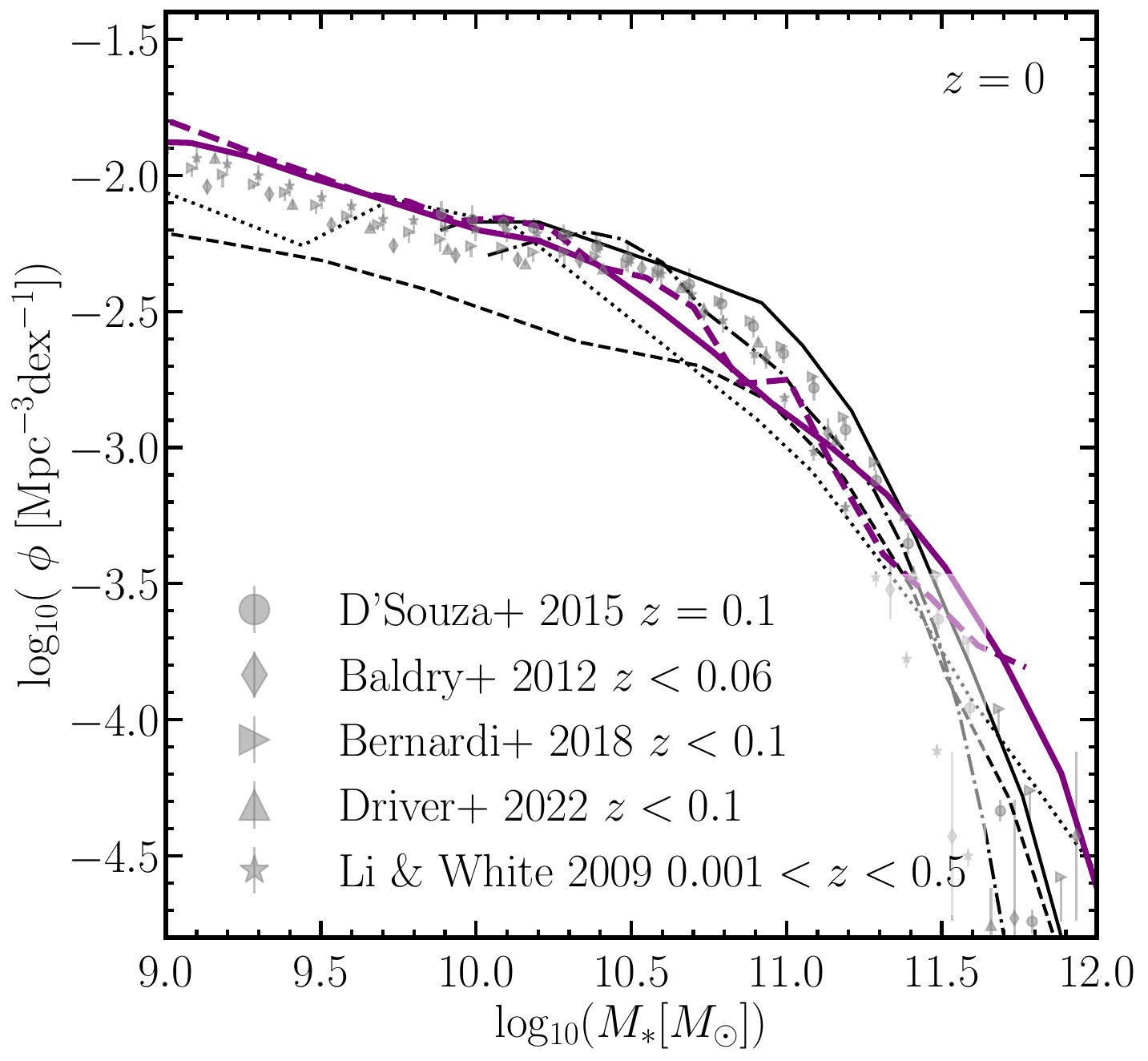}\hfill
\includegraphics[height=5.5cm,keepaspectratio]{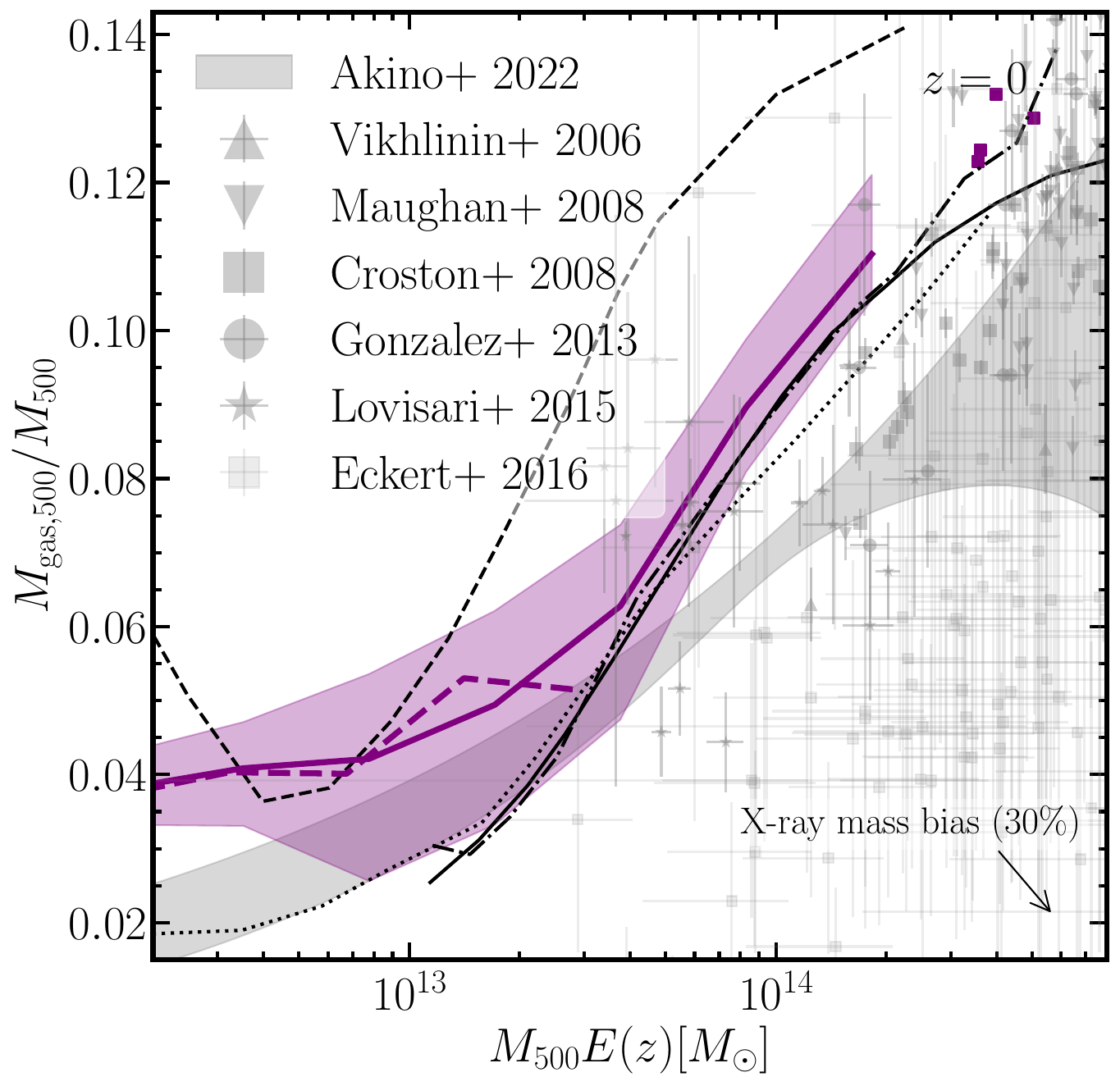}
\caption{The $z=0$ properties of the FABLE-40 (dashed purple line) and FABLE-100 (solid purple line) simulations compared to observational constraints and other hydrodynamical simulations; FLAMINGO L1$\_$m9 (hereafter denoted as FLAMINGO) \citep[][black solid line]{Schaye2023, Schaller2024}, MTNG740 \citep[][black dashed line]{Pakmor2023}, SIMBA \citep[][black dotted line]{Dave:2019} and BAHAMAS \citep[][black dash-dotted line]{McCarthy2017}.  \textit{Left:} the matter power spectrum suppression due to baryonic effects, $P(k)/P_{\mathrm{DM only}}(k)$, compared to the $A_{\mathrm{mod}} = 0.858 \pm 0.052$ constraint \citep[blue, shaded][]{Preston2023}, and the constraint from the combined DES Y3 cosmic shear and ACT kinetic Sunyaev–Zel’dovich (WL+kSZ) analysis of \citet{Bigwood2024} (purple, shaded). \textit{Centre:} the galaxy stellar mass function for the above mentioned simulation projects and FABLE. We further plot observational measurements of \citet{DSouza2015} ($z=0.1$), \citet{Baldry2012} ($z<0.06$), \citet{Bernardi2018} ($z<0.1$), \citet{Driver2022} ($z<0.1$) and \citet{Li2009} ($0.001<z<0.5$) as the grey errorbars. \textit{Right:} the hot gas fraction measured within $r_{500}$ as a function of halo mass $M_{500}$ or the above mentioned simulation projects and FABLE. For FABLE-40 and FABLE-100, lines denote the median relation and the purple band denotes the quartiles of the distribution in FABLE-100.  For FABLE-100, we plot the most massive systems that cannot be binned due to poor statistics as individual datapoints. The grey datapoints are the measurements of \citet{Vikhlinin2006} ($z<0.25$), \citet{Maughan2008} ($0.1<z<1.3$), \citet{Croston2008} ($z<0.2$), \citet{Gonzalez2013} ($z<0.2$), \citet{Lovisari2015} ($z<0.4$), \citet{Eckert2016} ($0.05<z<1.1$) and the grey shaded regions show the 1$\sigma$ constraints of \citet{Akino_2022} ($z<1$). We demonstrate that despite each simulation showing reasonable fits to observations, the predictions for the suppression of the matter power spectrum at $z=0$ vary significantly.} 
\label{figure:comparesims}
\end{figure*}

Hydrodynamical simulations are increasingly sophisticated in their ability to reproduce realistic cosmic populations of galaxies, galaxy groups and clusters. Nevertheless, there is a lack of agreement in the predictions from the state-of-the-art simulations, with one particularly salient example being the impact of baryonic physics on the matter power spectrum \citep{Chisari2019,vandaalen:2020}. Weak lensing analyses (in addition to i.e. N$\times$2pt analyses involving tSZ and kSZ data and effective field theory analyses of galaxy clustering) require an accurate prediction of the suppression of the matter power spectrum due to feedback to infer cosmological parameters, especially if they are to use the (mildly) non-linear scales. Marginalising over this spread of possible predictions already dominates the systematic uncertainty \citep[e.g.][]{Bigwood2024}.
To maximise the statistical power of the surveys, a coherent and consistent picture of feedback's impact on the total matter distribution is critical.  

In this section, to demonstrate this problem, we discuss the $z=0$ predictions of the matter power spectrum suppression, galaxy stellar mass function and hot gas fraction in groups and clusters from a number of state-of-the-art cosmological hydrodynamical simulations and compare them with available observations. We show the latter two observables as these are typically the key properties used to calibrate the feedback parameters in simulations. Indeed, simulations have shown a remarkable ability to reproduce a wide range of other observables when (largely) calibrated to these two key measurements \citep{McCarthy2017, Henden2018, Schaye2023}.  

Figure~\ref{figure:comparesims} shows these properties as measured in the fiducial FABLE $(40~h^{-1}\mathrm{Mpc})^3$ box (hereinafter FABLE-40) \citep{Henden2018}, as well as a larger $(100~h^{-1}\mathrm{Mpc})^3$ box we run, employing the fiducial FABLE physics model (hereinafter FABLE-100)\footnote{We note that in this work, we run all simulations, including FABLE-40, using a different 
random seed that determines the initial Gaussian density field to that utilised for the $(40~h^{-1}\mathrm{Mpc})^3$ presented in  \citet{Henden2018} and \citet{Martin-Alvarez2024}. We therefore find a small difference in the measured matter power spectrum suppression, which lies within the span of the scatter in the suppression due to cosmic variance (see Appendix~\ref{app:seeds}).}.  We refer the reader to Section~\ref{sec:methods} for an introduction to the FABLE simulation properties.  We compare to FLAMINGO L1$\_$m9 \citep[][$(1~\rm{Gpc})^3$ box with gas mass resolution of $10^9$ ${\mathrm M}_{\rm \odot}$]{Schaye2023, Schaller2024}, MillenniumTNG \citep[][(740~Mpc)$^3$ box with gas mass resolution of $3.1\times 10^7 {\mathrm M}_{\rm \odot}$]{Pakmor2023}, SIMBA \citep[][$(100~h^{-1}\mathrm{Mpc})^3$ box with gas mass resolution of $1.82\times10^7$ ${\mathrm M}_{\rm \odot}$]{Dave:2019}, and BAHAMAS \citep[][$(400~h^{-1}\mathrm{Mpc})^3$ box with AGN feedback parameter $\Theta_{\rm AGN}=7.8$ and gas mass resolution of $7.66\times10^8$ ${\mathrm M}_{\rm \odot}/h$]{McCarthy2017}.  

\subsection{Comparison of cosmological hydrodynamical simulations: the matter power spectrum suppression}\label{sec:pkcompare}
The left panel of Figure~\ref{figure:comparesims} shows the predicted suppression of the matter power spectrum from each simulation at $z = 0$. We compare to the predicted suppression required to reconcile the DES Y3 cosmic shear $S_8$ constraint with the \textit{Planck} $\Lambda$CDM model, $A_{\mathrm{mod}} = 0.858 \pm 0.052$ \citep{Preston2023} (blue band)\footnote{The DES Y3 lensing kernel, which defines the redshift sensitivity of the sample, peaks at $z \sim 0.4$ \citep{AmonDES2022}. Since the best-fit $A_{\mathrm{mod}}$ constraint has no explicit redshift dependence, a comparison to simulations should ideally be done at $z \sim 0.4$, but as simulation predictions are not readily available, we plot all simulation results at $z = 0$. For the FABLE-only analysis that we present later (see Figure~\ref{fig:mods_pk}), we discuss the redshift dependence.}. We also compare to the observational constraint from the joint weak lensing and kinetic Sunyaev–Zel’dovich (WL+kSZ) analysis presented in \citet{Bigwood2024} (purple band).

All of the hydrodynamical simulations predict suppression of the matter power spectrum on scales $k\gtrsim 0.5\, h\, \mathrm{Mpc}^{-1}$. However, there is no consensus on the amplitude or extent of suppression: at $k\sim1\, h\, \mathrm{Mpc}^{-1}$ the suppression predicted by the simulations displayed spans 1-5\%, and at $k\sim5\, h\, \mathrm{Mpc}^{-1}$ the range increases to 5-20\%. FLAMINGO, FABLE \citep[see also a recent study by][]{Martin-Alvarez2024} and MTNG740 predict a mild suppression, which is not consistent with the $A_{\mathrm{mod}}$ or WL+kSZ constraint, suggesting that if these simulations capture a realistic feedback scenario, baryonic effects are unable to resolve the $S_8$ tension. The simulations are not consistent with the larger-scale suppression constrained by the data at $k \lesssim 2\, h\, \mathrm{Mpc}^{-1}$, except for SIMBA. To avoid overcrowding Figure~\ref{figure:comparesims} we do not plot the Magneticum \citep{Steinborn2015} or Horizon-AGN \citep{Dubois2014} simulations, but we note Magneticum predicts a matter power spectrum suppression closely following BAHAMAS, and Horizon-AGN predicts a suppression close to that measured in MTNG740.  

We note that the FABLE predictions for the two box sizes explored here are in good agreement with each other at $k<3\, h\, \mathrm{Mpc}^{-1}$, and both show a maximum suppression at $k\sim10\, h\, \mathrm{Mpc}^{-1}$. FABLE-100 shows a slightly larger peak suppression of $\sim$13\%, compared to $\sim$10\% in the FABLE-40 box. This result is consistent with the impact of box size found in \citet{Springel2018}. We additionally found that there was no systematic difference between the two FABLE boxes when investigating the matter power spectrum suppression with redshift. We conclude that the greater peak suppression in the FABLE-100 box at $z=0$ likely results from the stochastic nature of radio-mode feedback in massive haloes. 

\subsection{Comparison of cosmological hydrodynamical simulations: the galaxy stellar mass function}
The GSMF is sensitive to the baryonic processes governing star formation, including cooling, stellar and AGN feedback channels. As it is tightly constrained by data at $z=0$, it provides a good test of galaxy formation models and has been used to calibrate the above simulations. In the middle panel of Figure~\ref{figure:comparesims}, we show the GSMF measurements of \citet{DSouza2015}, \citet{Baldry2012}, \citet{Bernardi2013}, \citet{Driver2022} and \citet{Li2009}. We compare these measurements with the FABLE simulations, as well as FLAMINGO, MTNG740, SIMBA and BAHAMAS.  

Generally, each simulation is in good agreement with the observations\footnote{Note that MTNG740 predictions for galaxies with stellar mass $\log_{10}(M_* [{\rm M_{\odot}}])<11$ largely stem from the effective mass resolution of the simulations, with higher resolution TNG results in much closer agreement with the data \citep[for further details see][]{Pakmor2023}.}. We note that both FABLE boxes show similarly good agreement with observational data, with the larger FABLE-100 box being able to better sample rare high stellar mass galaxies ($\log_{10}(M_* [\mathrm{M_{\odot}}])\sim12$), and hence extend the GSMF tail\footnote{For massive galaxies, the GSMF is over-predicted by FABLE. Further refinements in baryonic `sub-grid' physics, and a different choice of the stellar mass aperture (such as the commonly adopted $30$~kpc fixed aperture), may improve the agreement, but this is beyond the scope of this paper.}. Nevertheless, it is notable that independent simulations, in similarly good agreement with GSMF observations (at least at $z = 0$) given the observed uncertainties, predict significantly different baryonic suppression of the matter power spectrum. 

\subsection{Comparison of cosmological hydrodynamical simulations: hot gas fractions in groups and clusters}
The mass fractions of gas and stars in simulated groups and clusters are very sensitive to the AGN feedback modelling. Furthermore, the total baryon fraction has been shown to be directly related to the matter power spectrum suppression \citep{vandaalen:2020, Salcido2023, Martin-Alvarez2024}. The right panel of Figure~\ref{figure:comparesims} shows the hot gas fraction in groups and clusters, where simulation predictions from FABLE, FLAMINGO\footnote{We note that the FLAMINGO suite has also explored more extreme feedback variants which exhibit gas mass fractions lower than the fiducial FLAMINGO L1$\_$m9 box we compare to (see Table~2 of \citet{Schaye2023} and \citet{Schaller2024}).}, MTNG740, SIMBA and BAHAMAS are plotted. For the FABLE-100 box, in addition to the median, we show the quartiles of the gas fraction distribution as the purple-shaded band\footnote{In Figure~\ref{figure:comparesims} and throughout the remainder of the work, the shaded bands showing the quartile regions finish at the midpoint of the highest bin.}. For comparison, we plot the X-ray derived measurements of \citet{Akino_2022}\footnote{We note that unlike the remainder of the observational datasets where we plot individual objects, we plot the best-fit relation of \citet{Akino_2022}, as the data is model-dependent on error correlation considerations.}, \citet{Vikhlinin2006} ($z<0.25$), \citet{Maughan2008} ($0.1<z<1.3$), \citet{Croston2008} ($z<0.2$), \citet{Gonzalez2013} ($z<0.2$), \citet{Lovisari2015} ($z<0.4$) and \citet{Eckert2016} ($0.05<z<1.1$).  We note that the cluster masses of \citet{Akino_2022} and \citet{Eckert2016} are derived via weak lensing estimates, whereas the remaining sources use X-ray hydrostatic cluster masses.  The latter are derived under the assumption of hydrostatic equilibrium and hence may under-estimates the true halo mass by 10-35\% (e.g., due to neglecting non-thermal pressure support), with the exact magnitude of the bias still debated \citep{Eckert2016,Hoekstra2013,Miralda1995,Mahdavi2013,Arnaud2007,Rasia2012, Braspenning2024}. For illustrative purposes, the arrow indicates the effect on observations to correct for a  30\% mass bias.

 The two FABLE boxes are in very good agreement for groups of mass $M_{500}<10^{14}~\mathrm{M_{\odot}}$, with the FABLE-100 box having a large enough sample of clusters to compute the gas fractions up to $M_{500}\sim5\times 10^{14}~\mathrm{M_{\odot}}$.  FABLE-100 displays a very good match to the data, as well as the predictions from FLAMINGO, BAHAMAS and SIMBA\footnote{We note again the exception of MTNG740, which is at the upper end of the observations and the other simulations displayed, and refer the reader to \citet{Pakmor2023} where this result was initially discussed.}.  However, the scatter in the observed data is significant. We, therefore, re-emphasise the point made in the previous section: AGN feedback models that produce reasonable gas fractions (within the large observed scatter) exhibit a large discrepancy in the matter power spectrum suppression for cosmological studies. 
 
 We further observe that the gas fraction-halo mass relation measured in a hydrodynamical simulation could lie up to $\sim3\sigma$ lower than the one measured in FABLE-100, whilst still remaining within the large scatter of the data. More powerful AGN feedback responsible for this greater expulsion of gas may, in theory, then produce a power spectrum suppression greater than that predicted by the simulations in the left panel of Figure~\ref{figure:comparesims}. Moreover, provided the observed scatter is real, it remains to be understood if simulations need to produce a larger variety of gas fractions at a given halo mass, which would point towards a more stochastic nature of AGN feedback and more extreme feedback for a sub-set of objects.  
 
 Motivated by these findings, we explore the possibility of AGN feedback that produces a more extreme matter power spectrum suppression, in better agreement with observational constraints, while preserving the match to the observed gas fractions and GSMF.

\section{Methodology}\label{sec:methods}

\subsection{Numerical code and basic simulation properties}
In this study, simulations are performed with the massively-parallel moving-mesh code \textsc{Arepo} \citep{Springel2010, Pakmor2016}. The TreePM approach is used for computing gravitational interactions and hydrodynamics is solved on a quasi-Lagrangian Voronoi mesh, which approximately moves with the local flow velocity.

As a starting point, we adopt the FABLE simulation model. Its key characteristics are described below and we refer the reader to \citet{Henden2018, Henden2019, Henden2020} for a more detailed discussion. In a nutshell, the FABLE project adopts the same sub-grid models for gas radiative cooling \citep{Katz1996, Wiersma2009a}, chemical enrichment \citep{Wiersma2009b} and star formation \citep{Springel2003}, subject to a spatially uniform UV background \citep{Katz1996, Faucher-Giguere2009}, as developed for the Illustris project \citep{Vogelsberger2013, Torrey2014}. While the Illustris simulation models stellar winds in a purely kinetic fashion at launch, in FABLE, one-third of the wind energy is thermal \citep{Marinacci2014, Henden2018}. The fiducial FABLE model adopts two modes for AGN feedback; a quasar-mode for black holes in the radiatively efficient accretion regime \citep{DiMatteo2005, Springel2005} and a radio-mode feedback for the radiatively inefficient accretion regime \citep{Sijacki2007}, as in the Illustris model \citep{Sijacki2015}. The quasar-mode thermally and isotropically couples a fraction of the available feedback energy to the surrounding gas, whereas the radio-mode injects hot bubbles at some distance from the black hole, mimicking the radio lobes inflated by `mechanical' feedback.  Compared to Illustris, the two main differences in FABLE stem from adopting a fixed duty cycle in the quasar-mode, instead of injecting thermal energy continuously \citep[see][]{Booth2009, Henden2018}, and from reducing the duty cycle of radio bubble inflation, which leads to a more frequent but less energetic radio-mode feedback. We note that stellar and AGN feedback in FABLE have been calibrated to reproduce the galaxy stellar mass function and the gas mass fractions of massive haloes in the local Universe \citep[see also][for a similar calibration strategy]{McCarthy2017, Schaye2023}. We perform uniform cosmological boxes and do not consider the zoom-in group and cluster simulations from the original FABLE suite. 

We build a suite of $40\,h^{-1}$~cMpc-side simulation boxes to explore the effect of AGN feedback modifications to the FABLE model and cosmic variance (see Appendix~\ref{app:seeds}). These boxes have $512^3$ dark matter particles and gas cells (approximately), corresponding to a dark matter particle mass $m_{\rm DM} = 3.4 \times 10^7 \, h^{-1} \mathrm{M}_{\odot}$ and mean target gas cell mass $\bar{m}_{\rm gas} = 6.4 \times 10^6 \, h^{-1} \mathrm{M}_{\odot}$. We set the gravitational softening length to $2.4 \, h^{-1}$~pkpc (physical coordinates) below $z = 5$ and fix it in comoving coordinates at higher redshifts by following the empirical recommendation of \citet{Power2003}.  To ensure we have a sufficient statistical sample of galaxy groups and (low mass) galaxy clusters, we run two further cosmological boxes with a side length of $100\,h^{-1}$~cMpc, both for the fiducial FABLE baryonic physics model (FABLE-100) and for one of our new modified AGN feedback models, which henceforth we denote as `XFABLE-100'. These larger boxes have the same mass and spatial resolution as the $40\,h^{-1}$~cMpc on-a-side boxes, tracking $1280^3$ dark matter particles and $\sim 1280^3$ gas cells. Boxes are evolved to $z=0$ and adopt initial conditions consistent with the cosmological parameters measured by \citet{Planck2018} ($\Omega_{\Lambda} = 0.6856$, $\Omega_{\rm M} = 0.3144$, $\Omega_{\rm b} = 0.0494$, $\sigma_8 = 0.8154$, $n_s = 0.9681$ and $H_0 = 67.32$~km s$^{-1}$~Mpc$^{-1}$). 

\subsection{Black hole accretion and feedback in FABLE-like simulation models} \label{AGNfeedback_methodology}

In this work we focus on modifications to the AGN feedback model, since it has been shown to have the dominant effect in causing suppression of the matter power spectrum \citep[see e.g.][]{Chisari2019, vandaalen:2020, Martin-Alvarez2024}. We first describe the fiducial FABLE black hole accretion and feedback model. Summaries of the key model parameters and their values in both the fiducial FABLE and the XFABLE models are listed in Table~\ref{tab:params}.  

\begin{table*}
\centering
\caption{The key parameters associated with black hole accretion and feedback in the FABLE and XFABLE models.}
\begin{tabular}{cccc}\label{tab:params}

\textbf{Parameter} & \textbf{Description} & \textbf{Value in FABLE} & \textbf{Value in XFABLE} \\
\hline
$\alpha$ & Dimensionless parameter boosting the black hole accretion rate (Equation~\ref{eq:BHL}). & 100 & 100 \\
$\chi_{\mathrm{radio}}$ & Accretion rate threshold in Eddington units separating quasar and radio-mode   & 0.01 & 0.1\\
& activity. The quasar-mode acts when $\dot{M}_{\mathrm{BH}}/\dot{M}_{\mathrm{Edd}}>\chi_{\mathrm{radio}}$, and the radio-mode &  \\

& when $\dot{M}_{\mathrm{BH}}/\dot{M}_{\mathrm{Edd}}<\chi_{\mathrm{radio}}$. &\\
$\epsilon_r$ & Radiative efficiency, determining the fraction of energy gained from mass  & 0.1 & 0.1\\
 &  accretion that may be converted to radiation  (Equation~\ref{eq:feedback_energy}).& \\
$\epsilon_f$ & Thermal coupling associated with the quasar-mode, determining the fraction of & 0.1 & 0.1 \\
& the bolometric luminosity to be converted to thermal energy (Equation~\ref{eq:quasar_mode}). &\\
$\Delta t$ [Myr] & Duty cycle of the quasar-mode: the time for which feedback energy is stored & 25 & 25  \\
& before it is released in a single feedback event. & \\
$\epsilon_{\rm m}$ & Efficiency of mechanical heating associated with the radio-mode (Equation~\ref{eq:radio_mode}). & 0.8 & 0.8\\
$\delta_{\mathrm{BH}}$ & Duty cycle of radio-mode; bubbles are injected after the mass gain of the  & 0.01 & 0.01\\
&  black hole has exceeded $\delta_{\mathrm{BH}}=\delta M_{\mathrm{BH}}/M_{\mathrm{BH}}$. & \\
$D_{\mathrm{bub}}$ [$h^{-1}$kpc] & Distance bubbles are injected from the black hole in the radio-mode. & Equation~\ref{eq:dbub}, with $D_{\mathrm{bub,0}}=30$ & 100 \\
$R_{\mathrm{bub}}$ [$h^{-1}$kpc] & Radius of the injected bubbles in the radio-mode. & Equation~\ref{eq:rbub}, with $R_{\mathrm{bub,0}}=50$ & 50 \\
\hline
\end{tabular}
\end{table*}

Black hole formation proceeds by placing seed black holes of mass $10^5$~$h^{-1}\, {\rm M}_{\rm \odot}$ into every halo of mass greater than $5\times 10^{10}$~$h^{-1}\,{\rm M}_{\odot}$, where halos are identified using a fast Friend-of-Friend (FoF) algorithm on-the-fly. Black holes are modelled as collisionless sink particles and are able to grow in mass through black hole mergers and gas accretion. 

The black hole accretion rate, $\dot{M}_{\mathrm{BH}}$, is given by the Bondi-Hoyle-Lyttleton formula, where a dimensionless parameter, $\alpha$, boosts the accretion rate as
\begin{equation}\label{eq:BHL}
\dot{M}_{\mathrm{BH}}=\frac{4\pi\alpha G^2M_{\mathrm{BH}}^2\rho}{c_s^3} \,,
\end{equation}
where $\rho$ and $c_s$ are the gas density and sound speed, respectively. Note that $\dot{M}_{\mathrm{BH}}$ is capped at the Eddington limit. In the radiatively efficient regime, the black hole bolometric luminosity, $L_{\mathrm{bol}}$, is given by 
\begin{equation}\label{eq:feedback_energy}
L_{\mathrm{bol}}= \epsilon_r \dot{M}_{\mathrm{BH}} c^2\,,
\end{equation}
where $\epsilon_r$ is the radiative efficiency and $c$ is the speed of light. 

Feedback occurs in one of two modes, solely determined by the ratio of the accretion rate of the black hole to the Eddington rate, $f_{\mathrm{BH}}=\dot{M}_{\mathrm{BH}}/\dot{M}_{\mathrm{Edd}}$. If the black hole is accreting efficiently and $f_{\mathrm{BH}}$ exceeds the threshold of $\chi_{\mathrm{radio}}$, the quasar-mode is operating. This is typically the dominant feedback process at high redshifts, where a copious gas supply maintains high black hole accretion rates. A fraction of the bolometric luminosity is coupled thermally and isotropically to the gas surrounding the black hole, $\epsilon_f$, resulting in the feedback energy, ${E}_{\mathrm{feed}}$, being given by,
\begin{equation}\label{eq:quasar_mode}
\dot{E}_{\mathrm{feed}}=\epsilon_f L_{\mathrm{bol}}\,.
\end{equation}

If the thermal energy injected into gas cells is unable to significantly raise the gas temperature (for example as the result of spreading the energy over a large gas mass), or is predominantly injected into high density gas, then the energy can be radiated away before impacting the environment \citep{Katz1996, Booth2009, Bourne2015}. To prevent this numerical `overcooling', the feedback energy is stored for the time period of the duty cycle, $\Delta t$, and the energy accumulated in this time period is released in a single feedback event \citep[following a similar approach to][]{Schaye2015, LeBrun2014}. 

For $f_{\mathrm{BH}}<\chi_{\mathrm{radio}}$, the radiatively inefficient radio-mode operates.  Hot bubbles of radius, $R_{\mathrm{bub}}$, are injected 
 at a random spatial position within a sphere of radius, $D_{\mathrm{bub}}$, from the black hole, to mimic injection by an unresolved AGN jet.  This results in a largely isotropic feedback once averaged over sufficient time. The bubbles are periodically injected after the gain in the black hole's mass has exceeded $\delta_{\mathrm{BH}}=\delta M_{\mathrm{BH}}/M_{\mathrm{BH}}$. The energy content of the resulting bubble is given by,
\begin{equation}\label{eq:radio_mode}
E_{\mathrm{bub}}=\epsilon_{\rm m}\epsilon_r \delta M_{\mathrm{BH}}c^2\,,
\end{equation}
where $\epsilon_{\rm m}$ is the efficiency of this `mechanical' heating. In the fiducial model, the bubble distance and radius are scaled with energy and ICM density, $\rho_{\mathrm{ICM}}$, according to
\begin{equation}\label{eq:dbub}
D_{\mathrm{bub}}=D_{\mathrm{bub,0}}\left(\frac{E_{\mathrm{bub}}/E_{\mathrm{bub,0}}}{\rho_{\mathrm{ICM}}/\rho_{\mathrm{ICM,0}}}\right)^{1/5}\,,
\end{equation}

\begin{equation}\label{eq:rbub}
R_{\mathrm{bub}}=R_{\mathrm{bub,0}}\left(\frac{E_{\mathrm{bub}}/E_{\mathrm{bub,0}}}{\rho_{\mathrm{ICM}}/\rho_{\mathrm{ICM,0}}}\right)^{1/5}\,,
\end{equation}
where $D_{\mathrm{bub,0}}$, $R_{\mathrm{bub,0}}$, $E_{\mathrm{bub,0}}$ and $\rho_{\mathrm{ICM,0}}$ are normalisation constants. This follows studies \citet{Heinz1998, Scheuer1974, Begelman1989}, which show that more energetic AGN jets will lead to larger lobes at a greater distance from the black hole, and a greater ICM density will have the inverse effect of confining the bubbles. 
 
Note that when we modify the FABLE AGN feedback model, we test removing the scaling of Equation~\ref{eq:dbub} and  Equation~\ref{eq:rbub} and fixing $R_{\mathrm{bub}}$ and $D_{\mathrm{bub}}$ to specific values.  In this scenario, the bubbles are injected at a random spatial position on a spherical shell (rather than within the sphere) of radius $D_{\mathrm{bub}}$.

\subsection{Comparison to observations: methodology}\label{sec:obsmethods}

In this section we describe the derivation of a number of observables that we use to differentiate and validate our feedback models.

\subsubsection{The matter power spectrum \& the $A_{\rm{mod}}$ model}\label{sec:amod}

We calculate the 3D matter power spectrum, $P_{\rm m}(k)$, using the routines of Pylians \citep{Pylians}.  We first calculate the overdensity field, $\delta(x)=\rho(x)/\bar\rho(x)-1$, on a discrete Cartesian grid with $512^3$ pixels for boxes with side length $40 \, h^{-1}$~Mpc and $1024^3$ pixels for boxes with sides of $100\, h^{-1}$~Mpc. Taking the coordinates in the simulation snapshots, we assign gas cells, black holes, stars and dark matter particles to the grid via the first order linear cloud-in-cell (CIC) scheme, weighting by their mass. Using fast Fourier transforms the Fourier modes of the density contrast field are computed, $\delta(k)$, and the effect of the smoothing from the CIC kernel is deconvolved. The power spectrum is then calculated as the mean power per $k$-mode, $P_{\rm m}(k)=\langle|\delta(k)|^2\rangle$. To clearly isolated the effect of `baryonic feedback' on the power spectrum, we calculate the ratio of the full matter power spectrum to the dark matter-only case, $P_{\rm m}(k)/P_{\mathrm{DMonly}}(k)$, where $P_{\mathrm{DMonly}}(k)$ is the power spectrum computed on a gravity-only FABLE box with identical initial conditions and box size to that used to calculate $P_{\rm m}(k)$.

The prediction for extreme suppression of the non-linear matter power spectrum as a viable solution to the $S_8$ tension was first proposed by \citet{Amon2022} and \citet{Preston2023} using a phenomenological model, $A_{\rm mod}$. In this simple model, the amplitude of the non-linear power spectrum is modified by the parameter $A_{\mathrm{mod}}$ according to;
\begin{equation}
    P_{\rm m}(k,z)=P_{\rm m}^{\mathrm {L}}(k,z)+A_{\mathrm{mod}}[P_{\rm m}^{\mathrm{NL}}(k,z)-P_{\rm m}^{\mathrm{L}}(k,z)]\,,
\end{equation}
where the superscripts L and NL refer to the linear and non-linear power spectra, respectively, with the latter assuming cold dark matter cosmology.  
We refer to this model throughout this work in our assessment of the plausibility of more extreme AGN feedback.

\subsubsection{Galaxy stellar mass function calculation}

In our simulation boxes, we define a galaxy as a subhalo found by the \textsc{Subfind} algorithm \citep{Davis1985,Dolag2009,Springel2001} which has more than 100 star particles. Defining the total stellar mass of the simulated galaxy as the sum of all the star particles bound to the subhalo can overestimate the GSMF at the high mass end \citep[for further details see e.g.][]{Henden2018}. As a result, in this work we follow \citet{Genel2014} and define the galaxy stellar mass as that measured within twice the stellar half-mass radius of the subhalo, as given in the Subfind catalogue. Note that following \citet{Henden2018}, when comparing to observations, for all stellar masses we assume a \citet{Chabrier2003} initial mass function (IMF), which involves subtracting 0.25~dex for a \citet{Salpeter1955} IMF and 0.05~dex for a \citet{Kroupa2001} IMF.

\subsubsection{Quasar luminosity function calculation}\label{sec:qlfcalc}
For black holes in radiatively-efficient regime we calculate the bolometric luminosities of black holes in our simulation boxes, $L_{\mathrm{bol}}$, according to Equation~\ref{eq:feedback_energy}. As, for example, discussed in \citet{Churazov2005}, the radiative luminosity of AGN accreting at low $f_{\mathrm{BH}}$, i.e. those in the radio-mode, may be significantly lower than the values obtained by naively using  Equation~\ref{eq:feedback_energy}.  

We, therefore, explore the impact on the quasar luminosity function of distinguishing radiatively efficient and radiatively inefficient AGN. AGN with $f_{\mathrm{BH}}\geq0.01$, i.e. those in the quasar-mode in FABLE, have bolometric luminosities calculated following Equation~\ref{eq:feedback_energy}. For AGN with $f_{\mathrm{BH}}<0.01$, we follow \citet{Churazov2005, Habouzit2022} and approximate the bolometric luminosities as follows;
\begin{equation}
     L_{\mathrm{bol}}=10 f_{\mathrm{BH}}\epsilon_r \dot{M}_{\mathrm{BH}} c^2\,.
 \end{equation}
We use a linear spline to smooth the transition region in $f_{\mathrm{BH}}$ between the two regimes.  By comparing bolometric luminosities calculated with this distinction to those calculated under the assumption that all AGN are radiatively efficient, we aim to somewhat bracket the viable range of the quasar luminosity function predicted by our models in comparison to observations.

\subsubsection{Gas and stellar mass fractions calculation}\label{subsec:gasfrac}

To calculate the mass fractions, we select groups and clusters as halos found using the FoF algorithm with mass $M_{500}>10^{12}~\mathrm{M_{\odot}}$. We define $M_{500}$ as the mass contained within a sphere of radius $r_{500}$, centred on the minimum potential of the halo, where the mean density is 500$\times$ the critical density of the Universe. The vast majority of gas fraction measurements in the literature are derived from X-ray emission from hot diffuse gas. Therefore, to compare our simulated results with these observations, we follow the approach of \citet{Henden2018} by excluding gas cells with a temperature below $T<3\times 10^4$~K and those above the
density threshold required for star formation, thereby assuming their contribution to the X-ray emission is negligible. We measure both gas and stellar masses within $r_{500}$, selecting cells within this radius using a K-D tree algorithm.  

\subsubsection{X-ray scaling relations calculation}\label{sec:xraymethods}

We use the ICM's bolometric luminosity in combination with other global cluster properties to derive the X-ray scaling relations. We take a more simplistic approach to that used in \citet{Henden2018}, which involved the generation of mock X-ray spectra to derive X-ray luminosities.  We follow \citet{Rybicki1985} to estimate the hot ICM X-ray luminosity measuredwithin , $L_{500}^{\mathrm{bol}}$. The Bremsstrahlung emissivity density $\epsilon^{\mathrm{ff}}$ is given as;
\begin{equation}
    \epsilon^{\mathrm{ff}} = 1.4\times 10^{-27} T^{1/2}n_e n_i Z^2 g_{\mathrm B}\ (\mathrm{erg} \mathrm{s}^{-1} \mathrm{cm}^{-3}) \,,
\end{equation}
where $T$ is the gas temperature, $n_e$ and $n_i$ are the electron and ion number densities respectively, and $g_{\mathrm B} = 1.2$ is the average Gaunt factor. Assuming a fully ionized primordial plasma so that $n_e n_i \approx 1.4\rho^2 / (\mu  m_p)^2$, we arrive at;
\begin{equation}\label{eq:lum}
    L_{500}^{\mathrm{bol}} = \frac{2.35\times 10^{-27}}{\mu^2m^2_p} \sum_i^{r_{500}}m_i \rho_i T^{1/2}_i \ (\mathrm{erg} \mathrm{s}^{-1})\,,
\end{equation}
where $m_i$, $\rho_i$ and $T_i$ are the mass, density and temperature of $i$-th gas cell, $m_p$ is the proton mass and $\mu=0.59$ is the mean molecular weight.  

We investigate the X-ray scaling relations between $L_{500}^{\mathrm{bol}}$, $M_{500}$, the gas mass within $r_{500}$, $M_{\rm gas}$, and the mass-weighted mean temperature within $r_{500}$, $T_{\rm 500,mw}$.  We note that $T_{\rm 500,mw}$ differs from the characteristic temperature of Equation~\ref{eq:chartemp}.  As with gas mass fractions, since we are comparing to X-ray observations from hot diffuse gas, we measure these quantities in the simulations using only the hot and non-star forming gas, following the cuts described in Section~\ref{subsec:gasfrac}.  Following \citet{Henden2019}, we make an additional cut excluding gas cells with a temperature greater than four times the virial temperature, i.e. $k_b T< 4 G M_{200} \mu m_p / 2 r_{200}$.  This upper threshold aims to exclude the AGN driven bubbles created by the radio-mode feedback model, which would contribute excessively high temperature gas to the scaling relations if a recent strong feedback event had occurred.  The simplistic radio-mode model does not capture non-thermal pressure support within bubbles, which means in observations the bubbles should not contribute to the scaling relations until thermalisation has occurred.  Removing the artificially hot gas created by the feedback model thus reduces bias with respect to the observations.  

Finally, we compute the X-ray proxy of the tSZ Compton $Y_{500}$ parameter, $Y_X$ \citep{Kravtsov2006}. $Y_X$ is the product of the mean X-ray spectroscopic temperature of a cluster and the gas mass measured within $r_{\rm 500}$, and is thus sensitive to the cluster's total thermal energy. We approximate the spectroscopic temperature as $T_{\rm 500,mw}$. Following Appendix~B2 of \citet{Henden2018}, which finds no systematic offset between temperatures derived from X-ray spectra versus mass-weighted temperatures, we calculate $Y_X$ as;
\begin{equation}\label{eq:yx}
    Y_X=T_{\rm 500,mw} M_{\rm gas}\,.
\end{equation}
As above, we exclude the cold and star forming gas (Section~\ref{subsec:gasfrac}), in addition to the high temperature AGN driven bubbles.

\subsubsection{Thermal Sunyaev–Zel’dovich Compton $Y$ parameter calculation}\label{sec:tsz}

The thermal Sunyaev–Zel’dovich effect probes the line-of-sight integrated electron pressure, $P$, and is typically parameterised via the Compton $Y_{500}$ parameter; 
\begin{equation}
    D^2_a(z)Y_{500}=\frac{\sigma_T}{m_e c^2}\int^{r_{500}}_0 P dV\,,
\end{equation}
where $D^2_a(z)$ is the angular diameter distance of the cluster, $\sigma_T$ is the Thomson cross-section and $m_e$ is the electron mass. The quantity integrates the electron pressure in, thus providing a measure of the cluster's thermal energy.   

We compare the simulation computed $Y_{500}-M_{500}$ relation with \citet{Planck2013} results, as well as the \citet{Wang2016} re-analysis, which uses weak-lensing calibrated halo masses. To facilitate comparison to the observations, we scale $Y_{500}$ to a fixed angular diameter distance of $500$~Mpc.  Furthermore, the \citet{Planck2013} analysis integrates the tSZ flux within a projected circular aperture of radius $5r_{500}$ (giving $Y_{5r_{500}}$), rather than $r_{500}$.  In their analysis a conversion of $Y_{500}=Y_{5r_{500}}/1.796$ was thus applied, which assumes the universal pressure profile of \citet{Arnaud2010} as the spatial template in their matched filter. Since the \citet{Arnaud2010} profile is not well constrained at $5r_{500}$, we avoid the dependency on the assumed modelling choices used in the \citet{Planck2013} analysis when deriving the inferred $Y_{500}$ from $Y_{5r_{500}}$, and follow \citet{Henden2018} in measuring the $Y$ parameter within $5r_{500}$ in the simulation boxes. We revert the $Y_{500}$ measurements of \citet{Planck2013} back to $Y_{5r_{500}}$ with the 1.796 multiplicative factor.

\subsubsection{The ICM profiles calculation}\label{sec:profiles}

To calculate the electron density, $n_e$ and temperature, $T$,  for groups and clusters in our simulation boxes, for a given halo, we select gas cells within $3r_{500}$ and divide the gas cells into ten concentric logarithmically-spaced radial bins. We calculate the volume-weighted mean electron density, $n_e$, and mass-weighted mean temperature, $T$.  Since ICM profiles in literature are typically derived from X-ray observations of hot and dilute plasma, we make the same exclusions of gas cells as described in Section~\ref{subsec:gasfrac}; i.e. ensuring gas cells have $T>3\times 10^4$~K and zero star formation rate.  In the volume-weighted case, we take the total volume of the bin as the sum of the gas cell volumes, in order to account for the exclusion of gas cells due to the temperature and star formation rate cut. 
To calculate pressure, $P$, and entropy, $K$, radial profiles, we take the product of individual halo $n_e(r)$ and $T(r)$ profiles according to;
\begin{equation}
    P(r)=k_{\rm B} n_e(r)T(r) \ \ \ \mathrm{and} \ \ \ K(r)=k_{\rm B} T(r)/n_e^{2/3}(r) \, ,
\end{equation}
where $k_{\rm B}$ is the Boltzmann constant.  To allow for comparison between haloes of different mass, we normalise the temperature, pressure and entropy profiles by the `characteristic' quantities $T_{500}$, $P_{500}$ and $K_{500}$. $T_{500}$ is defined as;
\begin{equation}\label{eq:chartemp}
    T_{\rm 500}=\mu m_{\rm p} GM_{\rm 500}/2 r_{\rm 500}\,,
\end{equation}
$K_{\rm 500}$ is defined as;
\begin{equation}\label{eq:chark}
    K_{\rm 500}= k_{\rm B}  T_{\rm 500}/n_{\rm e,500}^{2/3}\,,
\end{equation}
with $n_{\rm e,500}=500 f_{\rm b} \rho_c(z)/\mu_{\rm e} m_{\rm p}$,
where $f_{\rm b}=\Omega_{\rm b}/\Omega_{\rm m}$ and $\rho_{\rm c}=3(100h)^2/8\pi G$ are the cosmological baryon fraction and critical density corresponding to our simulation cosmology, respectively, and $\mu_{\rm e} = 1.14$ is the molecular weight per free electron. $P_{\rm 500}$ is defined as;
\begin{equation}\label{eq:charP}
    P_{\rm 500}=k_{\rm B} n_{\rm e,500} T_{\rm 500}\,.
\end{equation}

\section{Exploring AGN feedback in FABLE-like simulation models}\label{sec:modelsoverview}

In Section~\ref{sec:comparehydrosims}, we demonstrated that although many independent hydrodynamical simulations can attain suitable fits to measured GSMF and hot gas fractions in groups and clusters, the predicted matter power spectrum suppression from `baryonic feedback' can vary significantly. This raises the question: what is the maximum amount of non-linear suppression one can obtain in a hydrodynamical simulation whilst still maintaining good agreement with the observations? 

To address this issue, we ran over 40 different FABLE-like $(40~h^{-1}\mathrm{Mpc})^3$ simulation boxes, modifying various aspects of the AGN feedback model in order to study the resultant power spectrum suppression and compare to a number of galaxy, group and cluster observations. The AGN feedback parameters utilised for the full set of simulation boxes are listed in Table~\ref{tab:simsall}. We select four illustrative AGN feedback models to discuss throughout the remainder of the work.  These are the named boxes in Table~\ref{tab:simsall}, i.e. QuasarBoost$z$2-40, RadioBoost-40, RadioBoost$M_{\rm BH,radio}$-40 and XFABLE.  In the following sections, we detail the motivation and modifications made to the fiducial FABLE AGN feedback model for each of these variations.

\subsection{QuasarBoost$z$2-40: \textit{a quasar-mode boost before cosmic noon}}\label{sec:quasarmodemod}

First, we boost the quasar-mode at high redshift by increasing the dimensionless parameter $\alpha$ and the feedback coupling efficiency $\epsilon_f$ at $z>2$.  Increasing $\alpha$ will result in a given black hole accreting at a greater rate (Equation~\ref{eq:BHL}), up to the Eddington limit, powering a quasar with a larger bolometric luminosity and thus allowing more feedback energy to be available (Equation~\ref{eq:quasar_mode}). Increasing $\epsilon_f$ results in a greater fraction of a quasar's bolometric luminosity being converted to thermal energy (Equation~\ref{eq:quasar_mode}).  Given that the AGN feedback model in FABLE was calibrated on the observed GSMF and hot gas fractions in groups and clusters, one may expect that naive boosts to these parameters would result in poor agreement between the simulations and observations. Certainly, significantly increasing the thermal heating in the centre of galaxies would over-quench star formation and drive more powerful outflows reducing the hot gas fraction.  

It is important to note however that FABLE is calibrated to data at $z=0$.  Observationally measured properties of hot halos in groups and clusters (especially with masses comparable to FABLE objects) become sparse with increasing redshift, particularly beyond cosmic noon at $z\sim2$. We, therefore, test boosting both $\alpha$ and $\epsilon_f$ by a factor of 100 for $z>2$, while resetting parameters to their fiducial values $\alpha=100$ and $\epsilon_f=0.1$ at $z<2$.  We keep all other model parameters, including those associated with the radio-mode, at their fiducial FABLE values.  The threshold redshift of $z=2$ is chosen as it approximately corresponds to the peaks in star formation rate and black hole growth \citep{Madau2014}. Furthermore, the large boost of a factor of 100 is chosen to exemplify the interplay between the maximum attainable matter power spectrum suppression and the damage to the GSMF and hot gas fractions.  As noted in Section~\ref{sec:pkcompare}, the DES Y3 redshift distribution peaks at $z=0.4$, meaning it is at this redshift that a greater matter power spectrum suppression is observed.  Our modification could allow a more destructive feedback scenario which produces a greater matter power spectrum suppression at $z\gtrsim0.4$, yet allow galaxy, group and cluster properties to recover by $z\sim0$, remaining in good agreement with current observations, which is qualitatively in line with the galaxy cluster the pre-heating scenario \citep[see e.g.][]{Borgani2001, Voit2003}. 

ion{RadioBoost-40: \textit{a high redshift boost to the radio-mode}}\label{sec:radiomod}

Next we consider modifications to the radio-mode AGN feedback.  We note that the matter power spectrum suppression required by the $A_{\rm mod}$ and WL+kSZ models to resolve the $S_8$ tension shows the greatest discrepancy with FABLE and other hydrodynamical simulations in the mildly non-linear regime of $k\sim1\, h\, \mathrm{Mpc}^{-1}$ (see the left panel of Figure~\ref{figure:comparesims}), corresponding to relatively large spatial scales of $\sim10$~Mpc at $z=0$. This indicates that AGN feedback would be required to impact the matter distribution at greater distances from the central black hole than currently occurs in the FABLE model.  Motivated by this, our first radio-mode modification involves injecting the hot radio-mode bubbles at a greater distance from the black hole by increasing the $D_{\rm bub}$ parameter.  This mimics bubbles arising from AGN jets which have travelled further through the intervening ICM. This qualitatively shares some similarities with the hydrodynamical decoupling of AGN jets in the SIMBA model \citep{Dave:2019}, although the exact details of the implementation are considerably different. 

A major increase of $D_{\rm bub}$ from it's fiducial value at $z \sim 0$ would be expected to reduce the hot gas fraction measured within $r_{500}$, and possibly result in a poorer fit to the data. We therefore follow a similar approach to the previous Section~\ref{sec:quasarmodemod} and trial a redshift dependent $D_{\rm bub}$.  We test increasing $D_{\rm bub}$ at high redshifts, with the aim of redistributing matter on the larger scales and thus suppressing the mildly non-linear matter power spectrum. We then decrease $D_{\rm bub}$ with redshift at late time, with the purpose of ensuring gas can re-accrete and allow the simulated gas fractions to attain good agreement with low redshift observations. 

The RadioBoost-40 simulation demonstrates this model.  At $z>4$ we fix $D_{\rm bub}=500$~$h^{-1}$kpc. From $z=4$ to $z=0$ we then decrease $D_{\rm bub}$ linearly with redshift until it reaches the value of $D_{\rm bub}=30$~$h^{-1}$kpc at $z=0$.  We fix the bubble radius $R_{\rm bub}$ to its fiducial normalisation value of $50$~$h^{-1}$kpc.  Note that we therefore remove the re-scaling of $D_{\rm bub}$ and $R_{\rm bub}$ with the bubble energy and ICM density, given in Equation~\ref{eq:dbub} and Equation~\ref{eq:rbub}, to have more control over the model. Furthermore, to increase the proportion of black holes undergoing feedback in this modified radio-mode, we increase $\chi_{\rm radio}$ to $0.1$.  We keep all other aspects of the model, including the quasar-mode, as in fiducial FABLE. 

\subsection{RadioBoost$M_{\rm BH,radio}$-40: \textit{a boost to the radio-mode for the most massive SMBHs}}\label{sec:radioboostMbh}

The next modified AGN feedback model we test builds on the modifications to the radio-mode outlined in the previous Section~\ref{sec:radiomod}.  We discussed that increasing the matter power spectrum suppression on the larger, mildly non-linear scales, could be possible through an increase to the distance that the hot `AGN jet driven' bubbles are injected at from the black hole. We however cannot guarantee that gas can re-accrete sufficiently in late times to recover observed gas fractions, whilst maintaining the larger-scale power spectrum suppression we desire.  In this section, we therefore consider an alternate approach. 

The box exemplifying our new approach is labelled `RadioBoost$M_{\rm BH,radio}$-40'. Firstly, we increase the bubble distance to $D_{\rm bub}=100$~$h^{-1}$kpc at all redshifts. Furthermore, we allow the radio-mode to act in only the largest halos, motivated by the aim of preserving reasonable gas fractions in (small) groups, where feedback-induced gas expulsion occurs more easily for the lower mass systems with smaller binding energies. It has been shown that above a stellar mass of $10^{11}$ M$_{\odot}$, radio-loud AGN are possibly always `switched on' (see \citet{Hardcastle2020} and references therein). This also approximately corresponds to the threshold halo mass for which a sufficiently dense hot halo is in place such that it can confine the radio bubbles and ensure that a sufficient fraction of energy from bubbles can be transferred to the ICM.  

 We therefore allow radio-mode feedback to operate only in haloes that have a considerable `hot halo' component, corresponding roughly to $M_{500}\approx10^{13}$ M$_{\odot}$, $\log_{10}(M_* [\mathrm{M_{\odot}}])\approx11$ \citep[using the best fit][stellar mass - halo mass relationship]{Moster2010}, and $\log_{10}(M_{\mathrm{BH}} [\mathrm{M_{\odot}})\approx9$ \citep[using the best fit][stellar mass-black hole mass scaling relation]{Kormendy2013}. We retain the $\chi_{\rm radio}=0.1$ of the previous section, and we allow only halos accreting below this limit \textit{and} with black hole mass $\log_{10}(M_{\mathrm{BH}} [\mathrm{M_{\odot}}])>9$ to be in the radio-mode.  Black holes of mass $\log_{10}(M_{\mathrm{BH}} [\mathrm{M_{\odot}}])<9$ are only allowed to undergo quasar-mode feedback, regardless of their accretion rate. This model has some qualitative similarities to the adopted separation between the quasar and radio-mode in the IllustrisTNG simulation \citep{Weinberger2018}, but note that quantitative details and radio-mode implementation are different. 

\subsection{XFABLE: \textit{a pressure limited, boosted radio-mode for the most massive SMBHs}}\label{sec:xfable}

Finally, we consider an additional modification to the `RadioBoost$M_{\rm BH,radio}$-40' model described in the previous section.  The modifications to the fiducial FABLE radio-mode described thus far do not address the energy transferred to the ICM by the injected radio bubbles. In fact, in fiducial FABLE there is no physical cap on the pressure contrast between the inflating bubble and the ICM.  This can result in high Mach number shocks that are not typically observed around X-ray cavities \citep[see][for a review]{Fabian2012}.  Furthermore, dedicated high-resolution simulations of jets in galaxy clusters typically find that they are inflated in approximate pressure equilibrium \citep{Bourne2021,Hardcastle2013}.  Considering that the `RadioBoost$M_{\rm BH,radio}$-40' model may result in a more destructive feedback scenario since bubbles are injected further from the central black hole, we test additionally applying an upper limit on the energy of the bubble with respect to the ICM.  The box that exemplifies this modification is labelled `XFABLE', in which we ensure the energy content of the bubble is limited to $E_{\rm bub}/E_{\rm ICM}<20$.  This limit was chosen through a series of trial runs, with the aim of achieving a balance between injecting sufficient energy into the ICM to suppress the matter power spectrum and preventing thermodynamic profiles of the ICM deviating significantly from observations.

In addition to the  $(40~h^{-1}\mathrm{Mpc})^3$ volume ran with this model (XFABLE-40), we additionally run a $(100~h^{-1}\mathrm{Mpc})^3$ box for improved statistics of rare systems (XFABLE-100). 

\subsection{Visualisation of the FABLE simulation suite}

\begin{figure*}
	\centering
	\includegraphics[width=0.99\linewidth]{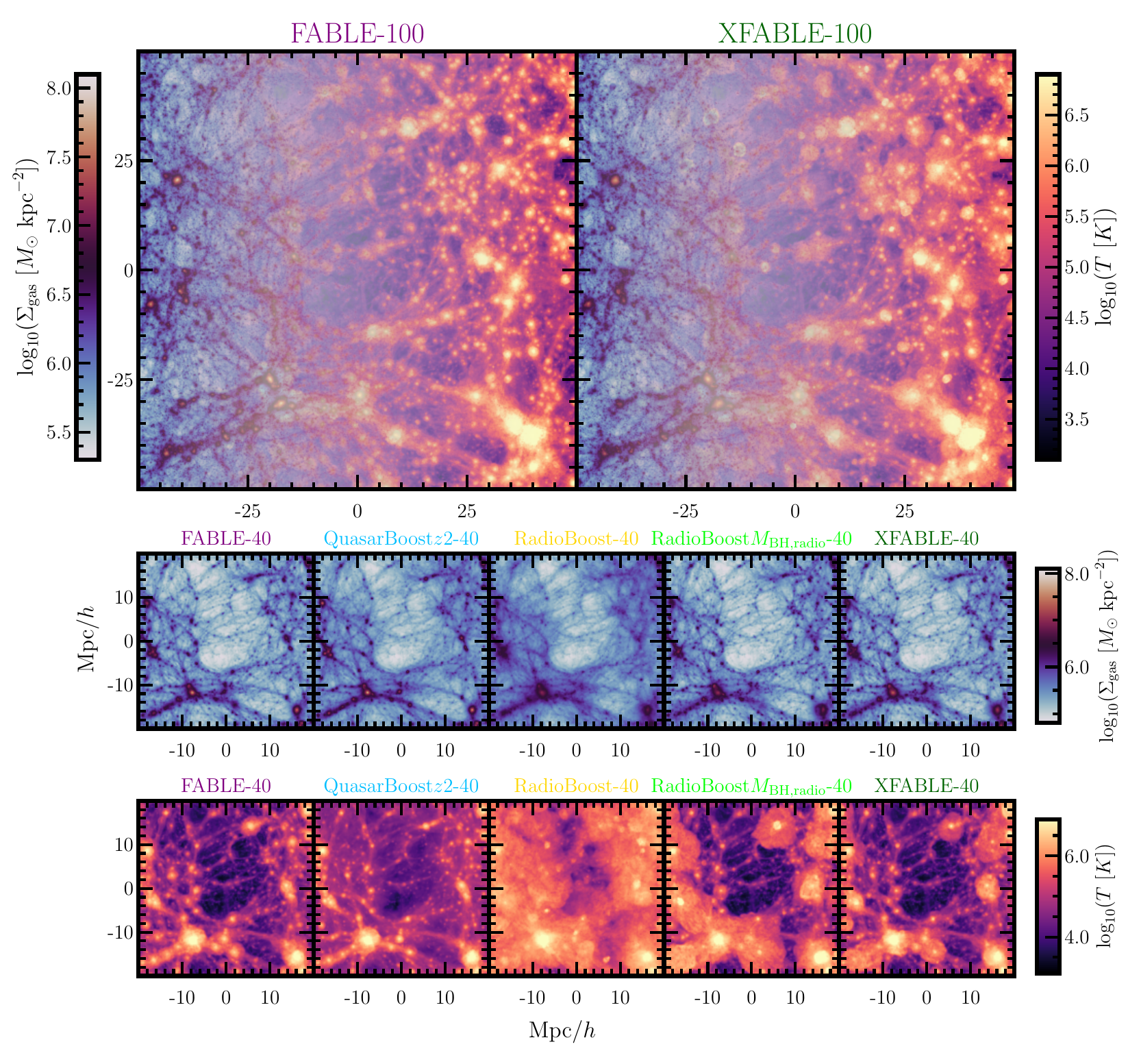} 
	\caption{Projections visualising the key simulation boxes analysed in this work.  The top panels display the blend of the gas surface density and mass-weighted temperature projections of the two $(100~h^{-1}\mathrm{Mpc})^3$ volumes: fiducial FABLE-100 and the modified AGN feedback model,  XFABLE-100. The middle panels show the gas surface density of the corresponding $(40~h^{-1}\mathrm{Mpc})^3$ FABLE-40 and XFABLE-40 boxes, in addition to the other modified AGN feedback models; QuasarBoost$z$2-40, RadioBoost-40 and RadioBoost$M_{\rm BH,radio}$-40. The lower panels show mass-weighted temperature projections for the same $(40~h^{-1}\mathrm{Mpc})^3$ volumes. All visualisations represent projections through the full depth of each simulation box at $z=0$.}    
\label{fig:projections}
\end{figure*}

Figure~\ref{fig:projections} shows visualisations of the $z=0$ large-scale structure formed in each of the simulation boxes introduced in this section.  Since the $(40~h^{-1}\mathrm{Mpc})^3$ and $(100~h^{-1}\mathrm{Mpc})^3$ boxes were ran with the same random seed for the initial conditions, we find that the cosmic web looks comparable between the volumes, with the largest clusters lying at approximately the same relative location between boxes. The $(100~h^{-1}\mathrm{Mpc})^3$ boxes however allows for rarer objects to form, with the three most massive clusters in FABLE-100 being of mass $M_{500}=4.05\times 10^{14}~ \mathrm{M_{\odot}}, 3.89\times 10^{14}~ \mathrm{M_{\odot}}, 3.20\times 10^{14}~ \mathrm{M_{\odot}}$, compared to $M_{500}=1.46\times 10^{14} ~\mathrm{M_{\odot}}, 1.01\times 10^{14}~ \mathrm{M_{\odot}}, 4.59\times 10^{13}~ \mathrm{M_{\odot}}$ in FABLE-40.

The surface gas mass density of the FABLE-100 and XFABLE-100 volumes look largely similar, albeit with XFABLE-100 displaying somewhat lower densities at the nodes.  In the $(40~h^{-1}\mathrm{Mpc})^3$ volumes we see greater variation in the surface gas density distribution between AGN feedback models.  In particular, the RadioBoost-40 and RadioBoost$M_{\rm BH,radio}$-40 visualisations reveal a more `fuzzy' gas distribution, with less defined filaments and nodes than fiducial FABLE-40. This largely arises from the choice of increased $D_{\mathrm{bub}}$ parameter in these AGN feedback models, which re-distributes gas to larger distances from the central SMBH.  The mass-weighted temperature projections further display the extremity of hot gas redistribution imposed by the RadioBoost-40 and RadioBoost$M_{\rm BH,radio}$-40 models.  The temperature projections also reveal clear deviations between the FABLE and XFABLE models, most notably that bubbles of hot gas around the largest clusters reach greater radii in XFABLE.

\section{Constraining AGN feedback models through comparison with observations}\label{sec: obs}

In this section, we test the ability of the models outlined in the previous section to reproduce a range of observed galaxy, supermassive black hole, galaxy group and galaxy cluster properties. We discuss models that can be ruled out as viable modifications to the FABLE feedback model, as well as the constraining power of specific observations to distinguish and/or exclude our theoretical models.  Throughout Section~\ref{sec: obs}, we plot results from $(40~h^{-1}\mathrm{Mpc})^3$ boxes using dashed lines and from the $(100~h^{-1}\mathrm{Mpc})^3$ FABLE and XFABLE boxes using solid lines. Observational measurements are shown in grey. 

\subsection{The suppression of the matter power spectrum}\label{sec:pkboxes}

\begin{figure*}
\centering
\includegraphics[width=0.38\textwidth,keepaspectratio]{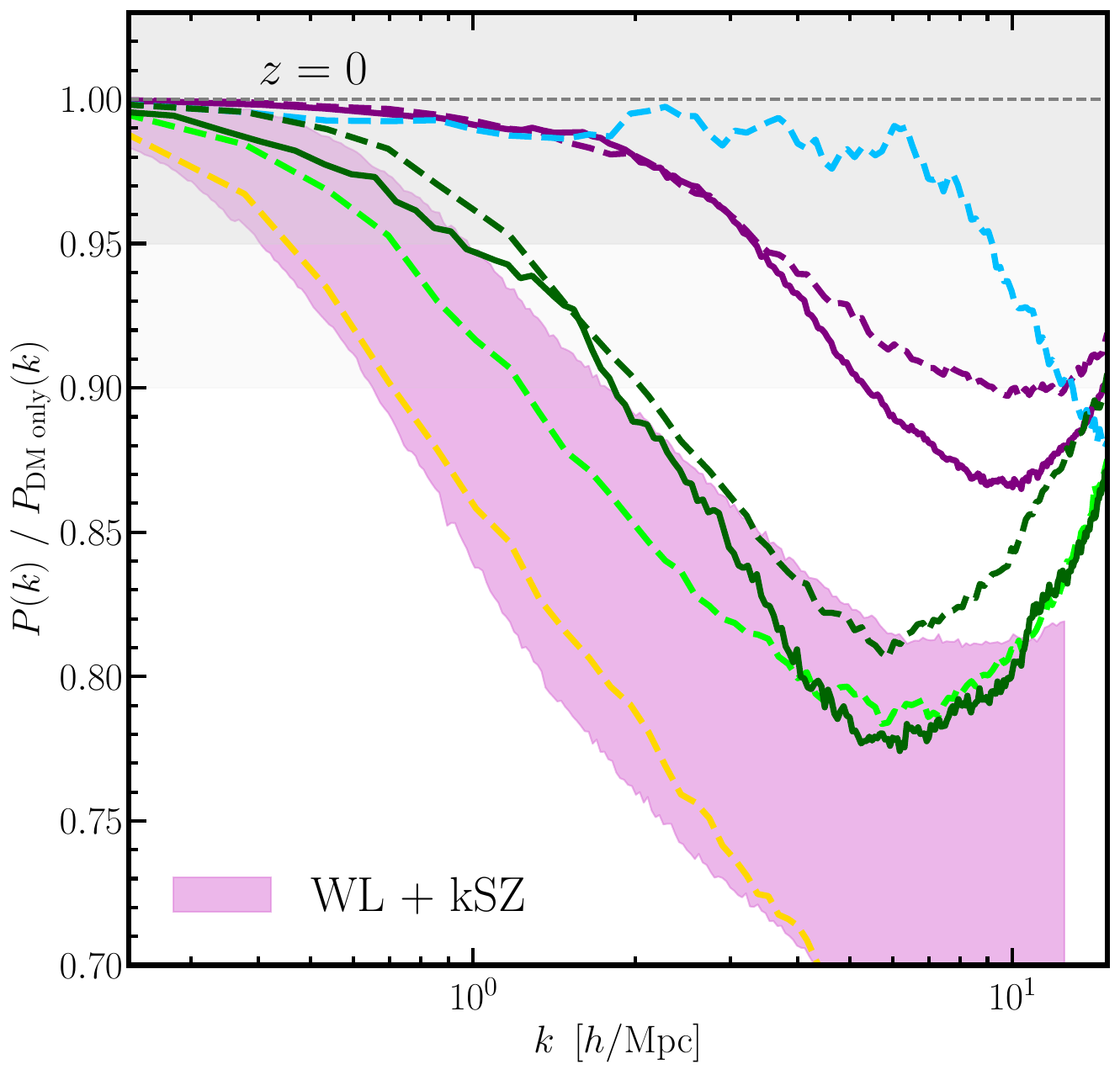}
\includegraphics[width=0.38\textwidth,keepaspectratio]{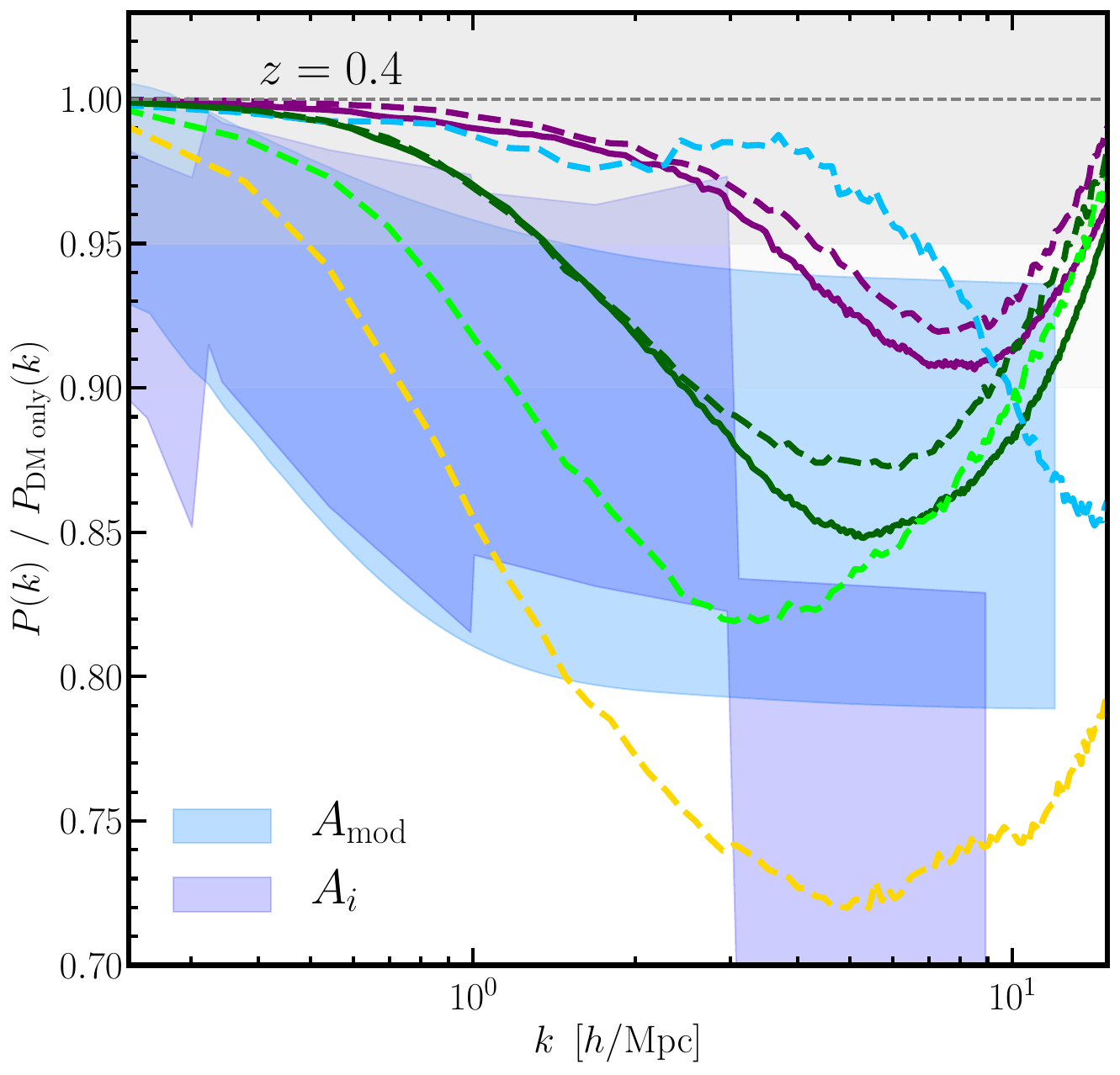}
\includegraphics[width=0.38\textwidth,keepaspectratio]{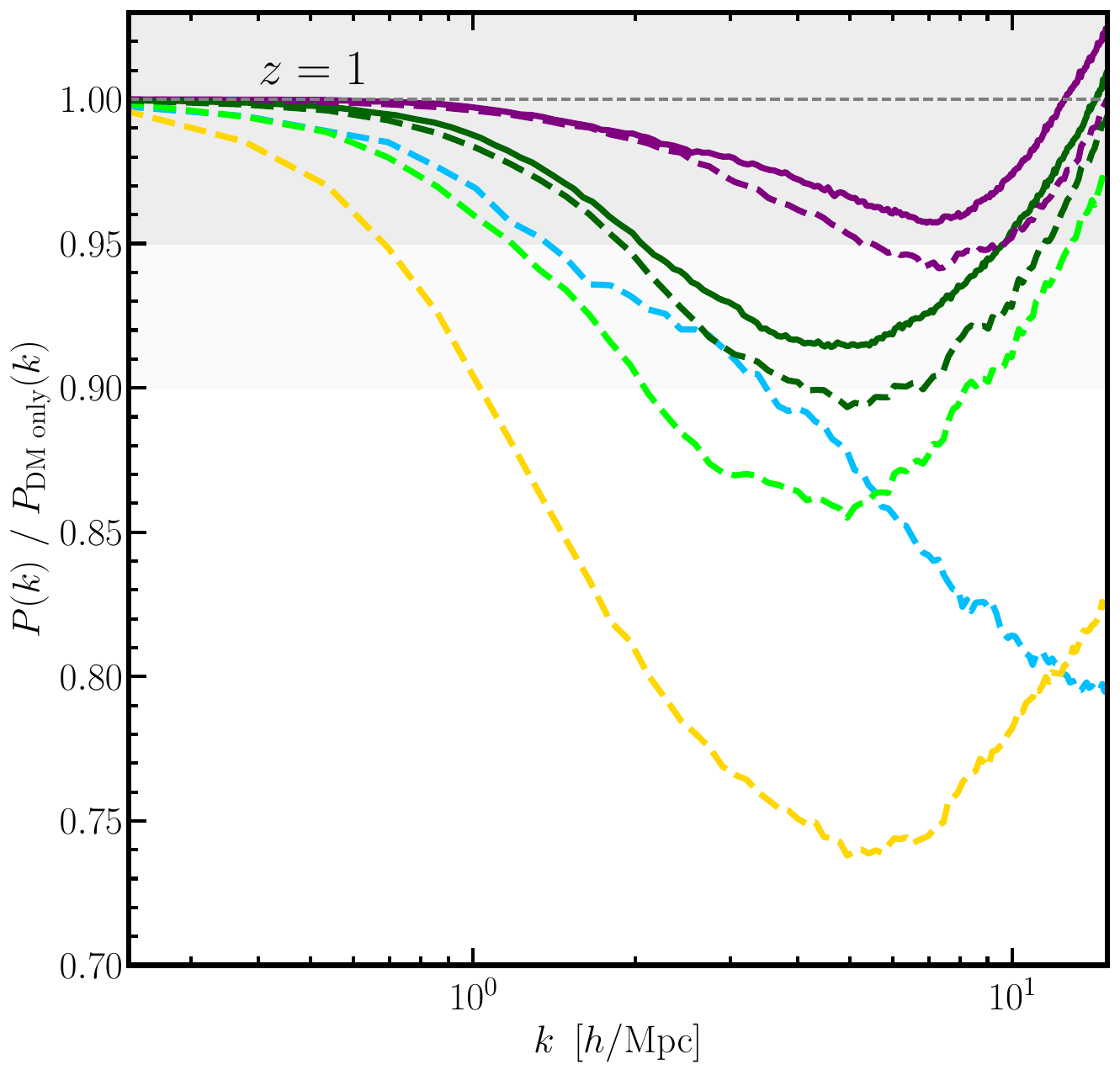}
\includegraphics[width=0.38\textwidth,keepaspectratio]{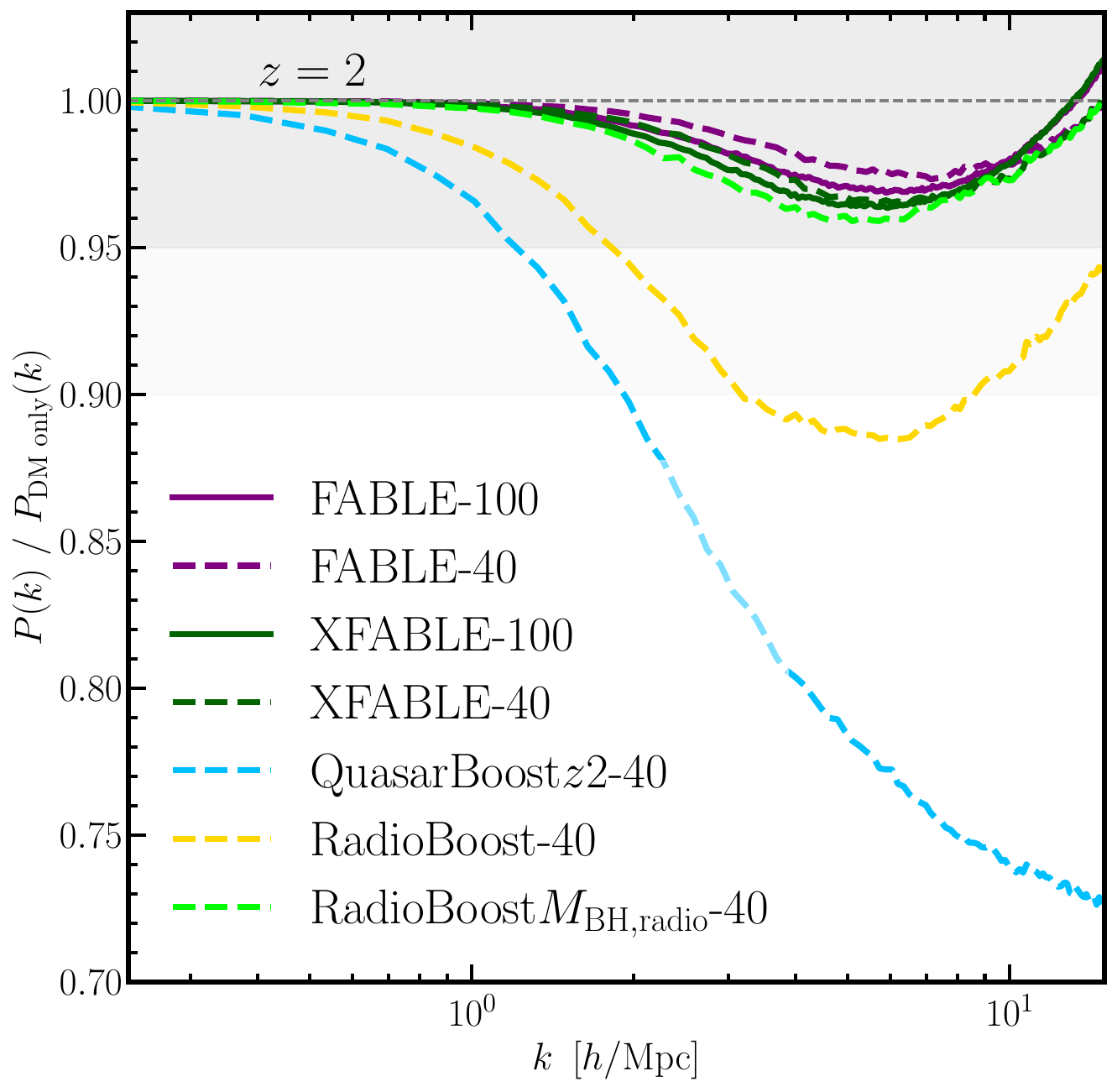}
\caption{The matter power spectrum suppression due to baryonic effects measured in each of our key simulation boxes, plotted at $z=0$ (upper left), $z=0.4$ (upper right), $z=1$ (lower left) and $z=2$ (lower right).  Dashed lines denote $(40~h^{-1}\mathrm{Mpc})^3$ boxes, and solid lines show $(100~h^{-1}\mathrm{Mpc})^3$ boxes.  We show the fiducial FABLE boxes (purple) and our key modified AGN feedback models; QuasarBoost$z$2-40 (light blue), RadioBoost-40 (yellow), RadioBoost$M_{\rm BH,radio}$-40 (light green) and XFABLE (dark green).  At $z=0$ we plot as the purple shaded region the constraints attained in a combined DES Y3 cosmic shear and ACT kinetic Sunyaev–Zel’dovich (WL+kSZ) analysis, presented in \citet{Bigwood2024}. At $z=0.4$ we plot as the light blue shaded region the $A_{\mathrm{mod}} = 0.858 \pm 0.052$ constraints of \citet{Preston2023} required to reconcile cosmology attained in a DES Y3 cosmic shear analysis with the \textit{Planck} best-fit $\Lambda$CDM model. The dark blue shaded region show the corresponding $1\sigma$ constraints when splitting the model into bins in wave-number, $A_i$.}
\label{fig:mods_pk}
\end{figure*}

We begin by discussing the baryonic suppression of the matter power spectrum with respect to a dark-matter only simulation, as predicted by FABLE and each of the modified feedback boxes. Figure~\ref{fig:mods_pk} shows the suppression at $z=0,1,2$ and at $z=0.4$, with the latter redshift plotted for comparison to the $A_{\rm mod}$ model, which we recall is the model required to produce enough suppression to be a viable solution to the $S_8$ tension (see Section~\ref{sec:pkcompare}).  We also plot at $z=0.4$ $A_{\rm mod}$ binned in $k$-space, $A_i$, to highlight the scale-dependence of the suppression \citep{Preston2023}. At $z=0$ we also compare to the constraints attained by the combined weak lensing and kinetic Sunyaev–Zel’dovich (WL+kSZ) analysis presented in \citet{Bigwood2024} (purple band). These constraints were computed at $z=0$, in contrast to the constraints attained using the $A_{\rm mod}$ model, which has no explicit redshift dependence, and hence is plotted at $z \sim 0.4$ where the total DES Y3 redshift distribution peaks.

In line with \citet{Martin-Alvarez2024}, we find that the suppression produced by the fiducial FABLE boxes (purple) increases with time, with the suppression pushing to larger $k$-scales and a greater amplitude with decreasing redshift.  This trend is observed in each of the modified feedback boxes, with the exception of QuasarBoost$z$2-40 (light blue). In this model, the high-redshift quasar-mode boost, and thus the increased thermal energy supplied to the gas, has significantly increased the power suppression at $z=2$ with respect to fiducial FABLE, as expected. Specifically, at $k=10\,h\,\mathrm{Mpc}^{-1}$ the suppression increases from $\sim2\%$ to $\sim25\%$ in QuasarBoost$z$2-40, and interestingly produces suppression on scales as large as $k\sim0.5\,h\, \mathrm{Mpc}^{-1}$.  We, however, find that the power suppression decreases in the QuasarBoost$z$2-40 model beyond $z<2$, and produces less suppression than fiducial FABLE on scales $1<k<10\,h\, \mathrm{Mpc}^{-1}$ at $z<0.4$.  Furthermore, the box is unable to reproduce the $A_{\rm mod}$ suppression required at $z=0.4$, and therefore this modification to the AGN feedback model cannot provide a solution to the $S_8$ tension, hinting at a need for sufficiently strong AGN feedback at lower redshifts as well.    

The models RadioBoost-40 (yellow), RadioBoost$M_{\rm BH,radio}$-40 (light green) and XFABLE (dark green) are all able to produce a greater non-linear suppression of the matter power spectrum than fiducial FABLE for $0.1<k<10\, h\, \mathrm{Mpc}^{-1}$ at $0<z<2$. In particular, RadioBoost-40 predicts the most extreme baryonic impact of all illustrative models shown, and at $z=0$ suppresses the power spectrum by $\sim25$\% at $k=3\, h\, \mathrm{Mpc}^{-1}$ (compared to a $\sim5$\% suppression in the fiducial FABLE box).  It lies consistent with both the $A_{\rm mod}$ and $A_i$ models at $z=0.4$, and the constraint attained from a weak lensing and kSZ combined analysis at $z=0$ \citep{Bigwood2024}.  Since the RadioBoost$M_{\rm BH,radio}$-40 model restricts a `boosted' radio-mode to act for only the most massive black holes in the box, we inevitably find that this model predicts a less extreme suppression at all $k$ than RadioBoost-40, at $z=0$ suppressing the power spectrum by $\sim17$\% at $k=3\, h\, \mathrm{Mpc}^{-1}$. RadioBoost$M_{\rm BH,radio}$-40 is however still able to lie within both the $A_{\rm mod}$ and WL + kSZ bands. The additional pressure limit on the AGN bubbles imposed in the XFABLE box further reduces the suppression measured at all $k$ and $z$ shown. XFABLE attains a $z=0$ suppression of $\sim13$\% at $k=3\, h\, \mathrm{Mpc}^{-1}$, approximately 2.5 greater than measured in FABLE.  This brings XFABLE to lie within the $1\sigma$ WL+kSZ constraints of \citet{Bigwood2024} at nearly all scales at which we expect the matter power spectrum to be suppressed due to feedback, i.e. within $0.1\, h\, \mathrm{Mpc}^{-1}<k<10\, h\, \mathrm{Mpc}^{-1}$. XFABLE also lies consistently within the $1\sigma$ $A_{\rm mod}$ and $A_i$ model predictions at $k\gtrsim1\, h\, \mathrm{Mpc}^{-1}$, lying within $2\sigma$ at the larger, mildly non-linear scales.

The default FABLE and QuasarBoost$z$2-40 models fail to match the the power suppression constrained by weak lensing and kSZ observations, while the RadioBoost-40, RadioBoost$M_{\rm BH,radio}$-40 and XFABLE models are consistent with the observations, and the predicted suppression to resolve the $S_8$ tension. It is interesting to note that our comparison between different AGN feedback models does not favour very strong, ejective AGN feedback at cosmic noon or higher redshifts (as advocated by basic pre-heating scenarios), given that there is sufficient cosmic time for this ejected gas to re-fall back within the galaxy groups and clusters at low redshifts, and hence significantly reduce the suppression of the matter power spectrum where we have the best observational constraints. Instead, AGN feedback that regulates their host properties seems to be required at lower redshifts as well, in accord with the observed presence of radio jets and lobes in local galaxy groups and clusters. 

\subsection{Galaxy and supermassive black hole population 
properties}

\subsubsection{The galaxy stellar mass function}\label{sec:gsmf_analysis}

\begin{figure*}
\centering
\includegraphics[width=0.99\textwidth]{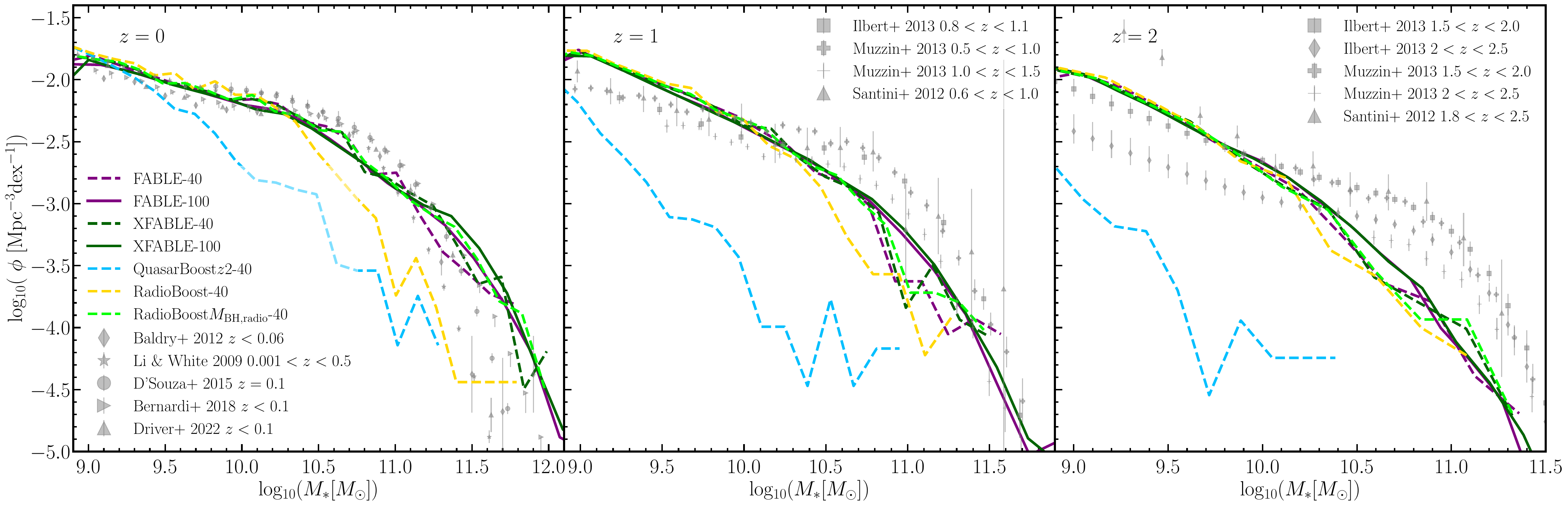}

\caption{The galaxy stellar mass function measured within twice the stellar half-mass radius in each of our key simulation boxes, plotted at $z=0$ (left), $z=1$ (centre), $z=2$ (right). Dashed lines denote $(40~h^{-1}\mathrm{Mpc})^3$ boxes, and solid lines show the $(100~h^{-1}\mathrm{Mpc})^3$ FABLE and XFABLE boxes.  We show the fiducial FABLE boxes (purple) and our key modified AGN feedback models; QuasarBoost$z$2-40 (light blue), RadioBoost-40 (yellow), RadioBoost$M_{\rm BH,radio}$-40 (light green) and XFABLE (dark green).  The grey datapoints show observational constraints.  At $z=0$ we plot the results of \citet{Baldry2012} ($z<0.06$), \citet{Li2009} ($0.001<z<0.5$), \citet{DSouza2015} ($z=0.1$)
\citet{Bernardi2018} $z<0.1$) and \citet{Driver2022} ($z<0.1$).  At $z=1$ we plot \citet{Ilbert2013} ($0.8<z<1.1$), \citet{Muzzin2013} (plotting both $0.5<z<1.0$ and $1.0<z<1.5$) and \citet{Santini2012} ($0.6<z<1.0$).  Similarly at $z=2$ we plot \citet{Ilbert2013} (plotting both constraints for $1.5<z<2.0$ and $2.0<z<2.5$), \citet{Muzzin2013} (plotting both constraints for $1.5<z<2.0$ and $2.0<z<2.5$) and \citet{Santini2012} ($1.8<z<2.5$).  Note that we convert all observational measurements to a \citet{Chabrier2003} IMF.  We show that XFABLE remains an equally good fit to the observations as FABLE, and that the QuasarBoost$z$2-40 and RadioBoost-40 boxes are ruled out by the data. }
\label{fig:gsmf}
\end{figure*}

Figure~\ref{fig:gsmf} shows the GSMF at $z=0,1$ and $2$ for FABLE and each of our four illustrative modified AGN feedback models. We compare to observational results measured using data attained from a number of surveys and fields; \citet{Baldry2012} (Galaxy And Mass Assembly, GAMA), \citet{Li2009} (Sloan Digital Sky Survey, SDSS), \citet{DSouza2015} (SDSS), \citet{Bernardi2018} (SDSS), \citet{Driver2022} (GAMA), \citet{Ilbert2013} (UltraVISTA), \citet{Muzzin2013} (COSMOS/UltraVISTA) and \citet{Santini2012} (Wide Field Camera 3, WFC3). The redshift ranges of the data plotted is shown in the caption. In line with \citet{Henden2018} and by construction, the fiducial FABLE boxes display a very good agreement with the observations at $z=0$.  At $z=1$ FABLE under-estimates the knee of the GSMF compared to the data, and at $z=2$ the simulated GSMF is systematically lower for $\log_{10}(M_{\rm *} [M_{\rm \odot}] > 10.3$ (also seen in \citet{Henden2018}), but maintains a broadly good qualitative agreement.

The QuasarBoost$z$2-40 model provides the poorest fit to the data, significantly underestimating the stellar mass throughout the galaxy population.  The boost to the quasar-mode increases thermal energy injected into the central galaxy at $z>2$, over quenching the star formation predicted at $z=2$ compared to observations. The GSMF begins to recover towards $z=0$ with the implementation of the fiducial quasar-mode parameters at $z<2$ due to gas fallback, however sufficient stellar mass cannot be formed to match the data at $z=0$. This strongly indicates (even considering significant changes in the stellar feedback sector) that strong, ejective central gas removal due to AGN feedback is disfavoured,  generating unrealistic star formation histories of the entire galaxy population.

At $z\leq1$, the RadioBoost-40 model also under-estimates the massive tail ($\log_{10}(M_{\rm *} [M_{\rm \odot}])>10.5$) of the GSMF. Recall, that this model imposes a linearly decreasing bubble distance with redshift, as well as ensuring that more black holes are in the radio-mode at a given time with an increased $\chi_{\rm radio}$. We compare to further observables to diagnose the source of the low GSMF (see Section~\ref{sec:gasfracanalysis} and Section~\ref{sec:tszanalysis}).  

Both the RadioBoost$M_{\rm BH,radio}$-40 and XFABLE models are in excellent agreement with the fiducial FABLE simulation and maintain the same level of agreement to observations at $z=0,1$ and $2$. This indicates that fixing $D_{\rm bub}=100$~$h^{-1}$kpc at all times, in addition to allowing the radio-mode to act in only the most massive halos, is able to prevent the over-heating/ejection of gas within the central galaxy and maintain realistic star formation. The addition of the AGN bubble pressure limit in XFABLE has no appreciable impact on the GSMF, likely because the radio-mode is heating only the outskirts of galaxies and therefore the stellar population remains largely unaffected. 

Comparison to observed GSMF can therefore rule out the QuasarBoost$z$2-40 and RadioBoost-40 models, as they lead to extreme stellar mass over-quenching, with the default FABLE, RadioBoost$M_{\rm BH,radio}$-40 and XFABLE models remaining in good agreement with the data. We conclude from this analysis that any significant variations of the AGN feedback which could lead to sufficiently large matter power spectrum suppression, need to largely act in galactic outskirts rather than galaxy central regions. Keeping the same stellar feedback model as in FABLE, it seems hard to reconcile the observed matter power spectrum suppression and GSMF data for strong, centrally-ejective AGN feedback models, hinting that the AGN feedback needs to be largely `preventative' and/or act on large scales. This point can be made even stronger, by noting that FABLE over-quenches galaxies at the massive end at $z \sim 2$, which implies that either stellar and/or AGN feedback acting in galaxy cores at early cosmic times before the cosmic noon is too powerful. Resolving this issue would either require less strong feedback overall at high cosmic times or AGN feedback which is more `preventative' and/or ejective but at large distances from galactic centres.  

\subsubsection{The bolometric quasar luminosity function}\label{sec:lumfunc}

\begin{figure*}
\centering
\includegraphics[width=0.99\textwidth]{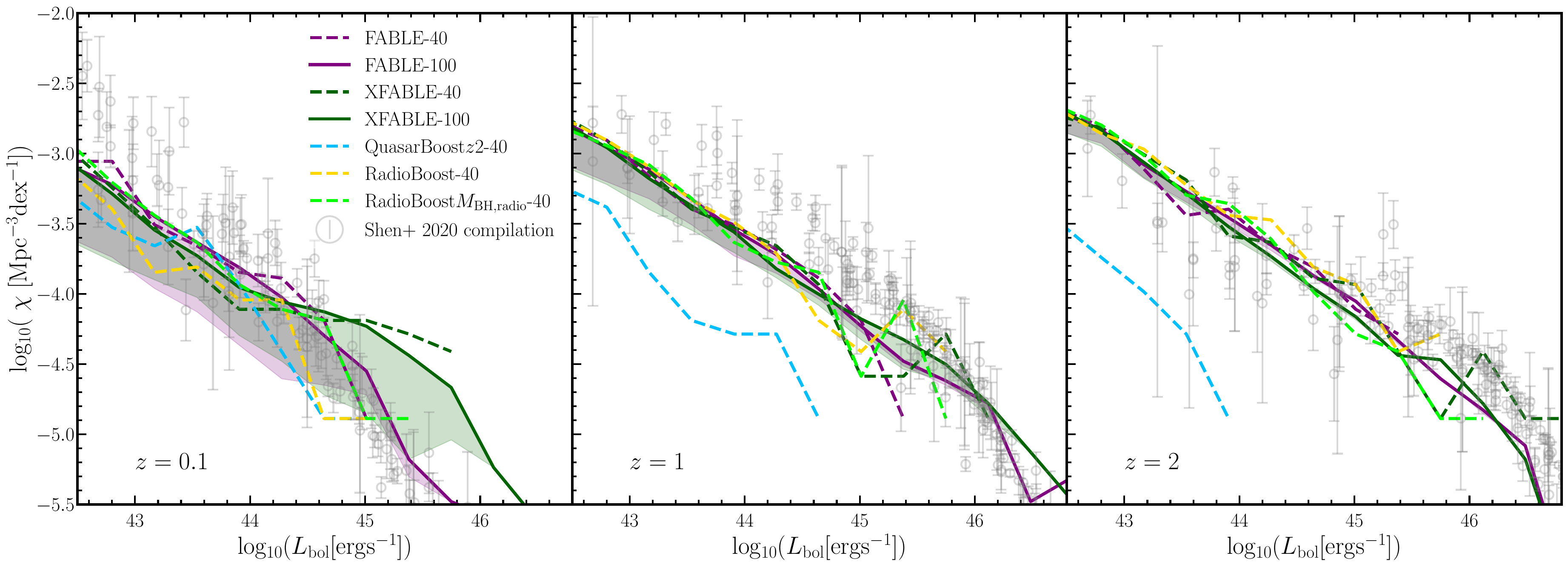}

\caption{The bolometric quasar luminosity function for each of our key simulation models, plotted at $z=0.1$ (left), $z=1$ (centre), $z=2$ (right). Dashed lines denote $(40~h^{-1}\mathrm{Mpc})^3$ boxes, and solid lines show the $(100~h^{-1}\mathrm{Mpc})^3$ FABLE and XFABLE boxes. We show the fiducial FABLE boxes (purple) and our key modified AGN feedback models; QuasarBoost$z$2-40 (light blue), RadioBoost-40 (yellow), RadioBoost$M_{\rm BH,radio}$-40 (light green) and XFABLE (dark green).  Lines show the quasar luminosity function computed under the assumption that all AGN are radiatively efficient. For the FABLE-100 and XFABLE-100 boxes, we add a shaded area that brackets the predicted luminosity function spanned by this assumption and that accounting for the radiatively inefficient AGN population at low Eddington ratios (see Section~\ref{sec:qlfcalc}). The grey datapoints show observational constraints from the \citet{Shen2020} compilation.  We demonstrate that the $z=0.1$ QLF is robust to the differences in the feedback model we show, but that the QuasarBoost$z$2-40 model is ruled out at $z\geq1$.   }
\label{fig:qlf}
\end{figure*}

Since quasars are the most luminous non-transient objects in the Universe, they can be detected and characterised to beyond $z>7$. As a result, the redshift evolution of the quasar luminosity function (QLF) provides a unique window into the growth of the active SMBH population and therefore is a key reference to compare our simulated SMBHs to. We compare our simulated quasar populations against the compilation of observational measurements by \citet{Shen2020}, which includes quasar samples measured in the optical/UV, X-ray and infrared bands. Figure~\ref{fig:qlf} plots the simulated and observed bolometric QLFs at $z=0.1$, $z=1$ and $z=2$, with the former redshift plotted due to the greater availability of data for comparison at $z=0.1$ rather than $z=0$.

At $z=0.1$, we find a good agreement between each of the modified AGN feedback models, fiducial FABLE and the observational data. This implies that the number density of low redshift quasars and their luminosities are relatively robust to the explored changes in AGN feedback prescriptions, and based on the QLF we cannot easily differentiate between models that predict a vastly different suppression of the matter power spectrum\footnote{Furthermore, we have analysed $M_{\mathrm{BH}}-M_*$ scaling relation for all of our AGN feedback models and compared them to the observationally derived relations of \citet{Reines2015} and \citet{Greene2020}. Due to the scatter in the available observational data, we find that we cannot exclude any of our modified feedback boxes using the $M_{\mathrm{BH}}-M_*$ scaling relation alone.}. We further note that at the bright end, XFABLE appears to overproduce the number density of the most luminous quasars in the box ($\log_{10}(L_{\mathrm{bol}} [\mathrm{erg}\,{\rm s}^{-1}])>45$) if we naively assume that all supermassive black holes are radiatively efficient. Accounting for a population of radiatively inefficient accretors at low Eddington ratios (with the shaded areas bracketing the predicted FABLE and XFABLE QLFs spanned by the assumption that all AGN are radiatively efficient (as is often assumed in luminosity functions derived from hydrodynamical simulation), and a calculation based on explicitly distinguishing between the luminosities of radiatively efficient and radiatively inefficient AGN (computed as detailed in Section~\ref{sec:qlfcalc})) largely removes this discrepancy, highlighting the importance of accurately computing radiative efficiencies. We finally note that all models somewhat under-predict the $z=0.1$ QLF at the faint end (($\log_{10}(L_{\mathrm{bol}} [\mathrm{erg}\,{\rm s}^{-1}])<43.5$), which indicates that the observed population is likely accreting more efficiently than in FABLE-like models in lower mass galaxies (we further caveat that we do not model X-ray binaries in this work). This intriguingly points towards a scenario of potentially greater feedback from these low luminosity AGN than modelled in the FABLE-like models (see also detailed discussion in \citet{Koudmani2022}). Note that the simulated QLFs should be seen in the context of current cosmological hydrodynamical simulations which show a significant uncertainty in predicting the bolometric QLF (see Figure 5 of \citet{Habouzit2022}).

At $z>1$ each of the radio-mode modifications, i.e. RadioBoost-40, RadioBoost$M_{\rm BH,radio}$-40 and XFABLE also do not deviate from the prediction by the fiducial FABLE model and agree very well with observational data. Hence, we can further infer that H's growth is not significantly impacted by the radio-mode bubbles acting far from the galaxy centre at these redshifts.  Despite recovering by $z=0.1$, the QuasarBoost$z$2-40 model significantly underestimates the quasar luminosity function at $z=1$ and $z=2$.  From this, we deduce that the outflows resulting from the increased thermal feedback drive too much gas away from central SMBHs, preventing their growth and therefore reducing the number density of luminous quasars. This same process overquenches the central galaxies as shown in Section~\ref{sec:gsmf_analysis}.

\subsection{Global properties of galaxy groups and clusters}

\subsubsection{Hot gas mass fractions}\label{sec:gasfracanalysis}

\begin{figure*}
\centering
\includegraphics[width=0.38\textwidth,keepaspectratio]{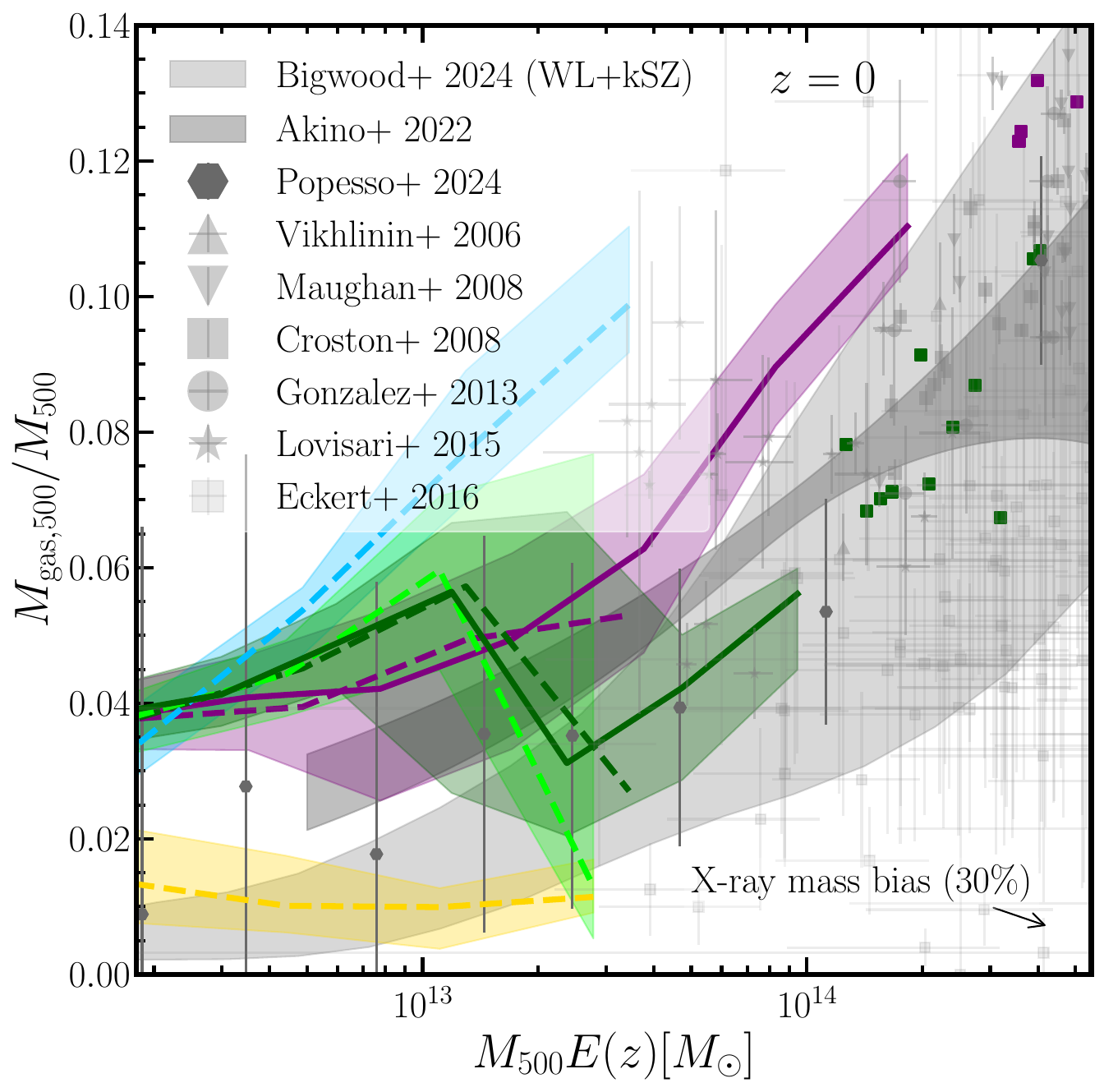}
\includegraphics[width=0.38\textwidth,keepaspectratio]{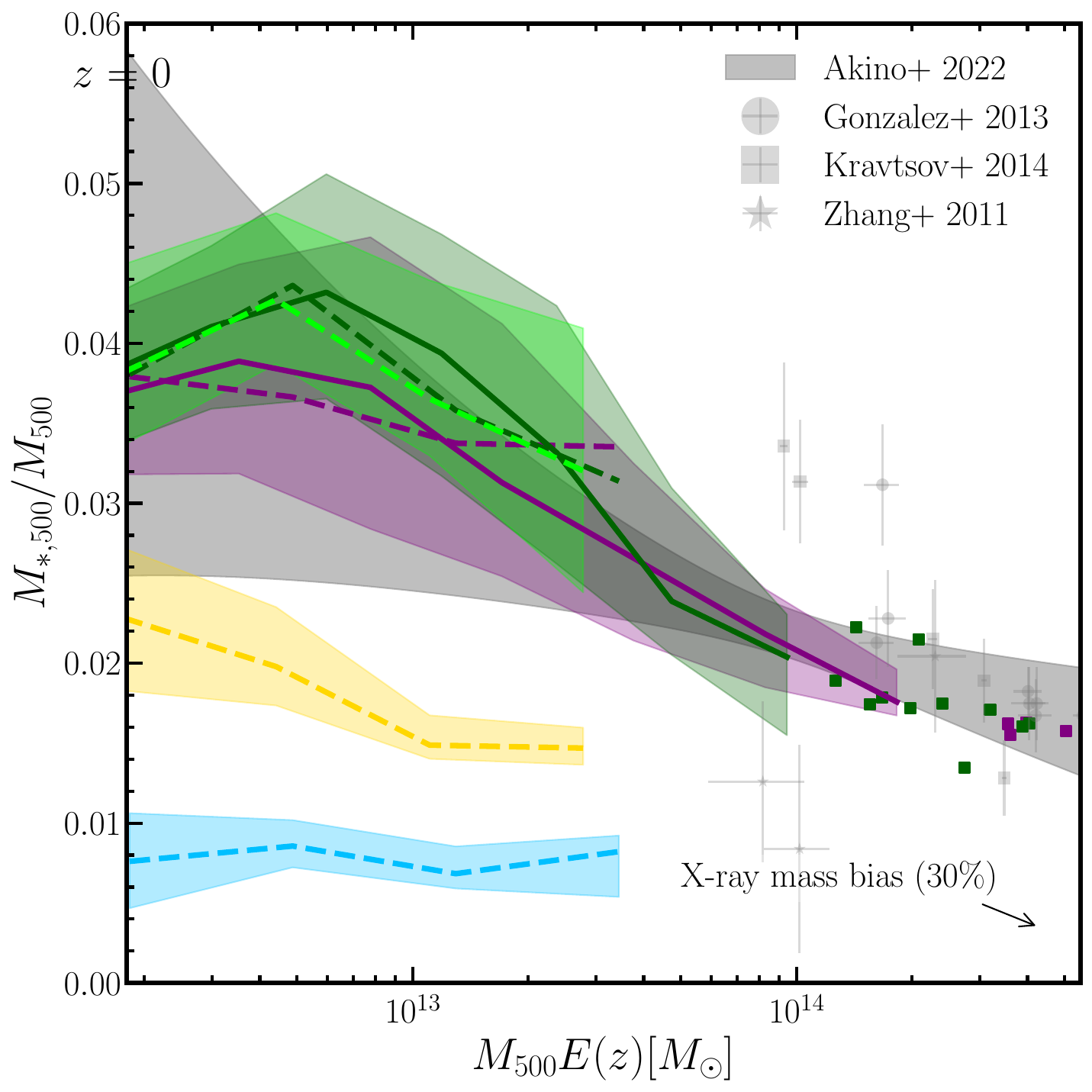}
\includegraphics[width=0.38\textwidth,keepaspectratio]{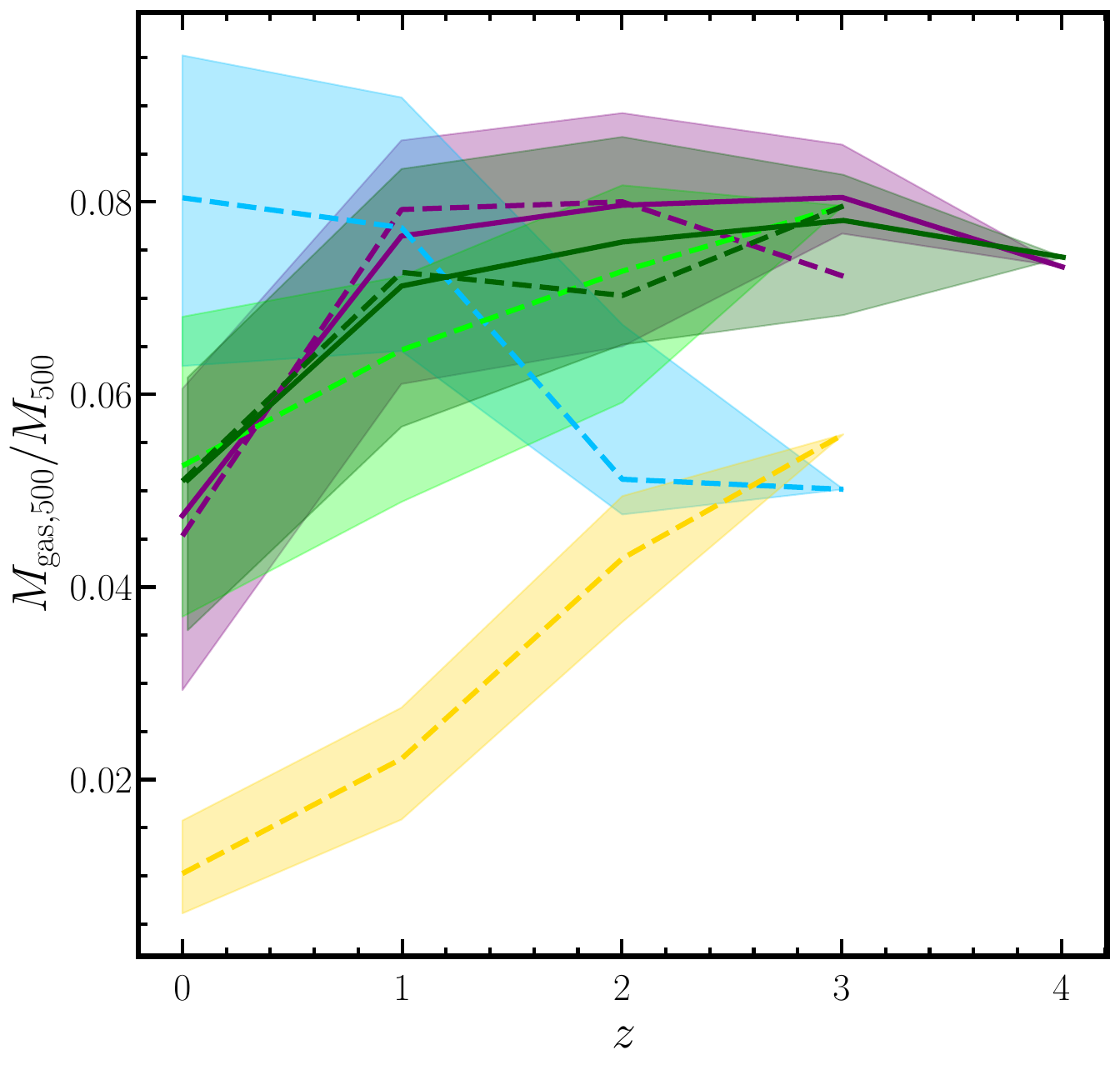}
\includegraphics[width=0.38\textwidth,keepaspectratio]{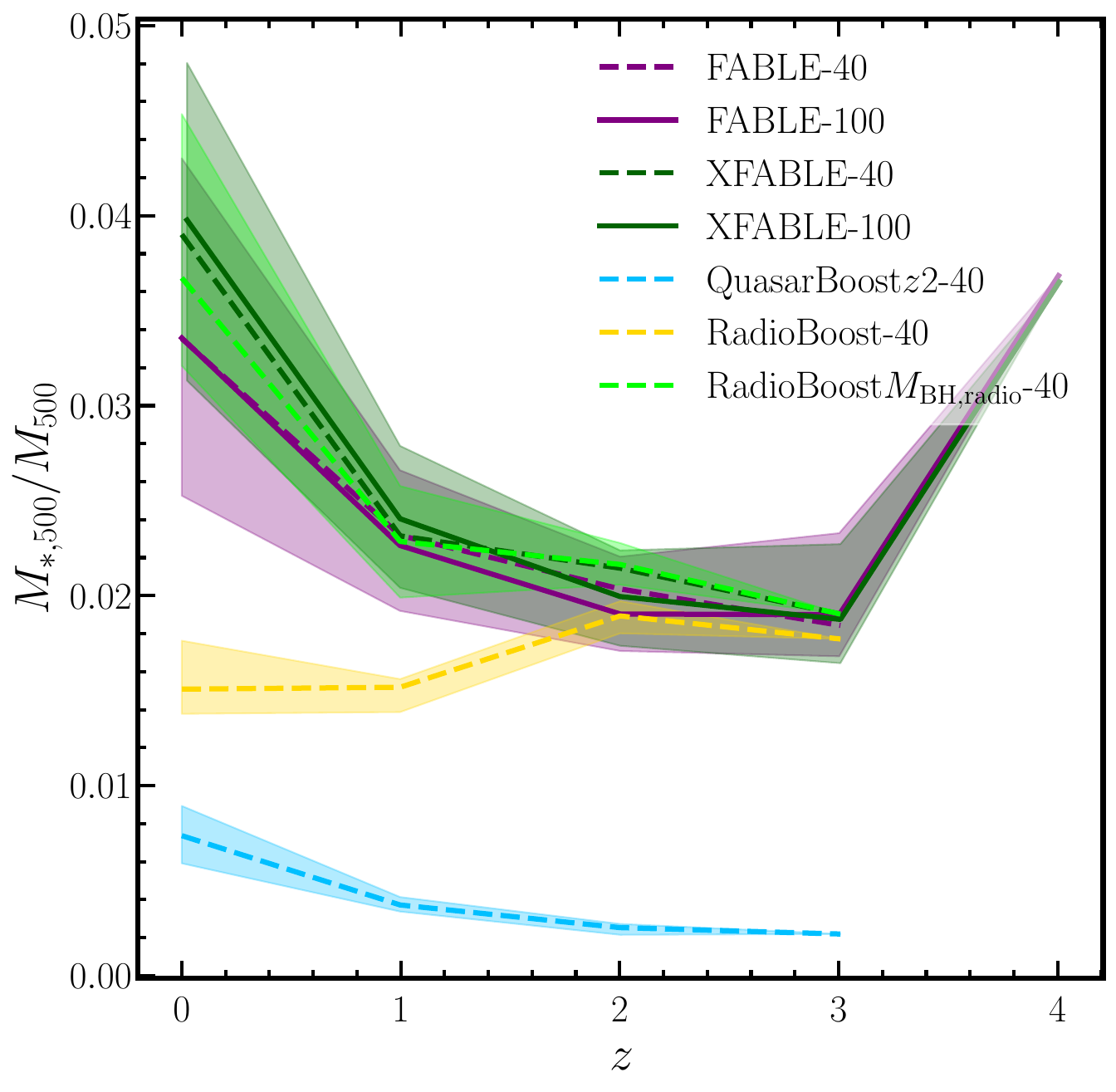}
\caption{The hot gas mass and stellar mass fractions in each of our key simulation models, measured within $r_{500}$.  Dashed lines denote $(40~h^{-1}\mathrm{Mpc})^3$ boxes, and solid lines show the $(100~h^{-1}\mathrm{Mpc})^3$ FABLE and XFABLE boxes. For each box, the solid/dashed lines denote the median relation, and the shaded regions span the upper and lower quartiles of the distribution.  We do not show the quartile regions of the FABLE-40 and XFABLE-40 boxes to avoid overcrowding the figure. For FABLE-100 and XFABLE-100, we plot the most massive systems that cannot be binned due to poor statistics as individual datapoints.  We show the fiducial FABLE boxes (purple) and our key modified AGN feedback models; QuasarBoost$z$2-40 (light blue), RadioBoost-40 (yellow), RadioBoost$M_{\rm BH,radio}$-40 (light green) and XFABLE (dark green). \textit{Upper left:} The hot gas mass fraction as a function of halo mass $M_{\rm 500}$ at $z=0$. The grey datapoints are the observation derived measurements of \citet{Popesso2024} ($z<0.2$), \citet{Vikhlinin2006} ($z<0.25$), \citet{Maughan2008} ($0.1<z<1.3$), \citet{Croston2008} ($z<0.2$), \citet{Gonzalez2013} ($z<0.2$), \citet{Lovisari2015} ($z<0.4$) and \citet{Eckert2016} ($0.05<z<1.1$).  The light grey shaded region shows the $1\sigma$ constraints derived from the joint weak lensing + kSZ analysis of \citep{Bigwood2024} and the dark grey shaded regions show the $1\sigma$ constraints of \citet{Akino_2022} ($z<1$). \textit{Upper right:} The total stellar mass fraction as a function of halo mass $M_{500}$ at $z=0$. The grey datapoints are the observationally derived measurements of  \citet{Gonzalez2013} ($z<0.2$), \citet{Kravtsov2018} ($z<0.1$) and \citet{Zhang2011} ($z<0.035$).  The grey shaded regions show the $1\sigma$ constraints of \citet{Akino_2022} ($z<1$).  An arrow is added to upper panels, indicating what would be the likely effect on X-ray-derived observations if one would correct for a hydrostatic mass bias of 30\%. \textit{Lower panels:} The redshift evolution of the hot gas mass fraction (left) and the total stellar mass fraction (right) for halos $M_{\rm 500}>5\times 10^{12}M_{\rm \odot}$.  We demonstrate that alike FABLE, XFABLE also shows a good agreement with the available data at $z=0$. }
\label{fig:fracs}
\end{figure*}

The upper left panel of Figure~\ref{fig:fracs} shows the $z=0$ hot gas mass fractions in the simulated groups and clusters for each illustrative AGN feedback model, in comparison to observations at $z\sim0$.  As in Figure~\ref{figure:comparesims}, we plot the gas mass-halo mass of \citet{Akino_2022} derived from the XXL X-ray selected sample, using the Hyper Suprime-Cam's photometry and weak-lensing mass measurements.  We also plot a range of X-ray observations from \citet{Vikhlinin2006} (Chandra), \citet{Maughan2008} (Chandra ACIS-I), \citet{Croston2008} (XMM-Newton REXCESS), \citet{Gonzalez2013} (XMM-Newton), \citet{Lovisari2015}  (XMM-Newton, selected using the ROSAT All-sky Survey) and \citet{Eckert2016} (XXL-100-GC clusters from XXM-Newton).  The redshift ranges of the data plotted is shown in the figure caption. We add an arrow to Figure~\ref{fig:fracs} indicating what would be the likely effect on X-ray-derived observations if one would correct for a hydrostatic mass bias of 30\%.

Recently, several observational studies have found evidence that gas mass fractions in group-mass systems may be lower than previous measurements derived using X-ray bright groups.  
These include the constraints of \citet{Popesso2024}, which measures the gas mass fractions in optically selected groups using eROSITA. Their optical selection aims to circumvent the potential biases that previous measurements of X-ray bright groups may have been susceptible to; namely that flux-limited X-ray selected samples can miss groups that have undergone feedback-induced gas removal and therefore have reduced X-ray luminosities.  This may result in previous measurements of X-ray bright groups overestimating gas mass fractions in group-mass systems \citep{Popesso2024}.  Independently, 
the joint weak lensing + kSZ analysis of \citet{Bigwood2024} also constrained gas mass fractions in groups to be lower than the existing X-ray measurements.  We therefore add these recent constraints to Figure~\ref{fig:fracs}.

In agreement with \citet{Henden2018}, the fiducial FABLE-100 box provides a good fit to the observational datapoints at $z=0$.  We note however that in light of the recent group-mass constraints of \citet{Popesso2024} and \citet{Bigwood2024} favouring lower gas mass fractions than previous measurements, the fiducial FABLE model may require re-calibration to match observations for the least massive groups at $M_{500} \lesssim 2-3 \times 10^{13}~{\mathrm M}_{\rm \odot}$.  We find that the RadioBoost$M_{\rm BH,radio}$-40 and XFABLE models also lie within the large scatter of the observations.  XFABLE displays a drop in the gas mass fractions at $M_{500} \sim 10^{13}~{\mathrm M}_{\rm \odot}$, owing to the simplistic nature of the modified sub-grid implementation of AGN feedback, which allows the boosted radio-mode to act solely in halos approximately above this mass\footnote{High-resolution simulations able to capture relativistic AGN jet propagation and bubble-inflation would self-consistently determine where and in which haloes the jet energy is thermalized, which would naturally lead to a scatter in the gas mass fraction relation, without introducing any sharp features.} (see Section~\ref{sec:xfable}).  We note that we also ran a box with an identical model to XFABLE, but with a lower minimum black hole mass at which radio-mode feedback is allowed to occur; decreasing the limit from $\log_{10}(M_{\mathrm{BH}} [\mathrm{M_{\odot}}])>9$ to $\log_{10}(M_{\mathrm{BH}} [\mathrm{M_{\odot}}])>8.5$ (see Table~\ref{tab:simsall}).  We found that the impact of allowing lower mass black holes to undergo radio-mode feedback was to lower the gas fractions with respect to XFABLE in halos with $5\times 10^{12}~{\mathrm M}_{\rm \odot} \lesssim M_{500} \lesssim 2\times 10^{13}~{\mathrm M}_{\rm \odot}$, and to cause a greater suppression of the matter power spectrum but only at $k>2\, h\, \mathrm{Mpc}^{-1}$, leaving the suppression on larger scales unchanged. Furthermore, it is also interesting to note that both the RadioBoost$M_{\rm BH,radio}$-40 and XFABLE models display significantly increased scatter in the predicted hot gas fractions for halo masses $M_{\rm 500} \gtrsim 10^{13}M_{\rm \odot}$ due to the more bursty nature of radio-mode, which is able to better reproduce the large observed scatter inferred from X-ray observations.

The RadioBoost-40 model predicts gas fractions up to a factor of four smaller than the fiducial FABLE model at $z=0$, with a constant median gas fraction of $M_{\mathrm{gas,500}}/M_{500}\sim 0.01$ across the group population. Since this modified AGN feedback model injects radio bubbles at a greater distance from the central black hole, we can infer that too much gas is ejected beyond $r_{500}$, placing gas mass fractions at the lower end of observations, especially for the most massive systems present in the box. This effect is amplified by the increased $\chi_{\rm radio}=0.1$, which forces more black holes be in the more efficient radio-mode.  We note however that the RadioBoost-40 model is in reasonable agreement with the recent group-mass constraints of \citet{Popesso2024} and \citet{Bigwood2024}.

The QuasarBoost$z$2-40 box lies above fiducial FABLE gas fraction at $z=0$ and at the very upper end of the observational scatter.  As discussed in Section~\ref{sec:pkboxes}, at $z>2$, significantly more thermal energy is supplied close to the black hole, driving powerful outflows that redistribute gas beyond $r_{500}$ and reduce the measured gas fractions.  This gas re-accretes only towards $z=0$, resulting in the higher gas fractions. This model also displays a steeper mass trend than fiducial FABLE, indicating that `tuned down' quasar feedback at lower redshifts is preferentially unable to prevent gas re-accretion in more massive haloes (see also \citet{Martin-Alvarez2024} for a discussion on the scale dependence of the feedback modes). 

The lower left panel of Figure~\ref{fig:fracs} shows the redshift evolution in the hot gas mass fractions for the same set of simulation models. Here we calculate the median and the quartiles in the hot gas fractions for the simulated halos satisfying $M_{\rm 500}>5\times 10^{12}~\mathrm{M_{ \odot}}$.  Unfortunately, currently there are no available observations overlapping with the resolvable range of FABLE group masses to benchmark these models at $z\geq1$, however we plot the redshift dependence to trace the evolution of gas ejection induced by our AGN feedback model variations. This predictions will turn very useful for the next generation SZ measurements, such as the Simons Observatory.   

We find that in the fiducial FABLE box, the gas mass fractions decrease with time from $z<3$.  This is due to late-time feedback in the radio-mode inducing gas expulsion beyond $R_{500}$ (see \citet{Martin-Alvarez2024} which demonstrated this is the dominant feedback mode at low redshift).  This trend is also observed in each of the modified radio-mode boxes; RadioBoost-40, RadioBoost$M_{\rm BH,radio}$-40 and XFABLE.  The RadioBoost$M_{\rm BH,radio}$-40 and XFABLE boxes display a redshift evolution in the hot gas mass fraction in good agreement with fiducial FABLE.  The RadioBoost-40 box however lies lower than FABLE at all redshifts shown, displaying gas fractions approximately half of those in FABLE already at $z=2$. This is due to the model's increased $D_{\rm bub}$ and $\chi_{\rm radio}$ parameters at all $z$, leading to the likely over-ejection of gas.  As with the matter power spectrum suppression, the QuasarBoost$z$2-40 modification displays the opposite redshift trend to the other boxes.  We find that the evolution is consistent with the previously discussed picture of over-expulsion of gas at $z>2$ and late-time re-accretion; the gas mass fractions increase with time between $0<z<2$, transitioning from lying below fiducial FABLE at $z=2$ to lying above at $z=0$.  

To summarise, even by taking the considerable scatter in the observed hot gas fractions at face value, the RadioBoost-40 and QuasarBoost$z$2-40 models are largely disfavoured. The RadioBoost$M_{\rm BH,radio}$-40 and XFABLE models can predict larger variations in hot gas fractions, and the new group-size data hint that radio-mode feedback could be even more effective that assumed in XFABLE for these low mass systems. Future X-ray and SZ data will be crucial to both constrain the required burstiness of radio-mode feedback and how `ejective' AGN feedback is likely to be as a function of cosmic time.

\subsubsection{Stellar mass fractions}\label{sec:stellarfrac}

\begin{figure}
\centering
\includegraphics[width=0.45\textwidth]{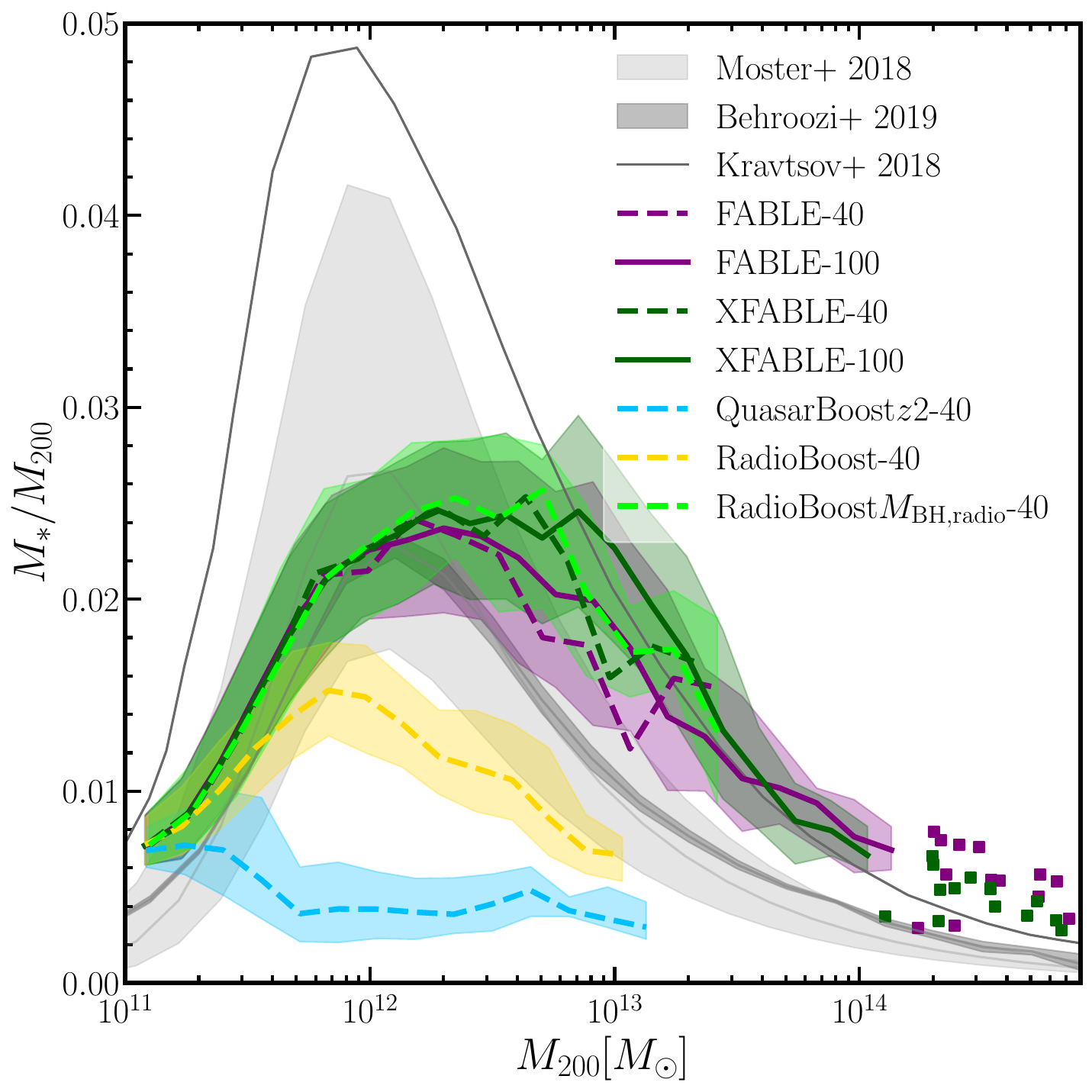}

\caption{The stellar mass fractions in each of our key simulation boxes, recomputed with $M_{200}$ as the halo mass and calculating the stellar mass within twice the stellar half-mass radius.  Dashed lines denote $(40~h^{-1}\mathrm{Mpc})^3$ boxes, and solid lines show the $(100~h^{-1}\mathrm{Mpc})^3$ FABLE and XFABLE boxes.  For each box, the solid/dashed lines denote the median relation, and the shaded regions span the upper and lower quartiles, calculated in bins of $M_{500}$.  We do not show the quartile regions of the FABLE-40 and XFABLE-40 boxes to avoid overcrowding the figure.  For FABLE-100 and XFABLE-100, we plot the most massive systems that cannot be binned due to poor statistics as individual datapoints. We show the fiducial FABLE boxes (purple) and our key modified AGN feedback models; QuasarBoost$z$2-40 (light blue), RadioBoost-40 (yellow), RadioBoost$M_{\rm BH,radio}$-40 (light green) and XFABLE (dark green).  We compare to the abundance matching models of \citet{Moster2018} (light grey), \citet{Behroozi2019} (medium grey) and \citet{Kravtsov2018} (dark grey).  We show that the FABLE and XFABLE models are in equally good agreement with the abundance matching models.}
\label{fig:smhm}
\end{figure}

In the upper right panel of Figure~\ref{fig:fracs} we plot the stellar mass fraction in groups and clusters at $z=0$, measured in the fiducial and modified FABLE boxes. We compute the total stellar mass within $R_{\rm 500}$ without differentiating the contributions from the brightest central galaxy (BCG), satellite galaxies and the intracluster light (ICL), and refer the reader to \citet{Henden2020} where the stellar mass content of the individual components are studied in detail.  We compare our results to the stellar mass-halo relation of \citet{Akino_2022} derived from the XXL X-ray selected sample, using the Hyper Suprime-Cam's photometry and weak-lensing mass measurements.  We also plot X-ray measurements derived from a number of surveys; \citet{Gonzalez2013} (XMM-Newton), \citet{Kravtsov2018} (XMM-Newton, Chandra and SDSS) and \citet{Zhang2011} (XMM-Newton, ROSAT and SDSS), with the redshift ranges of the data listed in the figure caption. The cluster masses of \citet{Akino_2022} are derived via weak lensing estimates, whereas the remaining sources use X-ray hydrostatic cluster masses. As in Figure~\ref{figure:comparesims},  we therefore add an arrow to Figure~\ref{fig:fracs} indicating the effect on observations of correcting for a hydrostatic mass bias of 30\%.

The fiducial FABLE-100 box displays a very good agreement with the observational data at $z=0$, consistent with the fit presented in \citet{Henden2018} for the FABLE-40 box. The RadioBoost$M_{\rm BH,radio}$-40 and XFABLE modifications also lie in excellent agreement with fiducial FABLE and the available observations.  As observed when discussing the GSMF (Section~\ref{sec:gsmf_analysis}), the QuasarBoost$z$2-40 model significantly overquenches star formation, with less than $\sim1$\% of the halo mass residing in stars at $z=0$. The stellar mass fraction in the RadioBoost-40 model also seems largely disfavoured by observational constraints at $z=0$, in line with our results for the GSMF.

The lower right panel of Figure~\ref{fig:fracs} shows the evolution in the stellar fractions for haloes with $M_{500}>5\times 10^{12}\,{\rm M}_{\rm \odot}$ in each of the illustrative models. As with the hot gas mass fractions, we lack $z\geq1$ observations of groups and clusters with masses overlapping those in FABLE, and therefore cannot at present directly benchmark the stellar fractions at higher redshifts, with our models providing useful predictions for future observations. FABLE, RadioBoost$M_{\rm BH,radio}$-40 and XFABLE all display a comparable redshift evolution of the stellar mass fraction. RadioBoost-40 displays a stellar fraction in good agreement with FABLE at $z\geq2$, however falls to a factor of $\sim2$ lower by $z=0$ due to the prevalence of the extreme radio-mode at late cosmic times. QuasarBoost$z$2-40 exhibits stellar fractions that are a factor of $\sim 4$ lower than FABLE at all redshifts shown. In this model we note a marginal recovery in the stellar mass fractions at $z<2$ as the quasar-mode feedback parameters return to the fiducial values, however this is insufficient to reach reasonable stellar fraction. 

Figure~\ref{fig:smhm} shows the stellar mass fractions recomputed with $M_{200}$ as the halo mass and calculating the stellar mass within twice the stellar half-mass radius, rather than within $r_{500}$. This facilitates comparison to the abundance matching models of, for example, \citet{Moster2018}, \citet{Behroozi2019} and \citet{Kravtsov2018}, allowing us to benchmark our lowest mass groups. We find once again that fiducial FABLE, RadioBoost$M_{\rm BH,radio}$-40 and XFABLE are all in good agreement with the abundance matching models for all group masses. At the massive end, our predicted stellar masses are somewhat too high with respect to the abundance matching models of \citet{Moster2018} and \citet{Behroozi2019}, but agree quite well with the estimates from \citet{Kravtsov2018},  indicating that the most massive galaxies in FABLE are reasonably realistic but perhaps not quite sufficiently quenched \citep{Henden2020}. As in the upper right panel of Figure~\ref{fig:fracs}, the underestimation of the stellar mass in the QuasarBoost$z$2-40 and RadioBoost-40 models lead to their agreement with the observations at $M_{200}>2 \times 10^{11} \,{\rm M}_{\rm \odot}$ and $M_{200}>5 \times 10^{11}\, {\rm M}_{\rm \odot}$, respectively, being poor.  

\subsubsection{X-ray scaling relations}\label{sec:xrayscaling}

\begin{figure*}
\centering
\includegraphics[width=0.38\textwidth,keepaspectratio]{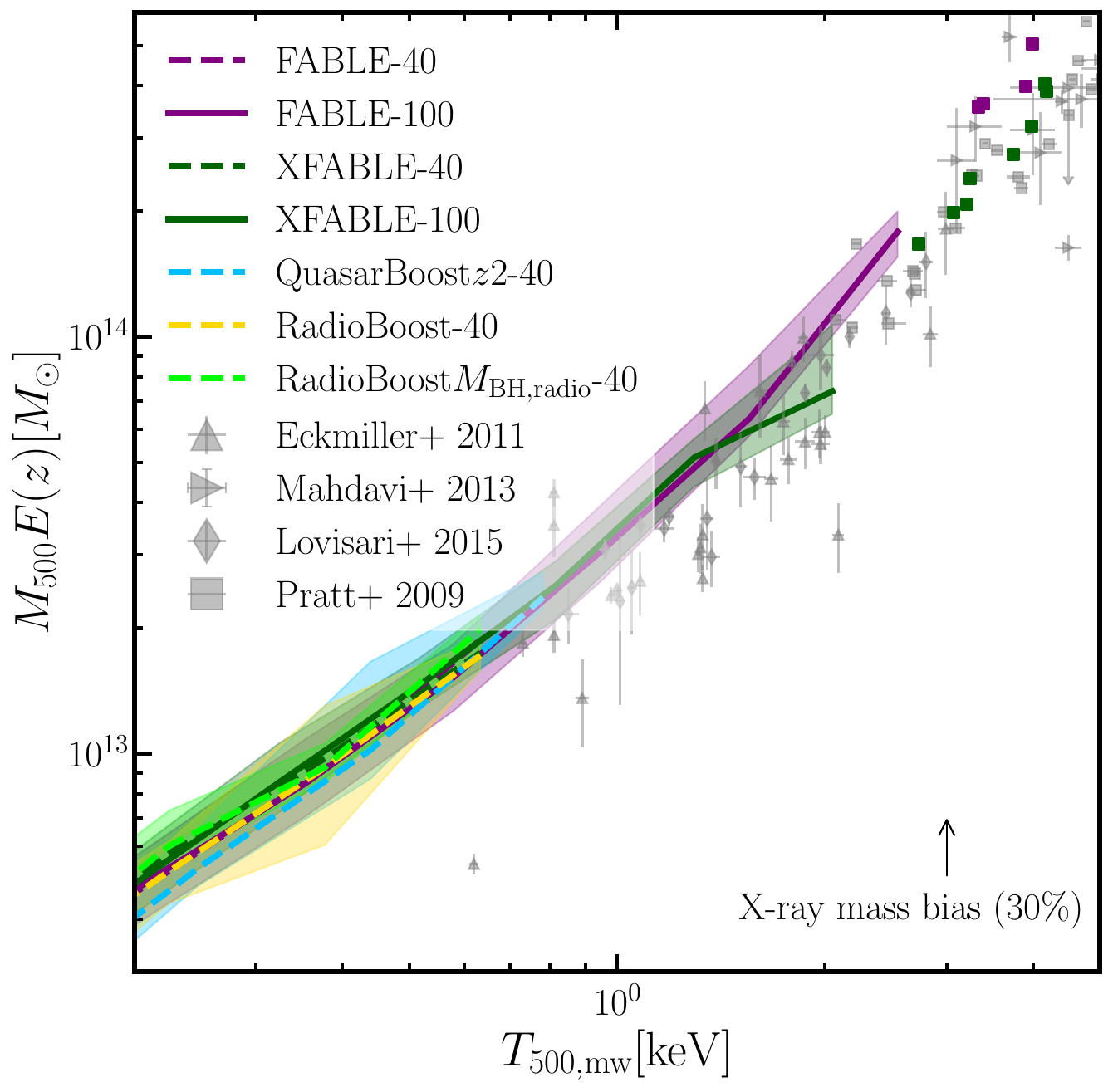}
\includegraphics[width=0.38\textwidth,keepaspectratio]{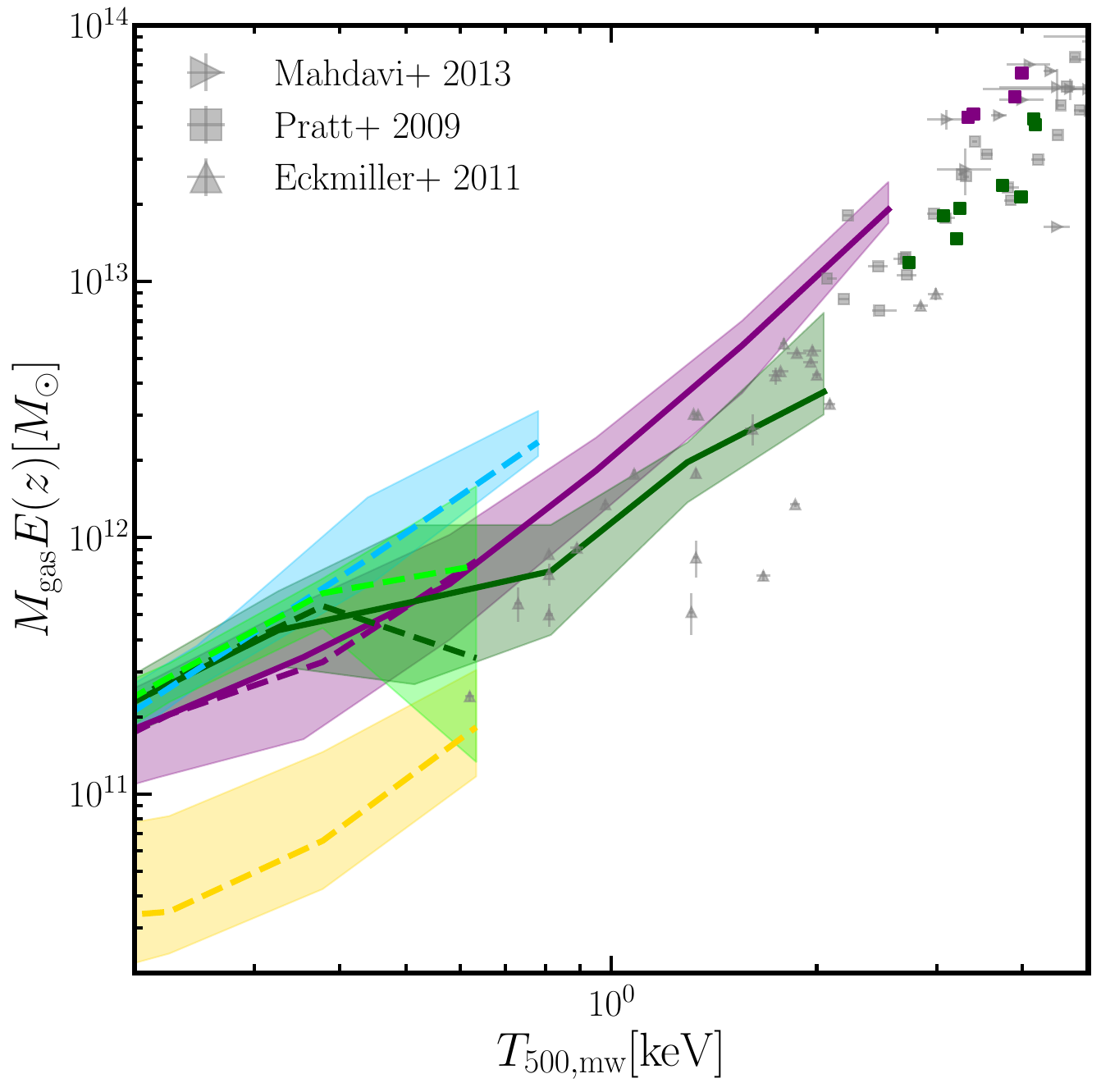}
\includegraphics[width=0.38\textwidth,keepaspectratio]{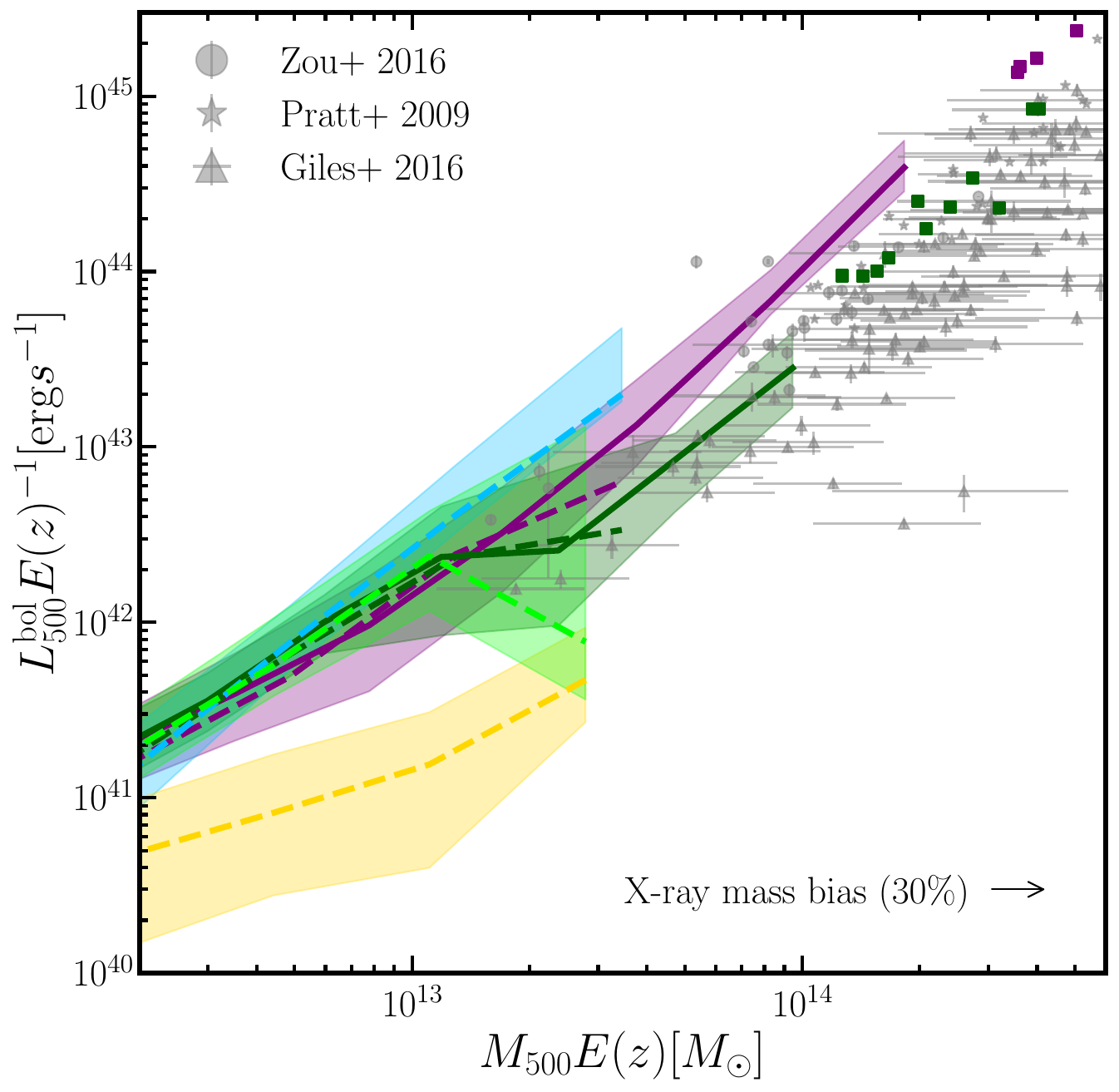}
\includegraphics[width=0.38\textwidth,keepaspectratio]{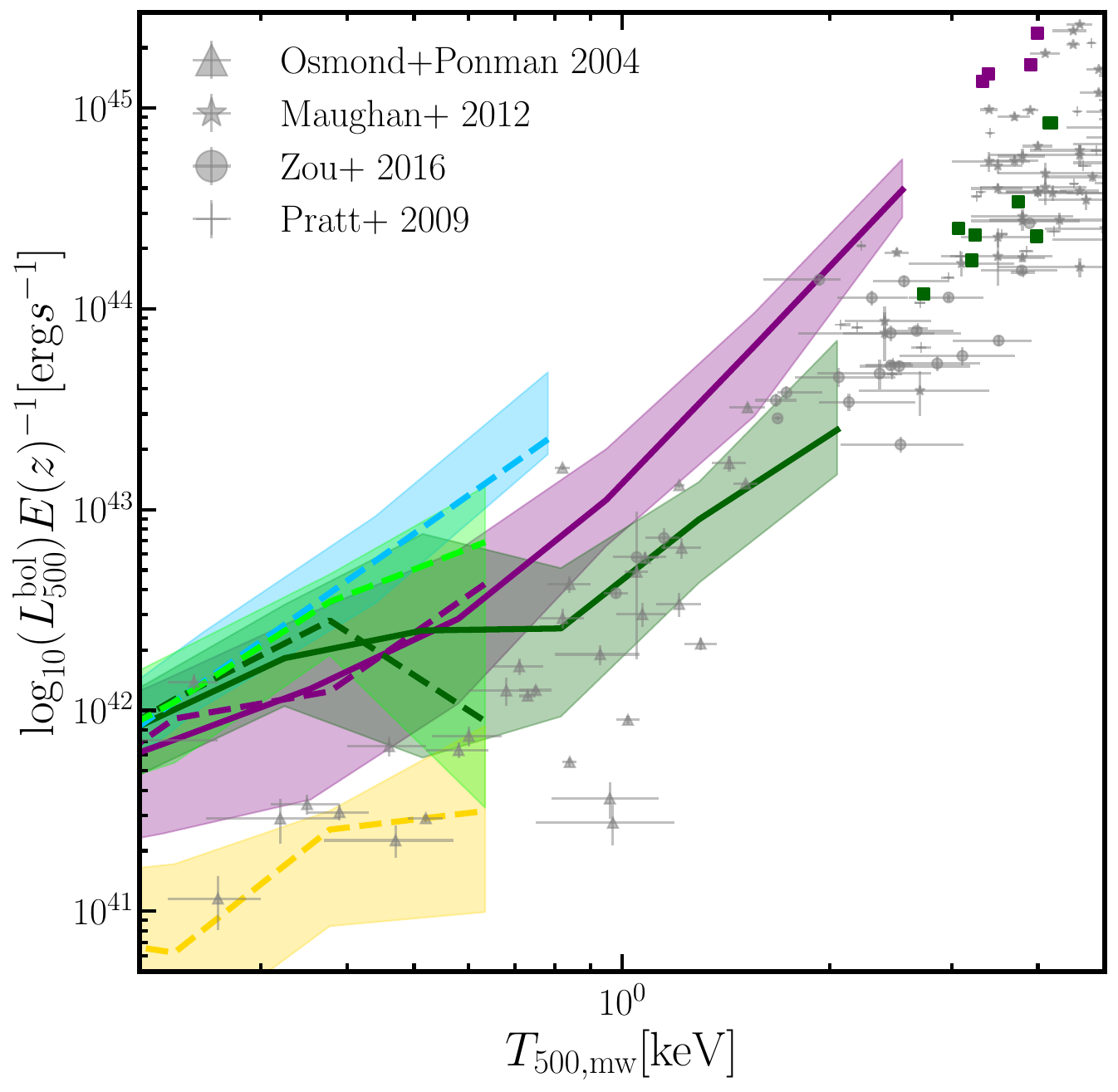}
\caption{The $z=0$ scaling relations between halo mass $M_{500}$, hot gas mass $M_{\rm gas}$, X-ray hot ICM luminosity $L_{500}^{\mathrm{bol}}$, and the mass-weighted mean temperature $T_{\rm 500,mw}$, measured in each of our key simulation models. All quantities are measured within $r_{500}$. Dashed lines denote $(40~h^{-1}\mathrm{Mpc})^3$ boxes, and solid lines show the $(100~h^{-1}\mathrm{Mpc})^3$ FABLE and XFABLE boxes. For each box, the solid/dashed lines denote the median relation, and the shaded regions span the upper and lower quartiles. We do not show the quartile regions of the FABLE-40 and XFABLE-40 boxes to avoid overcrowding the figure.  For FABLE-100 and XFABLE-100, we plot the most massive systems that cannot be binned due to poor statistics as individual datapoints. We show the fiducial FABLE boxes (purple) and our key modified AGN feedback models; QuasarBoost$z$2-40 (light blue), RadioBoost-40 (yellow), RadioBoost$M_{\rm BH,radio}$-40 (light green) and XFABLE (dark green).  We plot as the grey datapoints the $M_{500}-T_{\rm 500,mw}$ observational data of \citet{Eckmiller2011} ($z<0.5$), \citet{Mahdavi2013} ($z<0.6$), \citet{Lovisari2015} ($z<0.4$), \citet{Pratt2009} ($z<0.2$), the $M_{\mathrm{gas}}-T_{500}$ data of  \citet{Mahdavi2013} ($z<0.6$), \citet{Pratt2009} ($z<0.2$), \citet{Eckmiller2011} ($z<0.5$), the $L_{500}^{\mathrm{bol}}-M_{500}$ data of \citet{Zou2016} ($0.01<z<0.05$), \citet{Pratt2009} ($z<0.2$), \citet{Giles2016} ($z<1.1$), and the $L_{500}^{\mathrm{bol}}-T_{500}$ of \citet{Osmond2004}, \citet{Maughan2012} ($0.1<z<1.3$), \citet{Zou2016} ($0.01<z<0.05$) and 
\citet{Pratt2009} ($z<0.2$).  We show that XFABLE displays an improved fit to the scaling relations compared to FABLE. }

\label{fig:xrayscaling}
\end{figure*}

Thus far we have studied gas and stellar fraction of galaxy groups and clusters. Here, we extend our analysis of these objects by presenting scaling relations between global X-ray derived properties of groups and clusters for FABLE and our key modified feedback models. Comparing these to the wealth of observational data provides another benchmark that the AGN feedback model is able to produce a realistic cluster population, since the relations will be susceptible to independent model dependencies and systematics than cluster gas and stellar fraction measurements. We compare to a number of observational measurements from different surveys;  \citet{Eckmiller2011} (Chandra), \citet{Mahdavi2013} (Chandra and XMM-Newton), \citet{Lovisari2015} (XMM-Newton, with ROSAT-selected clusters), \citet{Pratt2009} (REXCESS XXM-Newton survey),  \citet{Zou2016} (Chandra), \citet{Giles2016} (XXL survey, XXM-Newton), \citet{Osmond2004} (GEMS and ROSAT) and \citet{Maughan2012} (Chandra). 

Figure~\ref{fig:xrayscaling} plots the relations $M_{500}-T_{\rm 500,mw}$ (upper left), $M_{\mathrm{gas}}-T_{\rm 500,mw}$ (upper right),  $L_{500}^{\mathrm{bol}}-M_{500}$ (lower left) and $L_{500}^{\mathrm{bol}}-T_{500}$ (lower right). We find that with the exception of the RadioBoost-40 and QuasarBoost$z$2-40 models, all the other AGN feedback models show scaling relations in very good agreement with fiducial FABLE, and with the observational data, with XFABLE displaying the best agreement for the most massive objects.  

Consistent with Section~\ref{sec:gasfracanalysis}, QuasarBoost$z$2-40 displays gas masses lying marginally higher than FABLE at all mean gas temperatures as well as increased ICM X-ray luminosities due to too high gas fractions at variance with observational data. The RadioBoost-40 model appears at variance with observations for most scaling relations examined, but the overlap with data is limited. 

Finally, we note that at low $T_{\rm 500,mw}$ all our simulation models overpredict X-ray bolometric luminosities, apart from the RadioBoost-40 model (see bottom right panel). This finding is interesting, as it likely indicates that our luminosities and hence gas fractions\footnote{This is unlikely an effect of unrealistic ICM temperatures, see detailed reasoning in \citet{Henden2018}.} may be too high in the lowest mass systems in agreement with the latest X-ray measurements from eROSITA \citep{Popesso2024}, as well as the indications from the kSZ effect \citep{Bigwood2024, Hadzhiyska2024} (see Figure~\ref{fig:fracs}).

\subsubsection{The tSZ $Y_{5r_{500}}-M_{500}$ relation}\label{sec:tszanalysis}

\begin{figure}
\centering
\includegraphics[width=0.45\textwidth]{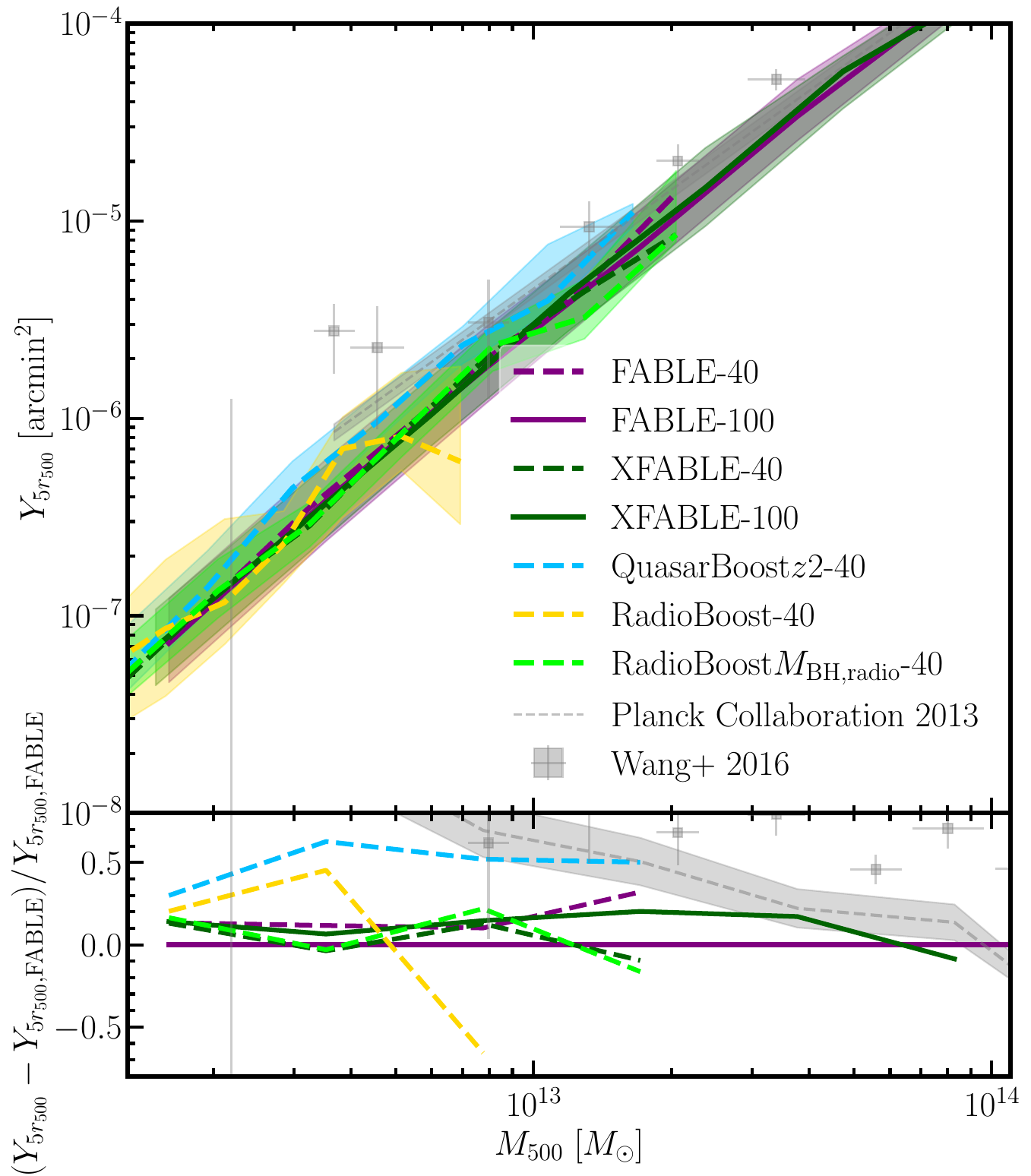}
\caption{The tSZ-halo mass relation, $Y_{5r_{500}}-M_{500}$, at $z=0$ measured in each of our key simulation models. We compute $Y_{5r_{500}}$ by measuring the Compton $Y$ parameter within a spherical aperture of $5r_{500}$ (as motivated in Section~\ref{sec:tsz}) and rescaling to a fixed angular diameter distance of 500~Mpc. Dashed lines denote $(40~h^{-1}\mathrm{Mpc})^3$ boxes, and solid lines show the $(100~h^{-1}\mathrm{Mpc})^3$ FABLE and XFABLE boxes.  For each box, the solid/dashed lines denote the median relation, and the shaded regions span the upper and lower quartiles.  We do not show the quartile regions of the FABLE-40 and XFABLE-40 boxes to avoid overcrowding the figure. We show the fiducial FABLE boxes (purple) and our key modified AGN feedback models; QuasarBoost$z$2-40 (light blue), RadioBoost-40 (yellow), RadioBoost$M_{\rm BH,radio}$-40 (light green) and XFABLE (dark green).  We compare to the observationally derived best-fit scaling relation of \citet{Planck2013} (grey dashed line and shaded region) and the re-calibration of the \citet{Planck2013} data by \citet{Wang2016} (grey square datapoints).  We do not extrapolate the \citet{Planck2013} relation to lower halo-masses than the last well constrained data-point.  To examine the differences between the predicted  $Y_{5r_{500}}-M_{500}$ in each box with greater clarity, the lower sub-panel shows the fractional difference between the median $Y_{5r_{500}}-M_{500}$ relation as measured in the fiducial FABLE box and the remaining models, ($Y_{5r_{500}}-Y_{5r_{500},\mathrm{FABLE}})/Y_{5r_{500},\mathrm{FABLE}}$.  We similarly re-scale the observational data in the lower panel.  We demonstrate that modified AGN feedback boxes produce a reasonable fit to the data, albeit being somewhat low, with some variation in the predictions between models. }
\label{fig:tsz}
\end{figure}

X-ray measurements of the ICM typically probe the most massive, and thus X-ray luminous, clusters in the Universe (but see recent work by \citet{Popesso2024}). SZ measurements provide a unique and complementary window into feedback as a result of its sensitivity to higher redshift systems, as well as groups and low mass clusters, which are believed to be more affected than massive clusters by AGN-driven gas ejection due to their shallower gravitational potential wells. Comparison of observed tSZ scaling relations with simulation predictions therefore provide insights into feedback's impact on a different cluster population to that studied with X-rays, with measurements prone to an independent set of systematics. Figure~\ref{fig:tsz} shows the $Y_{5r_{500}}-M_{500}$ relation calculated in FABLE and each of the key AGN feedback simulation models, with the lower subplot displaying the residual $Y_{5r_{500}}$ with respect to fiducial FABLE. We compare the simulated results to the \citet{Planck2013} $Y_{r_{500}}-M_{500}$ measurements, as well as the re-analysis of \citet{Wang2016} with weak-lensing calibrated masses. As discussed in Section~\ref{sec:tsz}, we multiply $Y_{r_{500}}$ in both observational datasets by a factor of $1.796$ in order to obtain $Y_{5r_{500}}$ and avoid the modelling assumptions used in the \textit{Planck} analyses.  

We find that each of the modified feedback boxes, with the exception of QuasarBoost$z$2-40 and RadioBoost-40, show little deviation to FABLE in the slope and amplitude of the $Y_{5r_{500}}-M_{500}$ relation.  Since $Y_{5r_{500}}$ provides a measure of a group's thermal energy, this indicates that the RadioBoost$M_{\rm BH,radio}$-40 and XFABLE feedback models likely do not overheat the systems, as was previously reflected in the GSMF (Section~\ref{sec:gsmf_analysis}), which analogously showed that there was sufficient cool gas in these models for realistic star formation. The QuasarBoost$z$2-40 model displays a higher amplitude of the $Y_{5r_{500}}-M_{500}$ relation, which is consistent with higher gas fraction due to low redshift gas re-accretion in this model.  It also shows an improved agreement over FABLE, when compared to the observed relation of \citet{Planck2013} and \citet{Wang2016}, but for wrong reasons given that higher amplitude is driven by a too large amount of ICM gas.  

The RadioBoost-40 model measures a low $Y_{5r_{500}}$ for group-scale halos.  This indicates that the feedback has expelled sufficient gas beyond $r_{500}$ to significantly reduce the thermal energy of groups.  This, as previously hinted at in the X-ray scaling relations (Section~\ref{sec:xrayscaling}), informs us that the low GSMF at $z=0$ discussed in Section~\ref{sec:gsmf_analysis} results from the feedback model leading to galaxies being gas-poor, rather than effective star-formation being prevented through over-heating.  We verify this picture in Section~\ref{sec:profiles}, where we gain insight into the local thermodynamic processes acting within groups through the radial profiles. 

\subsubsection{The X-ray $Y_{X, 500}-M_{500}$ relation}

\begin{figure}
\centering
\includegraphics[width=0.45\textwidth]{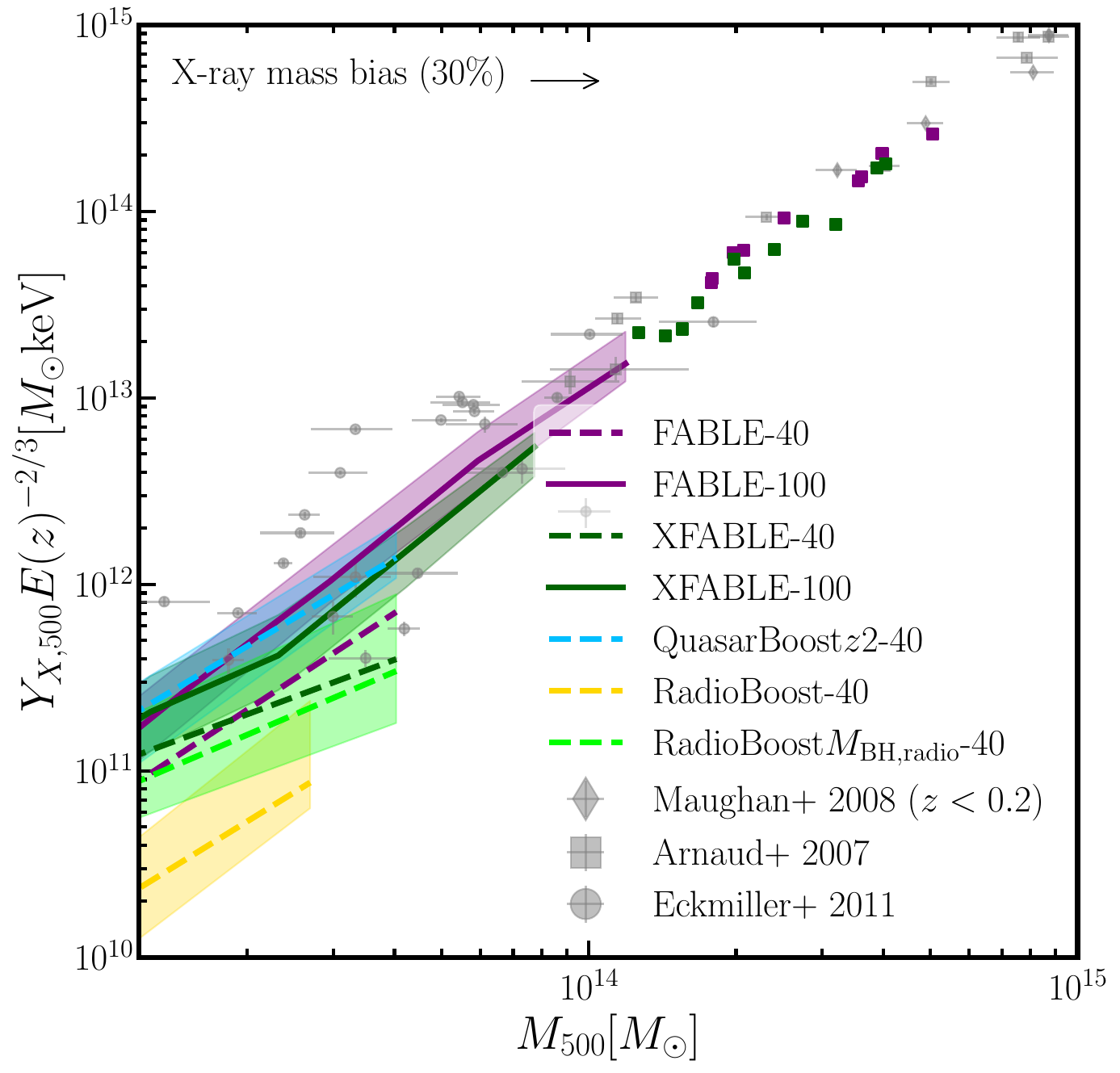}
\caption{The $Y_{X, 500}-M_{500}$ measured in each of our key simulation boxes. Here $Y_{X, 500}$ is the X-ray proxy of the tSZ Compton $Y_{500}$ parameter, measured within a spherical aperture of radius $r_{500}$ according to Equation~\ref{eq:yx}. Dashed lines denote $(40~h^{-1}\mathrm{Mpc})^3$ boxes, and solid lines show the $(100~h^{-1}\mathrm{Mpc})^3$ FABLE and XFABLE boxes. For each box, the solid/dashed lines denote the median relation, and the shaded regions span the upper and lower quartiles. We do not show the quartile regions of the FABLE-40 and XFABLE-40 boxes to avoid overcrowding the figure.  For FABLE-100 and XFABLE-100, we plot the most massive systems that cannot be binned due to poor statistics as individual datapoints. We show the fiducial FABLE boxes (purple) and our key modified AGN feedback models; QuasarBoost$z$2-40 (light blue), RadioBoost-40 (yellow), RadioBoost$M_{\rm BH,radio}$-40 (light green) and XFABLE (dark green).  We compare our simulation measured relations to the observational measurements of \citet{Maughan2008} (plotting only the clusters at $z<0.2$), \citet{Arnaud2007} ($z<0.2$) and \citet{Eckmiller2011} ($z<0.05$), shown as the grey datapoints.  We show that alike FABLE, XFABLE also lies in good agreement with the observational data. }
\label{fig:yx}
\end{figure}

The final global cluster property we explore in Figure~\ref{fig:yx} is the X-ray analogue of the Compton $Y_{r_{500}}$ parameter, $Y_{X, 500}$.  We compute the  $Y_{X, 500}-M_{500}$ relation for FABLE and each of our modified feedback boxes, and compare to the measurements of \citet{Maughan2008} (Chandra), \citet{Arnaud2007} (XMM-Newton) and \citet{Eckmiller2011} (Chandra).  

Alike $Y_{5r_{500}}$, $Y_{X, 500}$ is similarly sensitive to the thermal energy of groups and clusters within $r_{500}$ \citep{Kravtsov2006}.  We therefore find broadly the same relationship between the $Y_{X, 500}-M_{500}$ relation from our key simulation models as detailed in Section~\ref{sec:tszanalysis}; i.e.  the relation measured in FABLE and each of the modified feedback models display similar slopes, however the amplitude of the RadioBoost-40 model lies low.  For groups of mass $M_{500}\sim10^{13}~\mathrm{M_{\odot}}$, $Y_{X, 500}$ is approximately half the value in the RadioBoost-40 box compared to FABLE and can be ruled out by the observations.  Both the FABLE-100 and XFABLE-100 models remain in agreement with the observations, with the two models showing little deviation in their predicted relation.

\subsection{Thermodynamic profiles of the ICM}

\begin{figure*}
\centering
\includegraphics[width=0.38\textwidth,keepaspectratio]{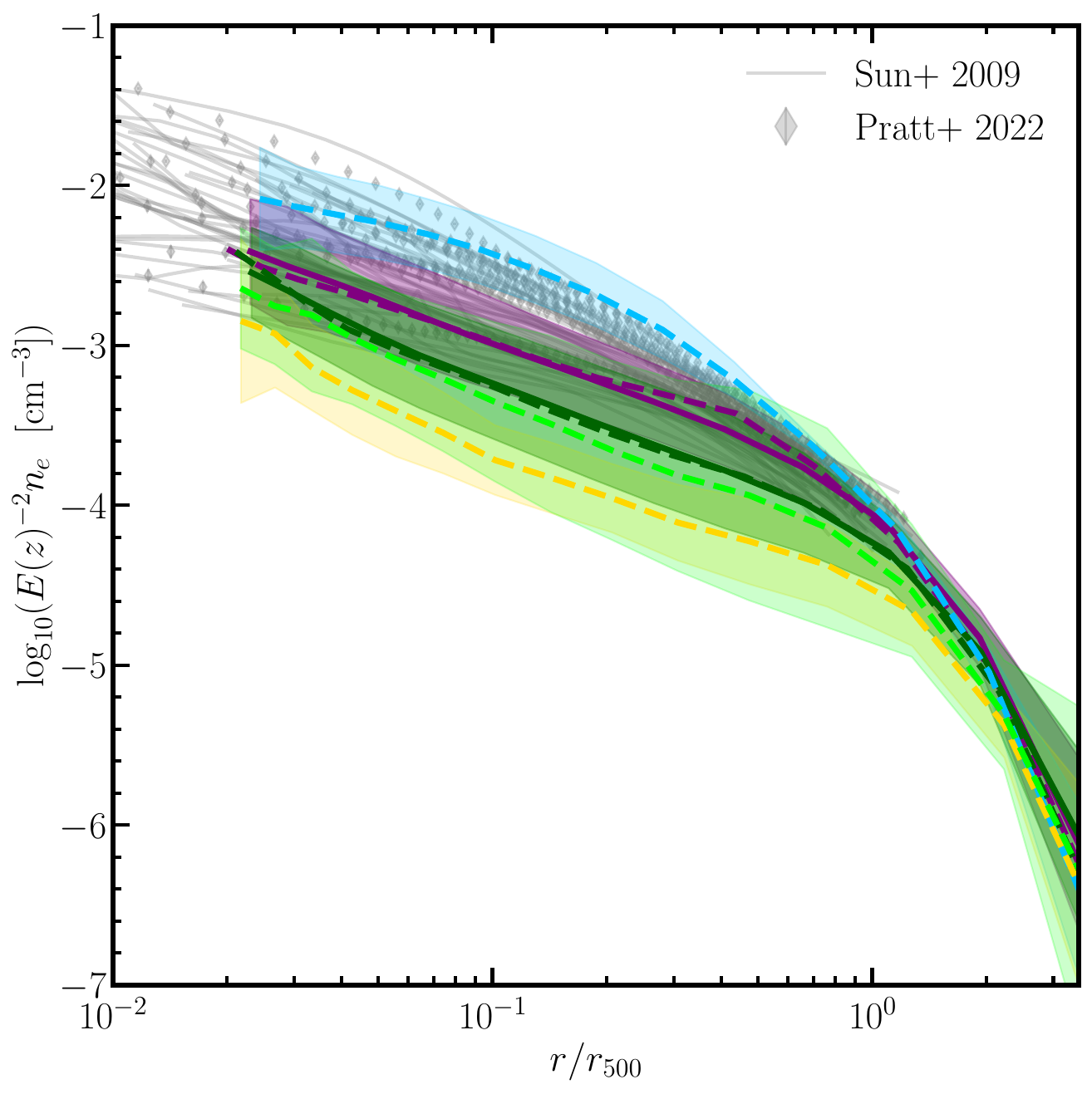}
\includegraphics[width=0.38\textwidth,keepaspectratio]{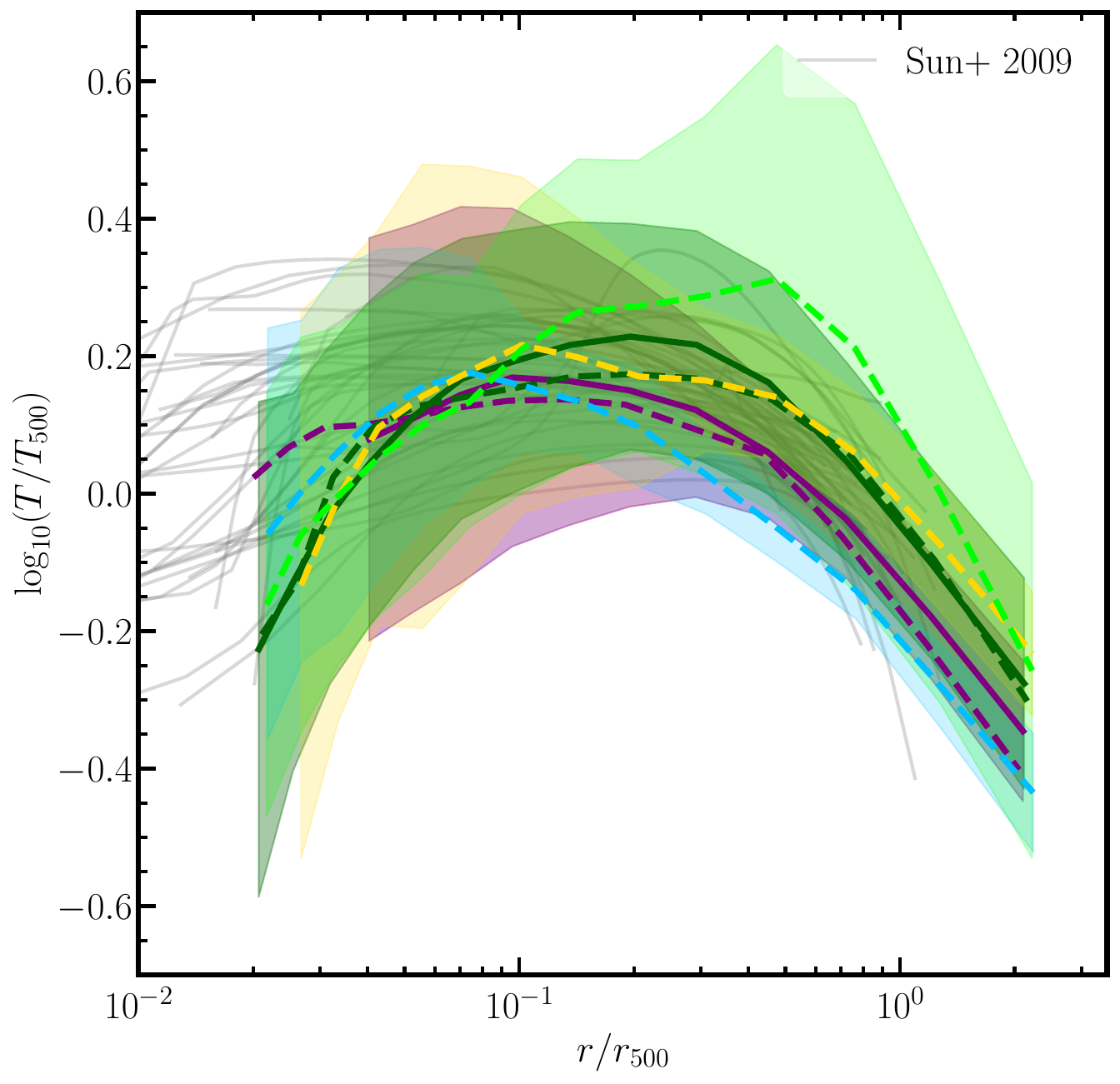}
\includegraphics[width=0.38\textwidth,keepaspectratio]{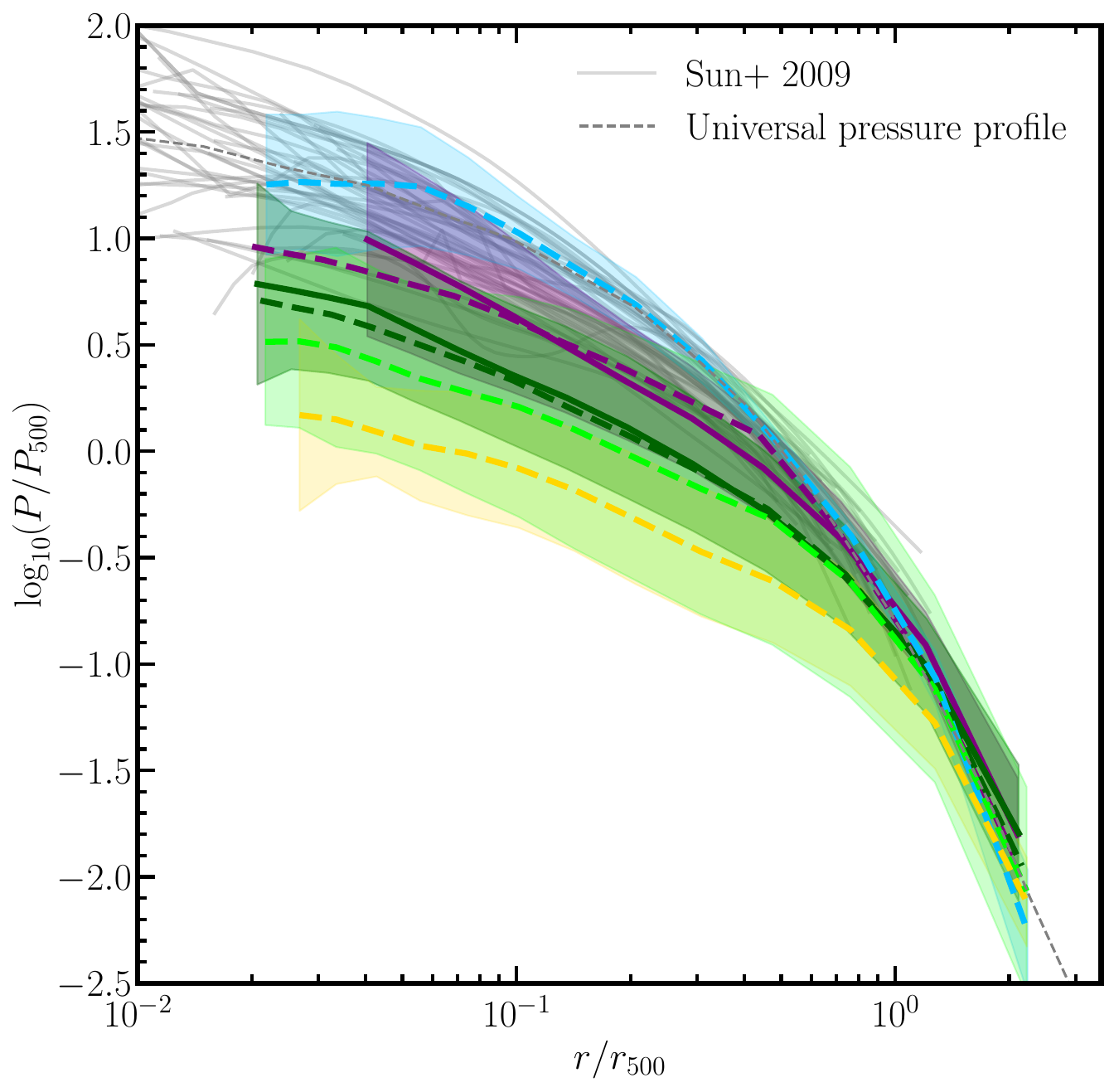}
\includegraphics[width=0.38\textwidth,keepaspectratio]{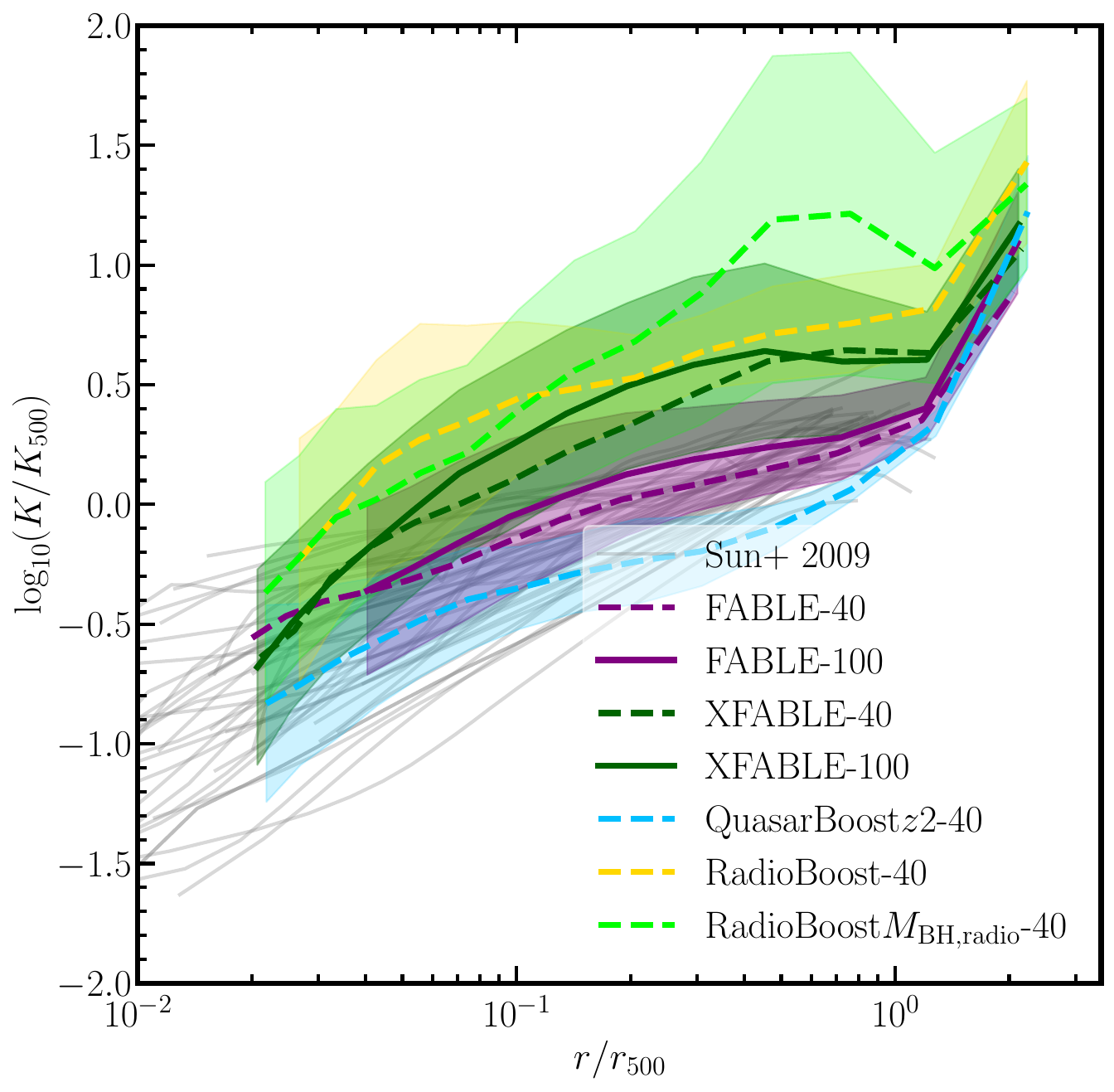}
\caption{Thermodynamic radial profiles of the hot ICM measured in each of our key simulation boxes at $z=0$. In each panel, we compare to the \citet{Sun2009} observationally derived ICM profiles (shown as the grey lines) and therefore compute profiles for simulated halos with masses lying within the range spanned by \citet{Sun2009} sample, $1.48\times 10^{13}< M_{500}\ [\mathrm{M_{\odot}}]< 1.49\times 10^{14}$. For the electron density profiles, we also compare to a sample of \citet{Pratt2022} profiles which include several heavier halos and therefore compute profiles for simulated halos with masses $1.48\times 10^{13}< M_{500}\ [\mathrm{M_{\odot}}]< 1.89\times 10^{14}$. For each box, the solid/dashed lines denote the mean profile weighted by halo mass and the shaded regions show the $1\sigma$ region spanned by the simulated profile sample. We do not show the $1\sigma$ region of the FABLE-40 and XFABLE-40 boxes to avoid overcrowding the figure. Dashed lines denote $(40~h^{-1}\mathrm{Mpc})^3$ boxes, and solid lines show the $(100~h^{-1}\mathrm{Mpc})^3$ FABLE and XFABLE boxes. We show the fiducial FABLE boxes (purple) and our key modified AGN feedback models; QuasarBoost$z$2-40 (light blue), RadioBoost-40 (yellow), RadioBoost$M_{\rm BH,radio}$-40 (light green) and XFABLE (dark green).
\textit{Upper left:} Electron number density profiles of the ICM. 
 \textit{Upper right:} Dimensionless temperature profile of the ICM, normalised by the characteristic temperature $T_{500}$ (Equation~\ref{eq:chartemp}). \textit{Lower left:} Dimensionless pressure profiles of the ICM, normalised by the characteristic pressure $P_{500}$ (Equation~\ref{eq:charP}).  For reference, we also plot the universal pressure profile of \citet{Arnaud2010}. 
 \textit{Lower right:} Dimensionless entropy profiles of the ICM, normalised by the characteristic entropy $K_{500}$ (Equation~\ref{eq:chark}). We compare to the measurements of \citet{Sun2009} ($0.012 <z< 0.12$) and \citet{Pratt2022} ($0.056 <z< 0.108$) when available.}
\label{fig:profiles}
\end{figure*}

In this section, we compute spherically-averaged radial profiles of the ICM in order to validate the local thermodynamical properties of our simulated groups and clusters. Figure~\ref{fig:profiles} shows the mean profile weighted by halo mass for halos within a given halo mass range matched to observations, as well as the $1\sigma$ region spanned by the sample of simulated profiles. We measure profiles of the electron number density ($n_e$), the dimensionless temperature ($T/T_{500}$), the dimensionless pressure ($P/P_{500}$) and the dimensionless entropy ($K/K_{500}$) in FABLE and each of our modified feedback boxes. We compare to the measured profiles of \citet{Sun2009} (Chandra), and the electron density profiles of \citet{Pratt2022} (XMM-Newton REXCESS sample).

The FABLE-100 box remains consistent with the FABLE-40 box, which was verified in \citet{Henden2018} to produce very good agreement with the measured thermodynamic profiles of \citet{Sun2009}. As anticipated from our previous analysis, and recalling that the RadioBoost-40 model allows for a larger fraction of black holes to be in the radio-mode, as well as injecting AGN-driven bubbles at a greater distance from the galaxy centre, it is not surprising that this model leads to an ICM with high-entropy outskirts, low-pressure inner regions, and in general reduced electron density, rendering the model incompatible with observations. The QuasarBoost$z$2-40 model predicts somewhat too high mean gas densities and too low mean temperatures for the intermediate range of spatial scales, $0.2 < r/r_{\rm 500} < 0.5$. The RadioBoost$M_{\rm BH,radio}$-40 modification, which we recall implements a boosted radio-mode in only the heaviest SMBHs, has thus far remained largely consistent against the galaxy, group and cluster observations we have tested it against.  The thermodynamic profiles however indicate that this model leads to too strong shock heating of the ICM. The mean density and pressure profiles are systematically lower than observational profiles at all radii, and the entropy and temperature profiles around $r/r_{\rm 500} \sim 0.5 -0.6$ display characteristic signatures of ICM overheating by too powerful AGN feedback, motivating the addition of a pressure-limit in the XFABLE model.  

The XFABLE model produces thermodynamic profiles that remain in good agreement with the observational measurements.  This highlights the importance of modelling AGN-driven bubble feedback as a `gentle' heating processes. Nevertheless, XFABLE predicts somewhat lower densities within groups and clusters, as well as lower gas pressures and higher gas entropy outskirts, which hint that this model is likely too effective at heating the ICM. Matching the observed ICM profiles from small groups to most massive clusters, as well as reproducing the cool core vs. non-cool core population remains one of the very important benchmarks for theoretical models of AGN feedback.

\section{Discussion and Conclusions}\label{sec:outlook}

Cosmological analyses using non-linear scales crucially rely on accurate theoretical predictions of the baryonic physics impact on the matter power spectrum. However, state-of-the-art cosmological galaxy formation simulations, such as FLAMINGO \citep{Schaye2023}, MillenniumTNG \citep{Pakmor2023}, SIMBA \citep{Dave:2019}, BAHAMAS \citep{McCarthy2017}, FABLE \citep{Henden2018}, Horizon-AGN \citep{Dubois2014} and Magneticum \citep{Steinborn2015}, currently do not provide a consensus view on this fundamental issue, as too large uncertainties persist in our understanding of modus operandi of stellar and AGN feedback processes. This astrophysical model uncertainty limits cosmological precision of weak lensing analyses \citep[e.g.][]{AmonDES2022, DESKIDS2023}, and it is possible that underestimating feedback effects can bias current constraints or mask the ability to test for signatures of models beyond $\Lambda$CDM \citep{Amon2022, Preston2023}. In the upcoming era of the Rubin Observatory Legacy Survey of Space and Time, the Euclid mission and the Nancy Grace Roman Space Telescope, pinning down the amplitude and extent of the suppression of the matter power spectrum due to `baryonic feedback' is critical.   

Recent studies indicate that `baryonic feedback' may be more extreme than state-of-the-art hydrodynamical simulations \citep{Preston2023, Bigwood2024, McCarthy2024}. Furthermore, recent eROSITA measurements that probe systems down to low mass groups point towards lower gas fractions \citep{Popesso2024}. Motivated by these findings, we have explored a range of AGN feedback models built around the FABLE project \citep{Henden2018, Henden2019, Henden2020} to understand the plausibility of stronger AGN feedback models. 

We have performed a large simulation suite that systematically explores AGN feedback models that act differently either as a function of cosmic time, host halo properties and/or spatial location where feedback energy is effectively coupled with the surrounding medium. Within this suite we found a viable AGN feedback model, XFABLE, which causes strong matter power spectrum suppression on large scales ($k \lesssim 1\, h\, \mathrm{Mpc}^{-1}$). To achieve this, AGN radio-mode feedback needs to: {\it (i)} act in larger population of black holes with respect to the FABLE model (i.e. below the Eddington accretion rate ratio of $\approx 0.1$); {\it (ii)} (at least) operate in halo hosts which have a well-developed `hot atmosphere' ($M_{\rm 500} \approx 10^{13}\, \mathrm{M_{\odot}}$); {\it (iii)} have jet lobes thermalizing at relatively large cluster-centric distances ($\approx 100$~$h^{-1}$kpc). Our main findings from our simulation suite are as follows:
\begin{itemize}
\item To produce sufficiently large matter power spectrum suppression consistent with the latest observational constraints \citep[e.g.][]{Bigwood2024}, AGN feedback needs to redistribute large amounts of gas towards outskirts of groups and clusters. This process cannot operate only at early cosmic times, as gas re-accretion onto the growing cluster's potential wells needs to be prevented at low redshifts. This process also needs to act {\it across a range of halo masses} up to the largest galaxy clusters probed by our simulations ($M_{\rm 500} \sim 5 \times 10^{14}\, \mathrm{M_{\odot}}$) to cause sufficient matter power spectrum suppression at low $k$ values, $k \lesssim 1\, h\, \mathrm{Mpc}^{-1}$ \citep[in agreement with findings from e.g.][]{vanLoon2024, Martin-Alvarez2024}. 

\item While AGN feedback needs to push sufficient amounts of gas to large cluster-centric distances, strong AGN feedback which removes the {\it central} gas reservoir is clearly disfavoured. Such modelling choice easily over-quenches central galaxies, with the observed GSMF at different redshifts providing stringent constraints on the amount of central, cold gas that is needed to build realistic galaxy stellar masses. 

\item Several of our key simulation models produce a reasonable cosmological population of supermassive black holes, with black hole-host galaxy scaling relations and redshift evolution of quasar luminosity function in agreement with observations. Comparisons to these observables do not allow us to constrain AGN feedback models which produce markedly different matter power spectrum suppressions.

\item Importantly, we found a novel, empirical AGN feedback model, XFABLE, that is able to produce large matter power spectrum suppression at low $k$-values, while maintaining a very good agreement with galaxy stellar mass function, gas fractions in groups and clusters, and all key galaxy cluster X-ray and tSZ scaling relations.  This indicates that there may exist a physically plausible galaxy formation model within the $\Lambda$CDM Universe, which is consistent with all current observational constraints from this diverse range of datasets, without the need to invoke alternative cosmological models.

\item Interestingly, both recent joint weak lensing + kSZ \citep{Bigwood2024} and X-ray constraints \citep{Popesso2024} indicate that the gas fraction in a few times $10^{12}-10^{13} \, \mathrm{M_\odot}$ systems may be even lower than in the XFABLE model, but we emphasize that to produce matter power suppression on large scales ($k \lesssim 1\, h\, \mathrm{Mpc}^{-1}$), accurately modelling AGN feedback effects in more massive haloes is the key. 

\item Unsurprisingly, radial profiles of key thermodynamical properties of the ICM provide crucial constraints on the nature and modus operandi of AGN feedback, facilitating to differentiate between the models that eject too much gas versus the models that overheat the gas at large radii. XFABLE retains good agreement with ICM radial profiles, but our detailed comparison clearly points towards the need to more self-consistently model AGN bubble inflation via jets and to account for the relativistic population within the jet lobes.    
\end{itemize}

The XFABLE model is deliberately constructed to impact gas at larger cluster-centric distances and allows for a larger population of supermassive black holes to act in radio-mode. However, if such an AGN feedback model is physically viable, the fundamental question remains for how it operates in detail. Tantalizingly, recent LOFAR observations have revealed AGN-driven radio jets spanning $\sim 7$~Mpc from the host galaxy with stellar mass of $\sim 5.5 - 6.7 \times 10^{11}\, \mathrm{M_\odot}$ \citep{Oei2024}, which would directly heat the IGM. LOFAR's LoTSS DR2 survey has recently revealed more than 10'000 of such giant radio galaxies \citep{Mostert2024}, with ILoTSS and SKA providing constrains on this rapidly rising population in the near future. These observational findings suggest that for a sufficiently large fraction of systems, radio lobes may well be inflated at large distances from the central galaxy proving a direct heating source at large-scales. Nonetheless, future observations constraining this population and detailed numerical simulations of magnetised, relativistic jets in full cosmological simulations will be needed to understand jet energetics and the likely thermalization of the surrounding medium \citep{Ehlert2018, Bourne2019, Bourne2021}.  

It is important to stress that a large range of AGN models explored in this work, which lead to markedly different matter power spectrum suppressions, still rely on the underlying FABLE-like baryonic physics modelling. Specifically, we have not explored any variations to star formation and associated stellar feedback models, and we have not explored alternative black hole accretion prescriptions or different heating channels, such as cosmic-ray-driven or radiation-pressure-driven outflows, for example. This highlights that the parameter space of baryonic physics modelling is much more vast than that explored here (or within most state-of-the-art large cosmological simulations) and that there may be even more degeneracies within models of these complex processes and the signatures they leave on the matter power spectrum. The range in the matter power spectrum suppression spanned by the various simulations (left panel of Figure~\ref{figure:comparesims}) may therefore significantly widen, before it is better constrained by observations. We stress that in cosmological analyses, it is critical that feedback mitigation strategies are flexible enough to capture the extent of the feedback modelling uncertainties, considering the potential degeneracies in the power suppression.

The future observational landscape to unravel how `baryonic feedback' operates is promising. eROSITA is already transforming our knowledge of galaxy clusters and groups, probing into low mass regime. Gas fraction measurements and detections of spatially-resolved ICM properties of group-sized objects both locally and at high redshifts, via X-rays with eROSITA, and the Athena X-ray observatory in future, will provide crucial constraints on the nature of AGN feedback as a function of cosmic time. X-ray detections of hot gas in filaments and of the warm-hot intergalactic medium will provide complementary constraints on the baryon cycle and gas re-distribution from groups and clusters.  Upcoming weak lensing surveys will also allow us to reconstruct the non-linear matter power spectrum, allowing us to distinguish between different models of `baryonic feedback' \citep{Preston2024}. Moreover, thermal and kinetic SZ observations with ACT and shortly with the Simons Observatory, both via stacking and characterization of individual systems, will independently constrain galaxy group and cluster gas properties, probing both lower mass systems and feedback effects at large cluster-centric distances. Furthermore, constrains on the stellar mass assembly of brightest cluster galaxies that large galaxy surveys such as LSST and Euclid will elucidate the role of `ejective' AGN feedback as well as constrain stellar feedback channels in these system. Finally, LOFAR has uncovered an unprecedented population of giant radio galaxies, and with ILoTSS and SKA on the horizon we will soon get a much better insight into the crucial issue of AGN jet energy transfer into the surrounding medium. Together, with the rapid advancement of galaxy formation simulations which are now starting to tackle complex physics of radiation effects on-the-fly, cosmic ray-driven feedback as well as black hole physics through realistic accretion discs and magnetised, relativistic jets, our ability to constrain baryonic physics for precision cosmology is within reach.

\section*{Acknowledgements}
We thank George Efstathiou for useful discussions throughout this work, and feedback on the manuscript. We also thank Joop Schaye for comments on the draft.  L.B, M.A.B. and D.S. acknowledge support from
the Science and Technology Facilities Council (STFC). V.I. acknowledges the support of the Kavli Foundation and PD51-INFN INDARK grant. M.A.B. acknowledges support from a UKRI Stephen Hawking Fellowship (EP/X04257X/1). The simulations were performed on the DiRAC Darwin Supercomputer
hosted by the University of Cambridge High Performance Computing Service (http://www.hpc.cam.ac.uk/), provided by Dell Inc. using
Strategic Research Infrastructure Funding from the Higher Education Funding Council for England and funding from the Science and
Technology Facilities Council.  Simulations were also performed using the COSMA Data Centric system
at Durham University, operated by the Institute for Computational
Cosmology on behalf of the STFC DiRAC HPC Facility.  This equipment was funded by a BIS National E-infrastructure capital grant
ST/K00042X/1, STFC capital grant ST/K00087X/1, DiRAC Operations grant ST/K003267/1 and Durham University.

%%%The data used in this work may be shared on reasonable request to
the authors.%%%%%%%%%%%%%%%%% REFERENCES %%%%%%%%%%%%%%%%%%


\bibliographystyle{mnras}
\bibliography{references} % if your bibtex file is called example.bib

%\begin{thebibliography}{99}
%\end{thebibliography}

%%%%%%%%%%%%%%%%%%%%%%%%%%%%%%%%%%%%%%%%%%%%%%%%%%

%%%%%%%%%%%%%%%%% APPENDICES %%%%%%%%%%%%%%%%%%%%%

\appendix

\section{The impact of cosmic variance}\label{app:seeds}
\begin{figure}
\centering

\includegraphics[width=0.45\textwidth]{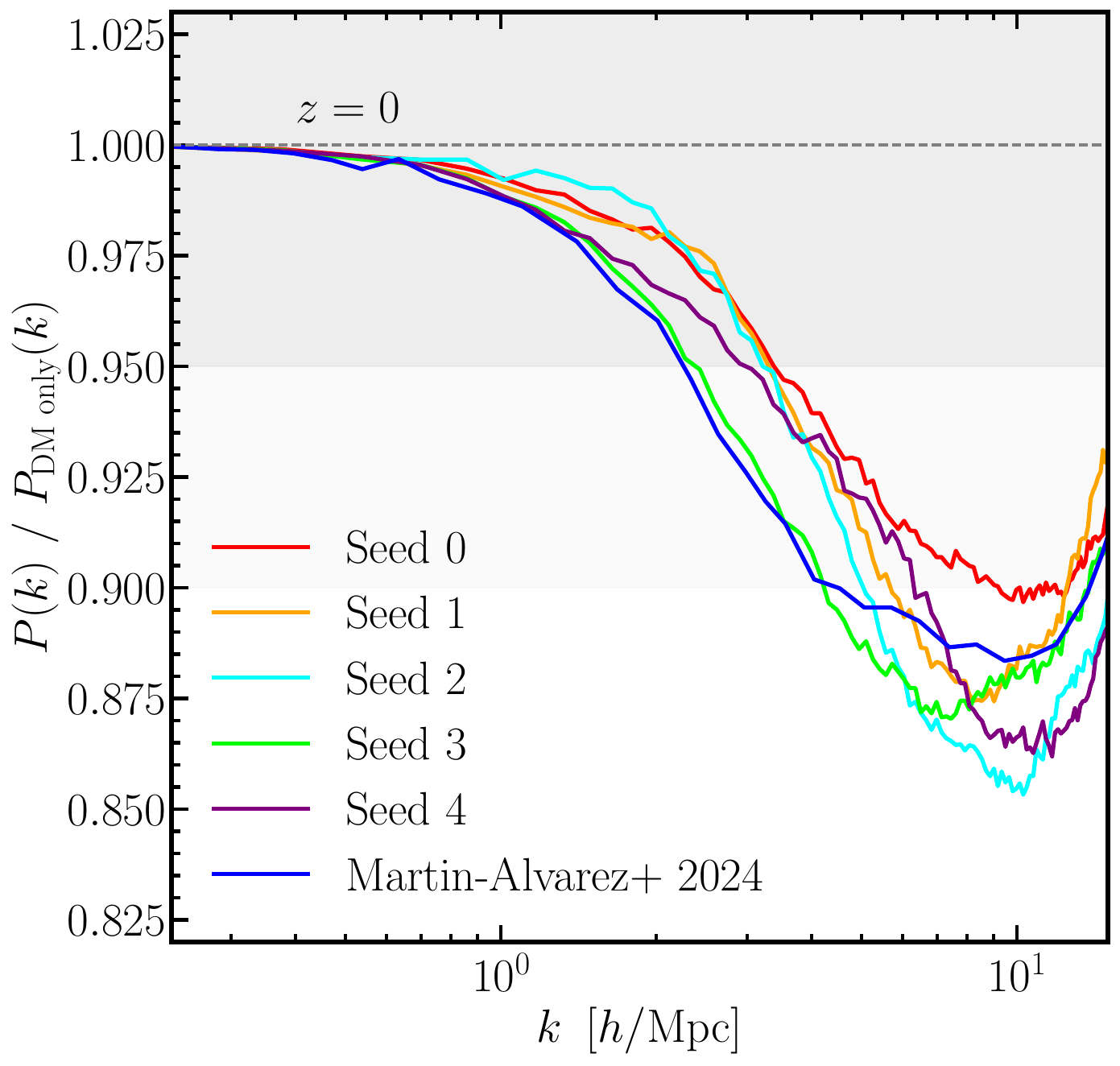}
\caption{The impact of cosmic variance on the $z=0$  matter power spectrum due to baryonic effects, $P(k)/P_{\mathrm{DM}}$, measured in $(40~h^{-1}\mathrm{Mpc})^3$ FABLE boxes.  The coloured lines show the measured $P(k)/P_{\mathrm{DM}}$ in $(40~h^{-1}\mathrm{Mpc})^3$ simulations ran with different values of random seed that
determines the initial Gaussian density field.  Seed 0 is the FABLE-40 box, ran with the random seed utilised for all boxes described in the main text.  We plot the suppression measured in the FABLE $(40~h^{-1}\mathrm{Mpc})^3$ box presented in \citet{Henden2018} and \citet{Martin-Alvarez2024}, which was also ran with a different random seed to that utilised in this work. }
\label{fig:cosmicvariance}
\end{figure}

In this Appendix, we investigate the effect of cosmic variance on the matter power spectrum suppression due to `baryonic feedback'.  We run five $(40~h^{-1}\mathrm{Mpc})^3$ simulations with the fiducial FABLE model, testing five different values of the random seed that determines the initial Gaussian density field (with `Seed 0' being the random seed utilised for all boxes described in the main text). 
 Figure~\ref{fig:cosmicvariance} demonstrates that the choice of random seed has a non-negligable impact on the matter spectrum suppression at $z=0$.  At $k=5\, h\, \mathrm{Mpc}^{-1}$ we see variations in the predicted suppression of $\sim 5\%$.  The maximum suppression varies by a similar amount, with the location of the maxima shifting marginally between boxes, lying in the range $k\sim7-10\, h\, \mathrm{Mpc}^{-1}$.  The difference in predictions between variations of the initial Gaussian density field results from the different numbers of the high-mass halos realised by $z=0$, exasperated by the limited box size.  Since AGN exhibiting the most extreme feedback live in these rare over-dense environments \citep{vandaalen:2020}, the suppression due to `baryonic feedback' effects is affected by cosmic variance due to the different numbers of powerful AGN realised.  Our results using the larger $(100~h^{-1}\mathrm{Mpc})^3$ boxes should be relatively robust to these effects, but even larger box sizes are desirable to fully test the convergence of the models to the box size \citep[see e.g.][]{vandaalen:2020, Pakmor2023, Schaller2024}.

\section{Full simulation suite}
Table~\ref{tab:simsall} lists the key parameter choices for the full suite of FABLE-like modified AGN feedback simulation boxes ran in this work.  

\begin{table*}
\setlength\extrarowheight{3pt}
\centering
\caption{The key AGN feedback parameters utilised in the full suite of FABLE-like $(40~h^{-1}\mathrm{Mpc})^3$ simulations ran for this work.  Parameters follow the definitions in Table~\ref{tab:params}, adding $M_{\rm BH, radio}$ as the black hole mass limit above which radio-mode feedback is allowed to occur (introduced in Section~\ref{sec:radioboostMbh}) and $E_{\rm{bub}}/E_{\rm{ICM}}$ as the limit on the energy content of radio-mode bubbles (introduced in Section~\ref{sec:xfable}). 
 A parameter that varies with redshift, $z$, as a step function is represented in the format $z < 2: 30$, $z > 2: 100$, where in this example $30$ is the value of the parameter for $z < 2$ and $100$ is its value for $z > 2$. A linear evolution in a parameter is represented in the format $z=0: 30$ $\rightarrow$ $z=4: 500$, where in this example the parameter is fixed to $500$ at $z>4$, and decreases linearly to $30$ by $z=0$.  \\
$^{a}$ Keeping other parameters the same, we also tested combinations of $i$) fixing $\alpha=100$ at all $z$,  $ii$) setting $\alpha$ to $1000$ for $z>2$, $iii$) fixing $\epsilon_f=0.1$ at all $z$ and $iv$) setting $\epsilon_f$ to $1$ for $z>2$. \\
$^{b}$Keeping other parameters the same, we also tested boxes with $i$) $\chi_{\mathrm{radio}}=0.1$, $\alpha=1000$, $\epsilon_f=1$ and $ii$) $\chi_{\mathrm{radio}}=0.1$, $R_{\mathrm{bub}}=20$. \\
$^{c}$Keeping other parameters the same, we tested combinations of $i$) $\chi_{\mathrm{radio}}=0.01, 0.05$, $ii$) a linear evolution of  $\epsilon_m$ as $z=0: 0.1$ $\rightarrow$ $z=4: 0.8$, $iii$) setting $D_{\mathrm{bub}}$ to 300 for $z>4$, $iv$) setting $D_{\mathrm{bub}}$ to $100$ at $z = 0$ and $v$) keeping $D_{\mathrm{bub}}$ fixed for $0<z<1$. \\
$^{d}$Keeping other parameters the same, we also explored $D_{\mathrm{bub}}= 30, 50, 150, 200$ and $M_{\rm BH, radio} = 0.01$. \\$^{e}$Keeping other parameters the same, we tested changing $\chi_{\mathrm{radio}}$ to $0.02$, and $D_{\mathrm{bub}}$ to $200, 500$.\\
$^{f}$We also explored $M_{\rm BH, radio} = 0.02$.\\ 
$^{g}$Keeping other parameters the same, we also tested $E_{\rm{bub}}/E_{\rm{ICM}}=2, 4, 6, 50, 100, 200, 500, 800$. \\
}
\begin{tabular}{cccccccccc}
\hline
\textbf{Simulation name} & $\chi_{\mathrm{radio}}$ & $\alpha$ & $\epsilon_f$ & $\Delta t$ & $\epsilon_{\rm m}$ & $D_{\mathrm{bub}}$  & $R_{\mathrm{bub}}$ & $M_{\rm BH, radio}$ & $E_{\rm{bub}}/E_{\rm{ICM}}$ \\
&&&&[Myr] &&[kpc/$h$]&[kpc/$h$] & [$10^{10}~\mathrm{M_{\odot}}/h$] &\\
\hline
\underline{\textbf{Fiducial box settings}}\\
\rowcolor[rgb]{0.8235, 0.6471, 0.8275}
FABLE & 0.01 & 100 & 0.1 & 25 & 0.8  & 30 & 50 &  - & - \\
\underline{\textbf{$z$-dependent quasar-mode}}\\
\rowcolor[rgb]{0.7961, 0.9451, 1.0000}
QuasarBoost$z$2 & 0.01 & $z<2: 100$$^{a}$, & $z<2: 0.1$$^{a}$, & 25 & 0.8  & 30 & 50 &  - & - \\
\rowcolor[rgb]{0.7961, 0.9451, 1.0000}
&&$z>2: 10^4$ & $z>2: 10$&&&&&&\\
\underline{\textbf{$z$-dependent $\epsilon_m$ in radio-mode}}\\
- & 0.01 & 100 & 0.1 & 25 & $z<2: 0.8$,  & 30 & 50 &  - & - \\
&&&&&$z>2: 8$\\
\underline{\textbf{$z$-dependent $D_{\mathrm{bub}}$ in radio-mode}}\\
- & 0.01$^{b}$ & 100$^{b}$ & 0.1$^{b}$ & 25 & 0.8 & $z<2: 30$,  & 50$^{b}$ &  - & - \\
&&&&&&$z>2: 100$\\
\rowcolor[rgb]{0.9961, 0.9686, 0.7961}
RadioBoost & 0.1$^{c}$ & 100 & 0.1 & 25 & 0.8$^{c}$ & $z=0: 30$ $\rightarrow$  & 50 &  - & - \\
\rowcolor[rgb]{0.9961, 0.9686, 0.7961}
&&&&&&$z=4: 500$$^{c}$&&&\\

\underline{\textbf{$M_{\mathrm{BH}}$-threshold for radio-mode}}\\
\rowcolor[rgb]{0.7961, 0.9961, 0.8000}
RadioBoost$M_{\rm BH,radio}$ & 0.1 & 100 & 0.1 & 25 & 0.8 & 100$^{d}$ & 50 &  0.06$^{d}$ & - \\
\underline{\textbf{further $E_{\rm{bub}}/E_{\rm{ICM}}$ limiter}}\\
\rowcolor[rgb]{0.6471, 0.7843, 0.6471}
XFABLE & 0.1$^{e}$ & 100 & 0.1 & 25 & 0.8 & 100$^{e}$ & 50 &  0.06$^{f}$ & 20$^{g}$ \\
%\underline{\textbf{$E_{\rm{bub}}/E_{\rm{IGM}}=20$ runs}}\\
%-& 0.1$^{f}$ & 100$^{f}$ & 0.1 & 25 & 0.8 & 100 & 50 &  0.06$^{g}$ & 20 \\
\hline
\end{tabular}\label{tab:simsall}
\end{table*}

%%%%%%%%%%%%%%%%%%%%%%%%%%%%%%%%%%%%%%%%%%%%%%%%%%

% Don't change these lines
\bsp	% typesetting comment
\label{lastpage}
\end{document}